\documentclass[a4paper,11pt]{article}

\usepackage{graphicx}
\usepackage{float}
\usepackage[T1]{fontenc}
\usepackage{epsfig}
\usepackage{color}
\usepackage{amsmath,jheppub,mathtools}
\usepackage[margin=2.0cm]{geometry}
\usepackage{subfloat}
\usepackage{amsfonts}
\usepackage{braket}
\usepackage{cleveref}
\usepackage{epstopdf}
\usepackage{caption}
\usepackage{comment}
\usepackage{subcaption}
\usepackage{enumitem}
\usepackage{array}
\usepackage{hyperref} 
\usepackage[numbers]{natbib}
\usepackage[titletoc,toc,title]{appendix}
\usepackage{hyperref}
\hypersetup{
	colorlinks=true,
	linkcolor=blue,
	filecolor=red,      
	urlcolor=blue,
	citecolor=blue
} 

\Crefname{figure}{Fig.}{Figs.}

\title{\boldmath Probing the Hierarchy of Genuine Multipartite Entanglement with Generalized Latent Entropy}
\author[a]{Byoungjoon Ahn,}
\author[c]{Jaydeep Kumar Basak,}
\author[c,d]{Keun-Young Kim,}
\author[a]{Gwon Bin Koo,}
\author[a,b]{Vinay Malvimat,}
\author[a,b,e]{Junggi Yoon}

\affiliation[a]{Department of Physics, College of Science, Kyung Hee University, Seoul 02447, Republic of Korea}
\affiliation[b]{Research Institute for Basic Sciences, Kyung Hee University, Seoul 02447, Republic of Korea}
\affiliation[c]{Department of Physics and Photon Science, Gwangju Institute of Science and Technology, \\
123 Cheomdan-gwagiro, Gwangju 61005, Korea}
\affiliation[d]{Research Center for Photon Science Technology, Gwangju Institute of Science and Technology, \\
123 Cheomdan-gwagiro, Gwangju 61005, Korea}
\affiliation[e]{International Center for Quantum Matter, Kyung Hee University, Seoul 02447, Republic of Korea}

\abstract{
We introduce generalization of the recently proposed \textit{ Latent Entropy} (L-entropy) \cite{Basak:2024uwc} as a refined measure of genuine multipartite entanglement (GME) in pure states of $n$-party quantum systems. Generalized L-entropy provides a natural ordering among $k$-uniform states maximizing for absolutely maximally entangled states (AME), effectively capturing the hierarchical structure of multipartite entanglement. We analyze the behavior of this measure for $n$-party Haar-random states and demonstrate that, in the large local-dimension limit, the maximal L-entropy saturates its upper bound for odd $n$, while for even $n$ it approaches the bound asymptotically. Furthermore, we apply this framework to examine multipartite entanglement properties of quantum states in several variants of the Sachdev--Ye--Kitaev (SYK) model, including SYK$_4$, SYK$_2$, mass-deformed SYK, sparse SYK, and $\mathcal{N}=2$ supersymmetric SYK model. The results demonstrate that the generalized L-entropy serves as a sensitive probe of multipartite entanglement, revealing how deformations influence quantum entanglement structure in such strongly interacting systems.}

\emailAdd{bjahn123@khu.ac.kr}
\emailAdd{jkb.hep@gmail.com}
\emailAdd{fortoe@gist.ac.kr}
\emailAdd{gugwonbin@gmail.com}
\emailAdd{vinaymalvimat@khu.ac.kr}
\emailAdd{junggi.yoon@khu.ac.kr}

\date{}

\begin{document}
	
	\maketitle

\pagebreak
\section{Introduction}\label{sec_intro}

Multipartite entanglement represents a captivating frontier in quantum research, steadily gaining prominence across diverse domains of physics. It not only offers profound insights into the foundations of quantum gravity and holography \cite{Susskind:2014yaa,Akers:2019gcv,Hayden:2021gno,Gadde:2022cqi}, but also underpins emerging technologies such as quantum batteries, quantum teleportation, quantum communication and quantum secret sharing \cite{Ma:2023ecg,Horodecki:2024bgc,Shi:2025gyu,Choi:2022lge,Hillery:1998yq,Cleve:1999qg,briegel2009measurement,Wong:2022mnv}. Its dual significance—bridging fundamental theory and practical realization—highlights its central role in advancing our understanding of quantum information science.

The notion of multipartite entanglement exhibits rich structure and is rather intricate. The archetypal examples: GHZ and W states which cannot be transformed into each other through any local operations and classical communication, regardless of whether deterministic or probabilistic strategies are employed \cite{Dur:2000zz}. This LOCC-incomparability underscores their distinct entanglement characteristics. 
In contrast to entanglement entropy, which serves as a canonical measure for bipartite entanglement, there is currently no universally accepted quantifier for multipartite entanglement. Furthermore, numerous notions and subtle intricacies exist regarding the very definition of genuine multipartite entanglement, making it an active and evolving area of research \cite{Vidal:1998re,Ma:2023ecg,Horodecki:2024bgc,Gadde:2024jfi}. In this work, we adopt the most widely accepted definition in quantum information theory wherein a state is considered genuinely multipartite entangled if it is not separable with respect to any bipartition \cite{Plenio:2007zz,Ma:2023ecg}. Any well-defined measure of genuine multipartite entanglement must satisfy three fundamental axioms established by the quantum information community:

\begin{itemize}
\item[($a$)] It must vanish for all biseparable states and take a strictly positive value for genuinely multipartite entangled states.
\item[($b$)] It must be invariant under local unitary (LU) transformations, reflecting the physical equivalence of locally rotated states.
\item[($c$)] It must be non-increasing on average under local operations and classical communication (LOCC), ensuring that genuine entanglement cannot be increased by local means.
\end{itemize}

Measures satisfying axioms (a)-(c) are referred to as genuine multipartite entanglement (GME) measures. However, recent work has identified a further operational requirement: any physically meaningful GME measure should reflect the relative utility of different entangled states in quantum information tasks \cite{PhysRevLett.127.040403}. Specifically, since GHZ states demonstrably outperform W states in applications like quantum teleportation, a proper GME measure must assign higher values to GHZ than to W states \cite{PhysRevLett.127.040403}. Measures meeting this additional criterion alongside axioms (a)-(c) are called proper GME measures. Ideally, a good quantifier of genuine multipartite entanglement should satisfy all of the properties ($a$)-($c$). However, it is not necessary for every useful measure or quantifier to obey all of them. For instance, quantities such as the tangle or the Markov gap may fail to detect certain classes of genuinely multipartite entangled states, so that they do not satisfy strict positivity on all GME states, but only positivity on the subclass of states they detect.  In particular, quantities that do not satisfy full LOCC monotonicity can still be operationally meaningful. At a minimum, one would require such a quantity to vanish for all biseparable states, to be non-negative, and to be invariant under local unitaries. These conditions ensure that it detects genuine multipartite entanglement, assigns no unphysical negative values, and is independent of local basis choices. In this paper, however, we will not distinguish between quantities satisfying all of ($a$)--($c$) and those satisfying only these minimal requirements, and for simplicity we will refer to both as GME measures whenever no confusion can arise.

Although several measures of multipartite entanglement have been proposed in the literature for finite-dimensional systems, most are limited by the number of parties or the local Hilbert space dimension \cite{Coffman:1999jd,Sabin:2008naa,Ma:2011yon,PhysRevA.81.012308,PhysRevLett.127.040403,PhysRevA.83.062325,Xie:2023dzw}. Only a few have been able to overcome these constraints and extend to systems with infinite degrees of freedom, such as quantum field theories (QFTs). Even among these, many remain restricted to Gaussian states and free field theories. In recent years, there has been a growing interest in exploring multipartite entanglement within the framework of holography. However, most of these quantities, do not obey above described axioms established in quantum information theory, and hence primarily serve as signals or witnesses of multipartite entanglement rather than bona fide measures (essentially means that they could be zero for some classes of entangled states) \cite{Harper:2024ker,Gadde:2023zzj,Gadde:2024taa,Gadde:2022cqi,Bao:2025psl,Balasubramanian:2025hxg,Balasubramanian:2024ysu}. 

In this context, some of the authors of the present work have sought to bridge the gap between axiomatic quantum information definitions and holographic constructions by proposing a computable measure that satisfies two of the above criteria and admits a natural holographic interpretation \cite{Basak:2024uwc}. This measure, termed \emph{latent entropy}, is constructed by exploring the upper bound of the reflected entropy \cite{Dutta:2019gen} obtained by purification of all 2-party systems, and has been shown to satisfy the aforementioned criteria for systems with up to $n=5$ parties. It was further demonstrated that this measure attains its maximum value for \emph{2-uniform states}, which are $n$-party pure states whose all two-party reduced density matrices are maximally mixed . In addition, it was shown that random five-party states exhibit maximal entanglement at leading order in an expansion in the local Hilbert space dimension. The measure discussed above, however, is limited to systems with up to five parties and cannot distinguish between $k=2$-uniform states and those with $k>2$. Furthermore, the maximal degree of uniformity achievable in an $n$-party system is $k_{\mathrm{max}} = \lfloor n/2 \rfloor$. The corresponding states, in which every reduction to $k_{\mathrm{max}}$ or fewer parties is maximally mixed, are known as absolutely maximally entangled (AME) states (see \cite{GISIN19981,HIGUCHI2000213,Arnaud_2013,Enriquez_2016,PhysRevA.86.052335} for detailed discussions on AME states as the maximally entangled multipartite states). 

In this work, we address both of the above mentioned limitations by proposing a generalization of the Latent Entropy to arbitrary $n$-party systems. The construction is based on examining the upper bound of the reflected entropy obtained through the purification of all $k$-party reduced density matrices, where $2 \leq k \leq \text{Max}\left[2,\lfloor n/2\rfloor\right] $. We demonstrate that, for a given $n$-party system, the generalized L-entropy, $\tilde{L}_{\mathrm{gen}}$,  attains its maximum value at $k = k_{\mathrm{max}} = \text{max}\left[2,\lfloor n/2\rfloor\right]$, corresponding to the absolutely maximally entangled (AME) states, as expected from quantum information theory. For systems with up to five parties, it reduces to the original L-entropy measure proposed in \cite{Basak:2024uwc}. Furthermore, it provides a natural ordering among lower $k$-uniform states in an $n$-party system, namely,
\[
\tilde{L}_{\mathrm{gen}}(\ket{\psi_k}) > \tilde{L}_{\mathrm{gen}}(\ket{\psi_{k-1}}),
\]
where $\ket{\psi_k}$ and $\ket{\psi_{k-1}}$ denote  $k$- and $(k-1)$-uniform states, respectively. We compile a comprehensive compendium of generalized L-entropy values for known $k$-uniform multi-qubit states, and demonstrate that all of them exhibit the expected ordering of generalized L-entropy. Even among states with the same degree of $k$-uniformity, we provide a possible explanation for the observed differences in L-entropy based on their proximity to $(k+1)$-uniform states.

We then investigate the behavior of the measure for Haar random states, making use of the reflected entropy results obtained in \cite{Akers:2021pvd}. Our analysis reveals a distinct difference between systems with an odd and an even number of parties. Specifically, when $n$ is odd, random states tend to be $\lfloor n/2 \rfloor$-uniform—corresponding to AME states in the large local-dimension limit. In contrast, when $n$ is even, the states typically exhibit $(n/2 - 1)$-uniformity and therefore do not qualify as AME states.

Subsequently, we compute the generalized L-entropy in various SYK model variants. Our numerical analysis shows that the saturation value of the $SYK_4$ model approaches the generalized L-entropy of AME states as $N$ increases, whereas the saturation value of $SYK_2$ remains consistently lower and appears to asymptote to a smaller constant. We further observe that the gap between the average saturation values of the two models increases with $N$ and possibly saturates at a finite value. We then extend our study to the mass-deformed SYK \cite{Banerjee:2016ncu,Garcia-Garcia:2017bkg,Nosaka:2018iat} and sparse SYK models \cite{Xu:2020shn,Orman:2024mpw} to examine the late-time growth and saturation behavior of the generalized L-entropy as the deformation and sparseness parameters are varied, respectively. In the mass-deformed case, the saturation values switch to that of $SYK_4$ as soon as the deformation parameter moves away from purely $SYK_2$ term. Furthermore, the saturation time and late-time approach depend sensitively on the deformation parameter. In contrast, for the sparse SYK model, we observe a smooth crossover of the saturation values of the generalized L-entropy as the sparseness is varied, accompanied by a qualitative change in the distribution of the generalized L-entropy across the eigenstates of the Hamiltonian. To further examine the role of supersymmetry in multipartite correlations, we perform analogous computations in the $\mathcal{N}=2$ supersymmetric SYK model \cite{Fu:2016vas} to understand the multipartite entanglement structure of fortuitous states~\cite{Chang:2024lxt}. The results demonstrate that the fortuitous states in the central $R$-charge sector exhibits a strong multipartite entanglement structure compared to the $Q$-exact states, and it is close to that of typical states in the same $R$-charge sector.

The organization of this paper is as follows. In \Cref{sec2}, we introduce the generalized Latent Entropy and discuss the appropriate normalization scheme. Section \ref{sec3} analyzes the properties of this measure, demonstrating that it satisfies the required axioms of a genuine multipartite entanglement measure and is sensitive to $k$-uniformity. In \Cref{sec4}, we present the computation of the generalized Latent Entropy for Haar-random states with varying numbers of parties and show that systems with an odd total number of parties tend to approach AME-like behavior. Section \ref{sec5} explores the behavior of L-entropy in the SYK model and its variants, including the mass-deformed, sparse and $\mathcal{N}=2$ SUSY SYK models. Finally, in \Cref{sec6} we summarize our findings and discuss possible future directions.

\section{\textbf{Generalized Latent Entropy} }\label{sec2}

The definition of the generalized L-entropy utilizes another quantity, reflected entropy, in the context of mixed states. In the bipartite setting, reflected entropy has proven to be a powerful probe, with a natural purification procedure that canonically doubles the Hilbert space. For a bipartite mixed state $\ket{\psi}_{AB}$, one can canonically purify the state as,
\begin{align}
   \rho_{AB} \;\longrightarrow\; \ket{\sqrt{\rho_{AB}}}_{ABA^*B^*}
\end{align}
where $A^*$ and $B^*$ are copies of $A$ and $B$, respectively. An example of such a purification is the TFD state. Now, the reflected entropy is defined as the von Neumann entropy of $AA^*$.
\begin{align}
   S_R(A:B)=S(AA^*) \;\leq\; 2 \min\{S(A),S(B)\}.
\end{align}
The reflected entropy satisfies the following bounds,
\begin{align}\label{refbd}
    2 \min\{S(A),S(B)\}\;\geq\;S_R(A:B) \;\geq\; I(A:B),
\end{align}
where $I(A:B)$ is the mutual information which measures the total correlation between $A$ and $B$. In \Cref{refbd}, the lower bound can be derived from the strong subadditivity property applied to the subsystems $A$, $A^\ast$, and $B$, while the upper bound follows from the subadditivity relations among the pairs of subsystems $(A,~A^\ast)$ and $(B,~B^\ast)$. Although the reflected entropy has been extensively studied in the contexts of quantum field theory and holography, it has been established that this quantity does not serve as a measure of either entanglement or correlation. Nevertheless, the bounds of the reflected entropy possess rich physical significance. The lower bound has been utilized to define a multipartite entanglement measure known as the Markov gap, which, however, fails to detect GHZ-type correlations, thereby disqualifying it as a faithful measure of genuine multipartite entanglement (GME). In contrast, the upper bound gives rise to a more robust GME measure—the Latent entropy—which provides a unifying framework connecting diverse domains of physics, including quantum information theory, quantum many-body systems, and high-energy physics. Furthermore, one can construct the generalized latent entropy, $L_{gen}$, as a hierarchical extension of the original L-entropy, allowing a systematic characterization of multipartite entanglement across arbitrary subsystems. To make the formalism transparent, we next present the explicit construction of the generalized L-entropy, $\tilde{L}_{gen}$, highlighting its recursive structure.

\begin{enumerate}
    \item \quad We begin with an $n$-party pure state and compute all of its reduced density matrices $\rho_k$ for
    \[
    2 \leq k \leq \max\!\left(2,\left\lfloor \frac{n}{2}\right\rfloor\right).
    \]
    Throughout this paper, we denote
    \begin{align}\label{kmax}
        k_{\max}=\max\!\left(2,\left\lfloor \frac{n}{2}\right\rfloor\right).
    \end{align}
    
    For each fixed $k$, there are $\binom{n}{k}$ such reduced density matrices, corresponding to the distinct choices of $k$ parties. We label a particular choice of subsystem by a superscript $(p)$ and denote the associated reduced density matrix by $\rho_k^{(p)}$.

    \item \quad For each reduced density matrix $\rho_k^{(p)}$, we construct its canonical purification,
    \[
    \rho_{A_1\cdots A_k}^{(p)}
    \;\longrightarrow\;
    \ket{\sqrt{\rho_{A_1\cdots A_k}^{(p)}}}_{A_1\cdots A_k A_1^*\cdots A_k^*}.
    \]

    \item  \quad Next, for each $\rho_k^{(p)}$, we compute all independent reflected entropies associated with bipartitions of the $k$ parties into two complementary subsets, one containing $\lfloor k/2\rfloor$ parties and the other containing the remaining $k-\lfloor k/2\rfloor$ parties. We refer to such bipartitions as \emph{balanced bipartitions}. For each balanced bipartition, indexed by $i$, we define
    \begin{align}\label{genell}
        \ell_{k,i}^{(p)}
        =
        2\min\{S(\mathcal{A}),S(\mathcal{B})\}
        -
        S_R(\mathcal{A}:\mathcal{B}),
    \end{align}
    where $\mathcal{A}$ and $\mathcal{B}$ form a balanced bipartition of the $k$ parties, with
    \[
    |\mathcal{A}|=\left\lfloor \frac{k}{2}\right\rfloor,
    \qquad
    |\mathcal{B}|=k-\left\lfloor \frac{k}{2}\right\rfloor.
    \]
    For instance, one may take
    \[
    \mathcal{A}=A_1A_2\cdots A_{\lfloor k/2\rfloor},
    \qquad
    \mathcal{B}=A_{\lfloor k/2\rfloor+1}\cdots A_k.
    \]
    These balanced bipartitions are singled out because they determine the tightest dimension-dependent upper bounds on the reflected entropy.

    \item \quad For a fixed reduced state $\rho_k^{(p)}$, we then define
    \[
    L_k^{(p)}=
    \left(\prod_i \ell_{k,i}^{(p)}\right)^{1/k_{\mathrm{tot}}},
    \]
    where $k_{\mathrm{tot}}$ denotes the total number of independent balanced bipartitions:
\begin{align}\label{ktot}
k_{\mathrm{tot}} &=
\begin{cases}
\dfrac{1}{2}\binom{k}{k/2}, & \text{if } k \text{ is even}, \\[8pt]
\binom{k}{\lfloor k/2\rfloor}, & \text{if } k \text{ is odd}.
\end{cases}
\end{align}
    Since each $\ell_{k,i}^{(p)}$ is bounded above by the corresponding dimension-dependent maximum,
    \begin{align}
        \ell_{k,i}^{(p)}
        \leq
        2\min\!\big[\log d_{\lfloor k/2\rfloor},\,\log d_{k-\lfloor k/2\rfloor}\big],
    \end{align}
    where $d_m$ is the Hilbert-space dimension of an $m$-party subsystem. In the case of identical local dimension $d$, this reduces to
    \begin{align}
        \ell_{k,i}^{(p)}
        \leq
        2\left\lfloor \frac{k}{2}\right\rfloor \log d.
    \end{align}
    Consequently, the same bound applies to $L_k^{(p)}$:
    \begin{align}
        L_k^{(p)}
        \leq
        2\left\lfloor \frac{k}{2}\right\rfloor \log d.
    \end{align}

    \item \quad Since there are
    \[
    n_k=\binom{n}{k}
    \]
    distinct $k$-party reductions, we define the \emph{$k$-partite $L$-entropy} as the geometric mean over all such reductions:
    \begin{align}
        L_k
        =
        \left(\prod_{p=1}^{n_k} L_k^{(p)}\right)^{1/n_k}.
    \end{align}
    It follows that
    \begin{align}
        L_k
        \leq
        2\left\lfloor \frac{k}{2}\right\rfloor \log d.
    \end{align}
    We therefore introduce the normalized quantity
    \begin{align}
        \tilde{L}_k
        =
        \frac{L_k}{2\lfloor k/2\rfloor \log d}
        \leq 1.
    \end{align}

    \item \quad Finally, we define the generalized $L$-entropy by combining all normalized $\tilde{L}_k$:
    \begin{align}\label{lgen2}
        \tilde{L}_{\mathrm{gen}}(A_1\cdots A_n)
        =
        \left(\prod_{k=2}^{k_{\max}} \tilde{L}_k\right)^{1/k_{\max}}.
    \end{align}
    Since each $\tilde{L}_k\leq 1$, the generalized $L$-entropy is also bounded above by 1. Moreover, because $\tilde{L}_{\mathrm{gen}}$ incorporates all of the quantities $\ell_{k,i}^{(p)}$, attaining its maximal value requires that all balanced bipartite $L$-entropies simultaneously saturate their respective upper bounds. This naturally singles out a distinguished class of states, namely absolutely maximally entangled (AME) states, which we discuss in the next section.
\end{enumerate}

\subsection*{Reduction}

For $n=3,4,5$, one has $k_{\max}=2$ (see \cref{kmax}). Thus, in these cases, the construction involves only the two-party reduced density matrices. Moreover, for $k=2$ there is only a single balanced bipartition, so that $k_{\mathrm{tot}}=1$ according to \cref{ktot}. As a result, the index $i$ in \cref{genell} becomes redundant, and we simply have
\begin{align}
    L_2^{(p)}=\ell_2^{(p)}.
\end{align}
It then follows that
\begin{align}
    L_2=\left(\prod_{p=1}^{\binom{n}{2}} \ell_2^{(p)}\right)^{1/\binom{n}{2}}.
\end{align}
Therefore, for $n=3,4,5$, the generalized $L$-entropy coincides exactly with the $L$-entropy proposed in \cite{Basak:2024uwc}.
\section{\textbf{Properties of generalized L-entropy ($\tilde{L}_{gen}$)}}\label{sec3}

The construction of the generalized L-entropy, $\tilde{L}_{gen}$, guarantees the fulfillment of the three essential conditions for a genuine multipartite entanglement (GME) measure, as outlined in the introduction. Besides, generalized L-entropy shows various other properties which are explained here. 

\subsection*{\textbf{A GME measure}}

To verify forward direction of condition $(a)$ stated in the Introduction, let us consider an $n$-party pure state that is separable across some bipartition. Then, for at least one choice of subsystem size $k \leq k_{\text{max}}$, there exists a reduced density matrix $\rho_k^{(p)}$ that is itself pure. For such a pure $k$-party state, any balanced bipartition into $\mathcal{A}$ and $\mathcal{B}$ behaves as an ordinary bipartite pure-state split. As a result, the corresponding reflected entropy saturates its upper bound $S_R(\mathcal{A}:\mathcal{B})=2S(\mathcal{A})=S(\mathcal{B})$. Therefore,
\[
   \ell_{k,i}^{(p)}=0 \quad\Rightarrow\quad L_{k}^{(p)}=0 \quad\Rightarrow\quad L_k=0 \quad\Rightarrow\quad \tilde{L}_{\text{gen}}=0.
\]
The only exception where $\rho_k^{(p)}$ is not pure is the when the full state possesses biseparability of the form for biseparable states of the form
\begin{align}
|\phi\rangle_{A_1}\otimes |\chi\rangle_{A_2\cdots A_n},
\end{align}
or, more generally, whenever one side of the bipartition consists of a single party. In this case, the above argument in terms of pure $k$-party reductions does not directly apply. However, for any $j\neq 1$, the two-party reduced density matrix factorizes as
\begin{align}
\rho_{A_1A_j}=\rho_{A_1}^{\textrm{pure}}\otimes \rho_{A_j},
\end{align}
which implies
\begin{align}
S_R(A_1:A_j)=0 \qquad S(A_1)=0.
\end{align}
Hence the corresponding two-party contribution vanishes, so that once again
\begin{align}
L_2=0 \qquad\Rightarrow\qquad \tilde{L}_{\text{gen}}=0.
\end{align}
Thus, every biseparable pure state has vanishing generalized $L$-entropy.

Note that the present analysis establishes only one direction of the implication: a fully separable state necessarily yields a vanishing balanced bipartite latent entropy, whereas the converse has not been analytically demonstrated. However, through extensive computations of the generalized latent entropy for multipartite states with different numbers of parties and local dimensions, we did not find any instance of a non-separable state exhibiting a vanishing generalized latent entropy. 

The verification of condition ($b$) requires applying local unitaries to the purified balanced bipartite reduced density matrix as
\begin{align}
\ket{\sqrt{\rho_{\mathcal{AB}}'}} = U_{\mathcal{A}} \otimes U_{\mathcal{B}} \otimes U_{\mathcal{A}^*} \otimes U_{\mathcal{B}^*} \ket{\sqrt{\rho_{\mathcal{AB}}}},
\end{align}
where $U_{\mathcal{A}^*}$ and $U_{\mathcal{B}^*}$ denote the copies of $U_{\mathcal{A}}$ and $U_{\mathcal{B}}$, respectively. Since the reflected entropy is defined as the von Neumann entropy of the subsystem $\mathcal{A A^*}$, and the von Neumann entropy is invariant under such transformations, the reflected entropy—and hence the balanced bipartite latent entropy defined from it—remains unaffected by these operations. The unitary operator on $\mathcal{A}$ may itself decompose into a tensor product of local unitaries acting on its constituent subsystems,
\begin{align}
U_{\mathcal{A}} = U_{A_1} \otimes U_{A_2} \otimes \cdots \otimes U_{A_{\lfloor \frac{k}{2} \rfloor}},
\end{align}
where each $U_{A_i}$ acts locally on $A_i$. This structure is precisely the one required for local unitaries acting on individual parties to satisfy condition ($b$).  The invariance under local unitaries acting on individual parties can be viewed as a special case of invariance under local unitaries acting on the composite subsystems $\mathcal{A}$ and $\mathcal{B}$ and the one common to all of them. The invariance established above therefore guarantees that the balanced latent entropies remain invariant under such local unitary transformations. 

\begin{figure}[h]
    \centering
    \begin{subfigure}[t]{0.32\textwidth}
        \centering
        \includegraphics[width=\textwidth]{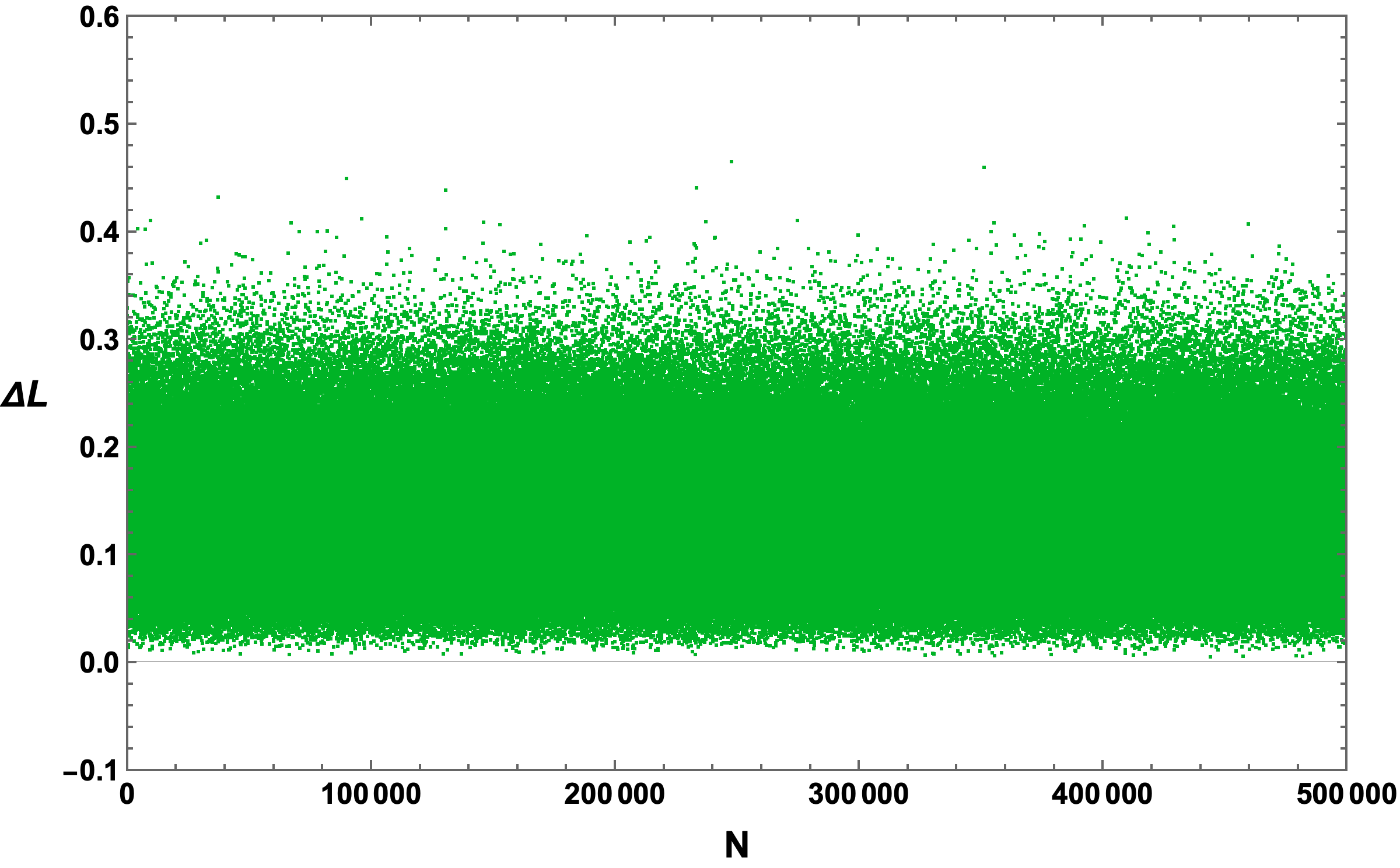}
        \caption{Six-qubits.}
        \label{fig:6p_locc}
    \end{subfigure}
    \hfill
    \begin{subfigure}[t]{0.32\textwidth}
        \centering
        \includegraphics[width=\textwidth]{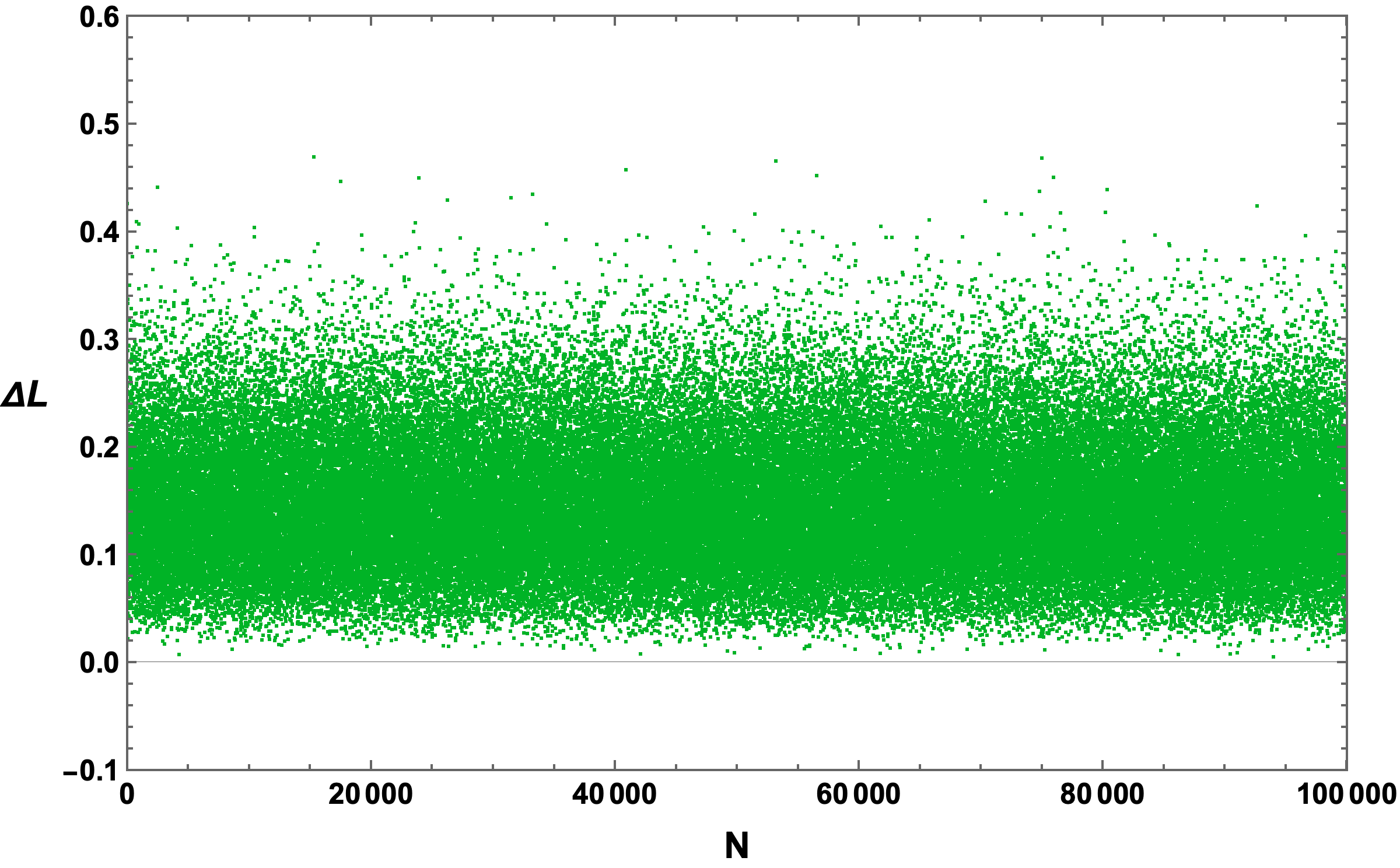}
        \caption{Seven-qubits.}
        \label{fig:7p_locc}
    \end{subfigure}
    \hfill
    \begin{subfigure}[t]{0.32\textwidth}
        \centering
        \includegraphics[width=\textwidth]{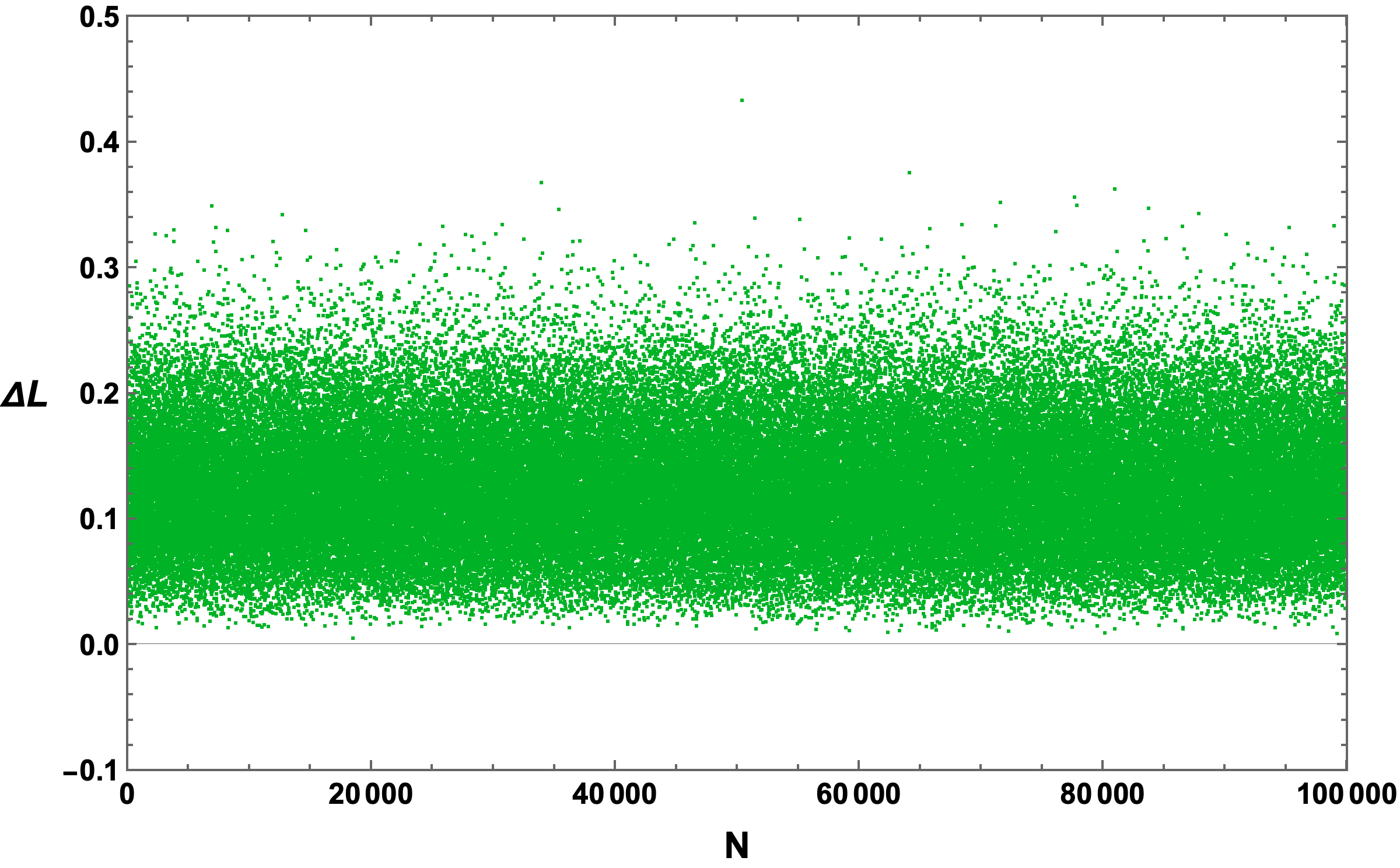}
        \caption{Eight-qubits }
        \label{fig:8p_locc}
    \end{subfigure}
    \caption{Numerical examination of the behavior of $L$-entropy under local random set of Kraus operations for random six-, seven-, and eight-qubit states random states. Across the samples considered, $\Delta L$ remains non-negative, and no decrease of $L$-entropy was observed.}
    \label{fig:locc_678}
\end{figure}

\begin{figure}[h]
    \centering
    \begin{subfigure}[t]{0.3\textwidth}
        \centering
        \includegraphics[width=\textwidth]{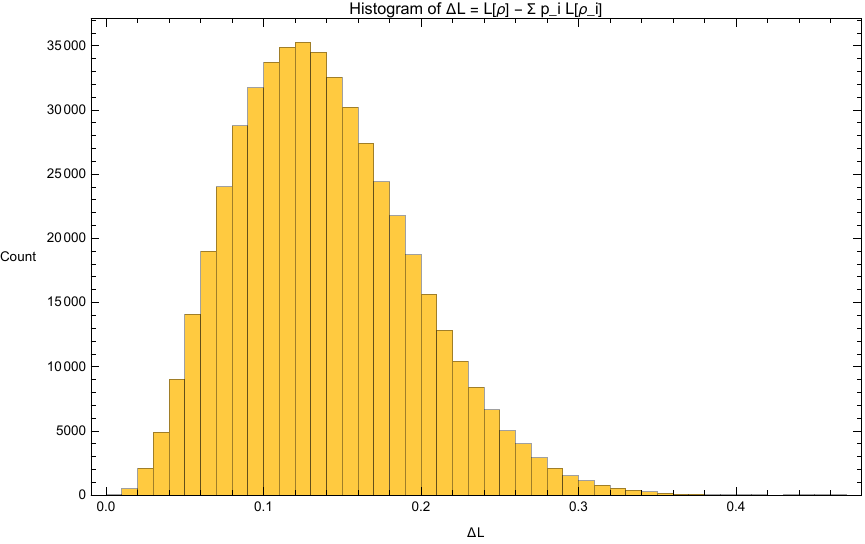}
        \caption{ Six-qubits }
        \label{fig:6p_hist}
    \end{subfigure}
    \hfill
    \begin{subfigure}[t]{0.3\textwidth}
        \centering
        \includegraphics[width=\textwidth]{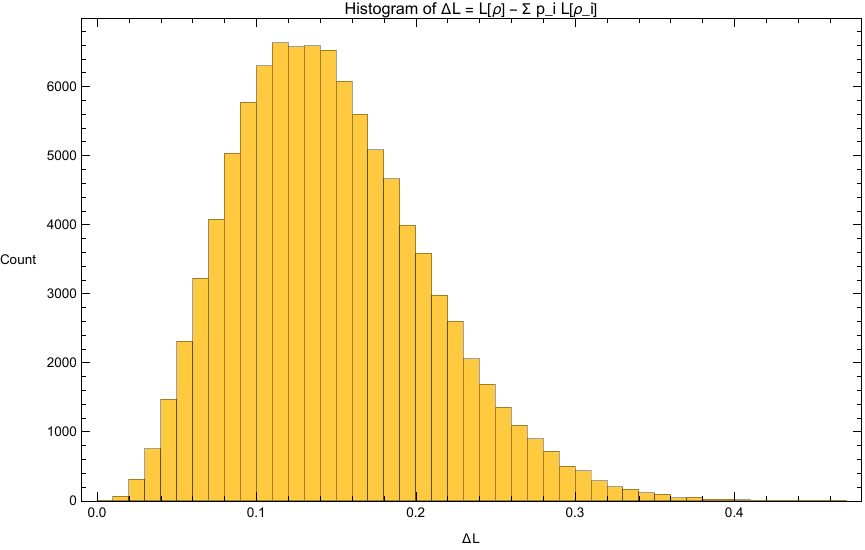}
        \caption{ Seven-qubits.}
        \label{fig:7p_hist}
    \end{subfigure}
    \hfill
    \begin{subfigure}[t]{0.3\textwidth}
        \centering
        \includegraphics[width=\textwidth]{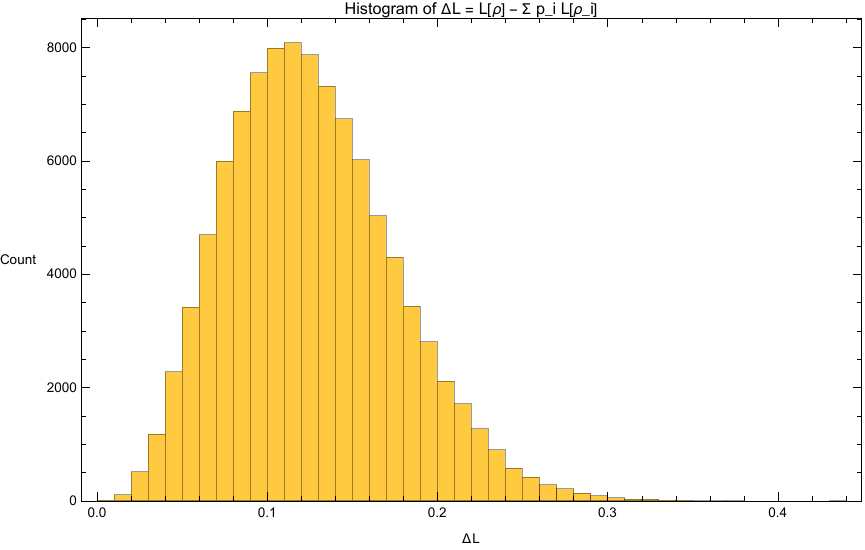}
        \caption{Eight-qubits.}
        \label{fig:8p_hist}
    \end{subfigure}
    \caption{Histograms of $\Delta L$ obtained from random six-, seven-, and eight-qubit states after local Kraus operations. The sampled data show no negative tail in $\Delta L$, providing numerical evidence for the non-decreasing behavior of $L$-entropy in these cases.}
    \label{fig:hist_678}
\end{figure}

As discussed in \cite{Basak:2024uwc}, proving the monotonicity of the $L$-entropy is hindered by the nonlinearity of the canonical purification map. Because of this nonlinearity, the canonical purification of a post-measurement state cannot generally be represented as the result of a local operation acting on the canonical purification of the original state. This obstruction persists in the present generalization as well. Due to the obstruction discussed above, an analytic proof of monotonicity is not presently tractable, and we therefore resort to numerical tests for when each party is a single qubit. As shown in Figs.~\ref{fig:locc_678} and \ref{fig:hist_678}, the sampled six, seven, and eight-qubit random states show no violations under a complete set of random local Kraus operators, providing numerical evidence that the $L$-entropy remains non-decreasing in these cases.

\subsection*{\textbf{Sensitivity to $k$-uniform states}}

An important feature of the generalized L-entropy is its sensitivity to the degree of uniformity of multipartite entangled states. Note that an $n$-party pure state is called $k$-uniform if all reductions to $k$ parties are maximally mixed. It can also be shown that a $k$-uniform state is necessarily an $m$-uniform state for any $m < k$. This hierarchical property implies that as $k$ increases, the state exhibits stronger and more evenly distributed multipartite correlations across all subsystems.
Consequently, the generalized L-entropy provides a quantitative signature of this hierarchy: for states with higher uniformity $k$, the constituent quantities $L_m$ with $m \le k$ attain their maximal values, while those with $m > k$ fall strictly below their maxima. This produces a monotonic ordering of the form
\begin{equation}
   \tilde{L}_{\text{gen}}^{(k+1\text{-uniform})} > \tilde{L}_{\text{gen}}^{(k\text{-uniform})},
\end{equation}
indicating that $\tilde{L}_{\text{gen}}$ increases systematically with the degree of uniformity. In particular, absolutely maximally entangled (AME) states---which are $\lfloor n/2 \rfloor$-uniform---achieve the upper bound $\tilde{L}_{\text{gen}} = 1$ and thus occupy the top of this entanglement hierarchy. Hence, the generalized L-entropy not only quantifies multipartite correlations but also consistently distinguishes between different classes of $k$-uniform states.

To substantiate the above theoretical discussion, we evaluate $\tilde{L}_{\text{gen}}$ for representative $n$-qubit states. As the number of qubits increases, it generally becomes difficult to classify the entanglement structure of quantum states into distinct families. Among these, we compare the generalized L-entropy for two of the most representative classes: the GHZ state and the W state. From the perspective of $k$-uniformity, the GHZ state is 1-uniform, while the W state is non-uniform. Accordingly, the computed L-entropy always yields a higher value for the GHZ state than for the W state (see Table~\ref{tab:entropy-GHZW}):
\begin{equation}
   \tilde{L}_{\text{gen}}^{(\mathrm{GHZ})} > \tilde{L}_{\text{gen}}^{(\mathrm{W})},
\end{equation}
demonstrating that $\tilde{L}_{\text{gen}}$ assigns a larger value to the GHZ state, which possesses stronger global correlations across all qubits, in agreement with the expected entanglement hierarchy.

\begin{table}[h]
\centering
\begin{tabular}{|c|c|c|c|}
\hline
Number of  qubits & GHZ state & W state& Cluster State \\
\hline
 3 & 0.5 & 0.1743 & 0.5 \\
 4  & 0.5& 0.2104 &0.7937\\
 5 & 0.5 & 0.2139 & 1\\
 6  & 0.5& 0.1441 &0.8705\\
 7  & 0.5& 0.1424 &0.9330\\
 8  & 0.3969& 0.1263 &0.8994\\
 9  & 0.3969& 0.1240 &0.9273\\
 10  & 0.3536& 0.1052 &0.8828\\
\hline
\end{tabular}
\caption{Generalized L-entropy values for GHZ, W \& Cluster states .}
\label{tab:entropy-GHZW}
\end{table}

\subsection*{Absolutely maximally entangled (AME) states}

Absolutely maximally entangled (AME) states are defined by the property that all $\lfloor n/2 \rfloor$-party reduced density matrices are maximally mixed. This implies that any balanced bipartition of a $k$-party reduced density matrix yields a maximal entanglement entropy for each subsystem in that bipartition which is consistent with the Hilbert space dimensions. Conversely, the reflected entropy between any two individual parties (or unions of parties) vanishes. Consequently, the upper bound of each $\ell_k^{(i)}$ is saturated. Interestingly, this property holds for all $k$ simultaneously which yields maximum value of all $\tilde{L}_k$ and subsequently $\tilde{L}_{gen}$ attains its global maximum. As we have utilized the a normalized version of L-entropy, we observe the value of $\tilde{L}_{gen}$ for AME states are 1.

As summarized in Table~\ref{tab:entropy-uniform}, we list the computed values of the generalized L-entropy for several known $k$-uniform states of different qubit numbers. The explicit forms of these states are provided in Appendix~\ref{app_A}. The numerical results clearly demonstrate that $\tilde{L}_{gen}$ faithfully reflects the degree of $k$-uniformity: for states with the same number of qubits, $\tilde{L}_{gen}$ increases monotonically with $k$. When $k$ reaches its maximal possible value, $k_{\max} = \lfloor n/2 \rfloor$, the generalized L-entropy attains its upper bound,
\begin{equation}
    \tilde{L}_{\text{gen}} = 1.
\end{equation}
Even among states belonging to the same $k$-uniform class, however, small variations in $\tilde{L}_{gen}$ can occur. This subtle distinction is most pronounced for 8-qubit 3-uniform states. Since $\tilde{L}_{gen}$ for an 8-qubit system incorporates contributions up to $L_4$, a 3-uniform state satisfies
\begin{equation}
   L_2 = L_3 = 1,
\end{equation}
while $L_4 < 1$. Accordingly, differences in $\tilde{L}_{gen}$ arise from $L_4$: the more mixed the four-party reductions, the closer the state is to a 4-uniform configuration, resulting in a larger $\tilde{L}_{gen}$.


\begin{table}[h]
\centering
\begin{subtable}{0.45\textwidth}
\centering
\begin{tabular}{|c|c|c|}
\hline
State & Uniformity & $\tilde{L}_{\text{gen}}$ \\
\hline
$|\psi^{(5)}_1\rangle$ & 1-uniform & 0.5 \\
$|\psi^{(5)}_2\rangle$ & 2-uniform (AME) & 1 \\
$|\psi^{(5)}_3\rangle$ & 2-uniform (AME) & 1 \\
\hline
$|\psi^{(6)}_1\rangle$ & 1-uniform & 0.5 \\
$|\psi^{(6)}_2\rangle$ & 2-uniform & 0.8706 \\
$|\psi^{(6)}_3\rangle$ & 2-uniform & 0.8706 \\
$|\psi^{(6)}_4\rangle$& 3-uniform (AME) & 1 \\
$|\psi^{(6)}_5\rangle$& 3-uniform (AME) & 1 \\
\hline
$|\psi^{(7)}_1\rangle$ & 1-uniform & 0.5 \\
$|\psi^{(7)}_2\rangle$ & 2-uniform  & 0.9330 \\
\hline
\end{tabular}
\caption{5- and 6- and 7-qubit states}
\end{subtable}
\begin{subtable}{0.45\textwidth}
\centering
\begin{tabular}{|c|c|c|}
\hline
State & Uniformity & $\tilde{L}_{\text{gen}}$ \\
\hline
$|\psi^{(8)}_1\rangle$ & 1-uniform & 0.3969 \\
$|\psi^{(8)}_2\rangle$ & 2-uniform & 0.8177 \\
$|\psi^{(8)}_3\rangle$ & 3-uniform & 0.9548 \\
$|\psi^{(8)}_4\rangle$ & 3-uniform & 0.9697 \\
\hline
$|\psi^{(9)}_1\rangle$ & 1-uniform & 0.3969 \\
$|\psi^{(9)}_2\rangle$ & 2-uniform & 0.8574 \\
$|\psi^{(9)}_3\rangle$ & 2-uniform  & 0.9547 \\
\hline
$|\psi^{(10)}_1\rangle$ & 1-uniform & 0.3536 \\
$|\psi^{(10)}_2\rangle$ & 2-uniform & 0.7676 \\
$|\psi^{(10)}_3\rangle$ & 2-uniform & 0.8856 \\
\hline
\end{tabular}
\caption{8- and 9- and 10-qubit states}
\end{subtable}
\caption{Generalized L-entropy values for $k$-uniform states.}
\label{tab:entropy-uniform}
\end{table}

\section{\textbf{Random states}}\label{sec4}
In this section we will analyze the behaviour of the generalized L-entropy in the context of random states. Here the computations will be done upto 9 party pure states where any such state can be written as $\rho_{\mathcal{A}\mathcal{B}~\mathcal{\overline{AB}}}$ where the reflected entropy between $\mathcal{A}$ and $\mathcal{B}$ can be expressed as \cite{Akers:2021pvd},
\begin{align}\label{refran}
    S_R(\mathcal{A}:\mathcal{B})\approx-p_0(q) \ln p_0(q)-p_1(q) \ln p_1(q)+p_1(q)\left(\ln d_\mathcal{A}^2-\frac{d_\mathcal{A}^2}{2 d_\mathcal{B}^2}\right).
\end{align}
Note that, $\mathcal{A}$, $\mathcal{B}$ can be constructed with multiple parties. In the above expression $q$ is the ratio of the dimensions of $\mathcal{A}\mathcal{B}$ and $\overline{\mathcal{AB}}$. The first two terms of \Cref{refran} have a similar structure as the Shannon entropy with the probability functions,
\begin{align}\label{piq}
   p_0(q)&\equiv \frac{C_{m / 2}\left(q^{-1}\right)^2}{C_m\left(q^{-1}\right)} \ ,\\p_1(q)&\equiv\frac{C_m\left(q^{-1}\right)-C_{m / 2}\left(q^{-1}\right)^2}{C_m\left(q^{-1}\right)},
\end{align}
where $C_s(x)$ is known as Catalan number,
\begin{align}
    C_s(x^{-1}) \equiv \begin{cases}\frac{1}{x} \hspace{0.1cm} {}_2 F_1(1-s,-s ; 2 ;\frac{1}{x}), & x \geq 1 \\ \frac{1}{x^s}\hspace{0.1cm} { }_2 F_1(1-s,-s ; 2 ; x). & x<1\end{cases}
\end{align}
The upper limit of $S_R(\mathcal{A}:\mathcal{B})$ involves the entanglement entropy of $\mathcal{A}$ and $\mathcal{B}$ which has the functional form,
\begin{align}\label{EErand}
S(\mathcal{\xi})=\log[d_\mathcal{\xi}]-\frac{d_\mathcal{\xi}}{2 d_{\overline{\mathcal{\xi}}}}\hspace{1cm}\text{for~~}\mathcal{\xi}=\mathcal{A},\mathcal{B},
\end{align}
where $d_\xi$ denotes the dimension of $\xi$. Considering our interest, we will utilize \Cref{refran} with $\mathcal{A}$ and $\mathcal{B}$ yielding the balanced bipartition. Let us consider an $n$-party system where $\lfloor \frac{N}{2}\rfloor=k$. The computation of generalized Latent entropy involves the multiplication of all $\widetilde{L}_k$ for $\text{Max}\left[2,\lfloor \frac{N}{2}\rfloor\right] \geq k\geq 2$. Let us consider a specific Latent entropy, $\widetilde{L}_{k^\prime}$ ($k\geq k^\prime$) where we have traced out all possible $N-k^\prime$ parties and purified the remaining $k^\prime$-party mixed state. A balanced bipartition divides the $k^\prime$ parties into $\lfloor \frac{k^\prime}{2}\rfloor$ parties and the remaining. Now following the definition of Latent entropy in \Cref{genell},
\begin{align}
    \ell_{k^\prime}=2\lfloor \frac{k^\prime}{2}\rfloor\log d-\frac{1}{d^{N-2\lfloor \frac{k^\prime}{2}\rfloor}}+p_0(q) \ln p_0(q)+p_1(q) \ln p_1(q)-p_1(q)\left(2\lfloor \frac{k^\prime}{2}\rfloor\ln d-\frac{1}{2 d^2}\right).
\end{align}
with $q\leq 1$ where the equality holds for even $n$. Considering 6, 7, 8 and 9 party random states with each party having dimension $d$, we will systematically obtain the behaviour of $\widetilde{L}_k$ and finally compute the generalized L-entropy in the limit of large $d$. 

\subsection*{\textbf{6 party system}}
Here, we only have to consider $\tilde{L}_2$ and $\tilde{L}_3$ which will results into the generalized latent entropy following \Cref{lgen2}. The two party purification yields $\tilde{L}_2$ in the large $d$ limit as,
\begin{align}
    \widetilde{L}_2=1-\frac{1}{2d^2}-\frac{1+4 \log 2 }{16~d^2 \log d}+\mathcal{O}\left(\frac{1}{d^3}\right),
\end{align}
which goes to its maximum value for large $d$. On the other hand, the three-party purification corresponds to $\tilde{L}_3$ as,
\begin{align}
    \widetilde{L}_3=a_0+\frac{a_1}{\log (d)}+\frac{a_2}{d^2 \log (d)}+\mathcal{O}\left(\frac{1}{d^3}\right),
\end{align}
where the constants are
\begin{align}
    a_0&=\left(8/3\pi\right)^2,\nonumber\\
    a_1&=\frac{1}{2}\left[\left(1-a_0\right)\log \left(1-a_0\right)+a_0 \log a_0\right],\nonumber\\
    a_2&=\frac{1}{4}\left(1-a_0\right)\nonumber.   
\end{align}
Note that, the value of $\tilde{L}_3$ goes to $a_0=0.7205$ which is less than the maximum value. It clearly indicates that the generalized latent entropy for the 6-qubit random system does not obtain the maximum value. However, the generalized latent entropy can be written in the large $d$ as,
\begin{align}
    \widetilde{L}_{gen}^{(6)}&=\sqrt{a_0}+\frac{a_1}{2 \sqrt{a_0} \log d}+\mathcal{O}\left(\frac{1}{d^2}\right),
\end{align}
where the maximum value is 
\begin{align}
    \widetilde{L}_{gen, max}^{(6)}&=\sqrt{a_0}=0.8488
\end{align}

\subsection*{\textbf{7 party system}}
Here we consider 7 party qudit system where we compute the generalized latent entropy which is constituted with $\widetilde{L}_{2}$ and $\widetilde{L}_{3}$ as
\begin{align}
    \widetilde{L}_2&=1-\frac{5}{8 d^3}-\frac{1+4 \log 2}{16 d^3 \log d}+\mathcal{O}\left(\frac{1}{d^4}\right),\nonumber\\
    \widetilde{L}_3&=1-\frac{3}{8 d}-\frac{1+2 \log 2}{8 d \log d}+\mathcal{O}\left(\frac{1}{d^2}\right)\nonumber.
\end{align}
Note that here both $\widetilde{L}_2$ and $\widetilde{L}_3$ reaches to the maximum value 1 in the large $d$ limit. Finally the generalized latent entropy can be written as,
\begin{align}
    \widetilde{L}_{gen}^{(7)}&=1-\frac{3}{16 d}-\frac{1+2\log 2}{16 d \log d}+\mathcal{O}\left(\frac{1}{d^2}\right).
\end{align}
Note that, unlike the case of 6 party qudit systems, here we observe the maximum permissible value for the generalized latent entropy in the large $d$ limit as
\begin{align}
    \widetilde{L}_{gen, max}^{(7)}=1
\end{align}

\subsection*{\textbf{8 party system}}
For the 8 party qudits, we sill follow a similar construction where $\widetilde{L}_{2}$, $\widetilde{L}_{3}$ and $\widetilde{L}_{4}$ will contribute in the computation of the generalized latent entropy. 
\begin{align}
    \widetilde{L}_2&=1-\frac{3}{4 d^4}+\frac{1+4 \log (2)}{16 d^4 \log (d)}+\mathcal{O}\left(\frac{1}{d^5}\right),\nonumber\\
    \widetilde{L}_3&=1-\frac{1}{2 d^2}-\frac{1+2 \log (2)}{8 d^2 \log (d)}+\mathcal{O}\left(\frac{1}{d^4}\right),\nonumber\\
    \widetilde{L}_4&=a_0+\frac{a_1}{2\log d}+\frac{1-a_0}{8 \log d}-\frac{1}{4 d^4 \log d}+\mathcal{O}\left(\frac{1}{d^5}\right).
\end{align}
In the above expressions, it can be observed that although $\widetilde{L}_2$ and $\widetilde{L}_3$ reach the maximum value in large $d$, $\widetilde{L}_4$ approaches $a_0$. Combining all of them, the generalized latent entropy can be written as,
\begin{align}
    \widetilde{L}_{gen}^{(8)}&=a_0^{1/3}-\frac{a_0-4 a_1-1}{24 a_0^{2/3} \log (d)}+\mathcal{O}\left(\frac{1}{d^2}\right).
\end{align}
where the maximum value of $\widetilde{L}_{gen}^{(8)}$ is given as,
\begin{align}
    \widetilde{L}_{gen, max}^{(8)}=a_0^{1/3}=0.8965
\end{align}
Notably, this value does not reach the maximum, suggesting that the genuine multipartite entanglement in an eight-party qudit random state remains below its maximal limit.

\subsection*{\textbf{9 party system}}
Similar to the 8 party case here we will compute $\widetilde{L}_2$, $\widetilde{L}_3$ and $\widetilde{L}_4$ for 9 party qudit random state as,
\begin{align}
    \widetilde{L}_2&=1-\frac{7}{8 d^5}-\frac{1+4 \log 2}{16 d^5 \log d}+\mathcal{O}\left(\frac{1}{d^6}\right),\nonumber\\
    \widetilde{L}_3&=1-\frac{5}{8 d^3}-\frac{1+2 \log 2}{8 d^3 \log d}+\mathcal{O}\left(\frac{1}{d^4}\right),\nonumber\\
    \widetilde{L}_4&=1-\frac{5}{16 d}-\frac{1+4 \log 2}{32 d \log d}+\mathcal{O}\left(\frac{1}{d^2}\right).
\end{align}
Finally the generalized latent entropy is computed as,
\begin{align}
    \widetilde{L}_{gen}^{(9)}&=1-\frac{5}{48 d}-\frac{1+4 \log 2}{96~ d~\log d}+\mathcal{O}\left(\frac{1}{d^2}\right)
\end{align}
with the maximum value,
\begin{align}
    \widetilde{L}_{gen, max}^{(9)}=1
\end{align}
Below we plot all the generalized latent entropies $\widetilde{L}_{gen}^{(6)}$, $\widetilde{L}_{gen}^{(7)}$, $\widetilde{L}_{gen}^{(8)}$ and $\widetilde{L}_{gen}^{(9)}$ for random states with respect to the dimension $d$.

\begin{figure}[H]
    \centering
    \begin{subfigure}[t]{0.45\linewidth}
        \centering
        \includegraphics[width=\linewidth]{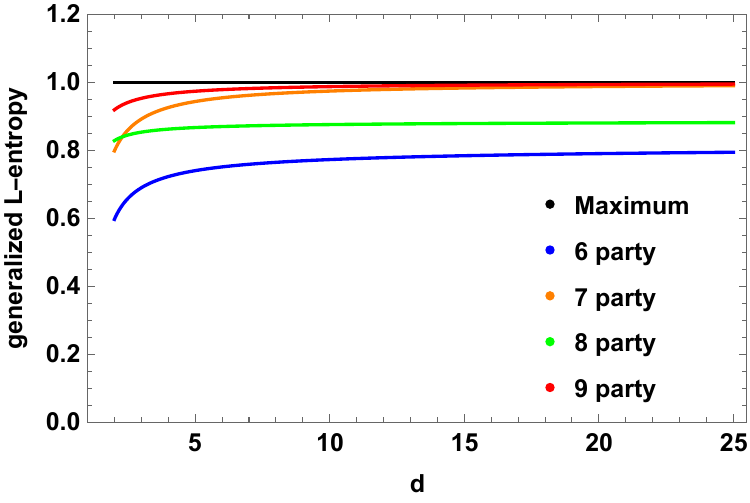}
        \caption{$\widetilde{L}_{gen}^{(6)}$ (blue), $\widetilde{L}_{gen}^{(7)}$ (orange), $\widetilde{L}_{gen}^{(8)}$ (green) and $\widetilde{L}_{gen}^{(9)}$ (red) are plotted with increasing dimension $d$. Note that, here all the parties possess same dimensional Hilbert space. The plot indicates that $\widetilde{L}_{gen}^{(7)}$ and $\widetilde{L}_{gen}^{(9)}$ reaches the maximum value where $\widetilde{L}_{gen}^{(6)}$ and $\widetilde{L}_{gen}^{(8)}$ converges to values less than the maximum. Analyzing other states it can be observed that $\widetilde{L}_{gen}^{(N)}$ will obtain maximum value only when $n$ is an odd number.}
        \label{random_l_analytic}
    \end{subfigure}
    \hfill
    \begin{subfigure}[t]{0.45\linewidth}
        \centering
        \includegraphics[width=\linewidth]{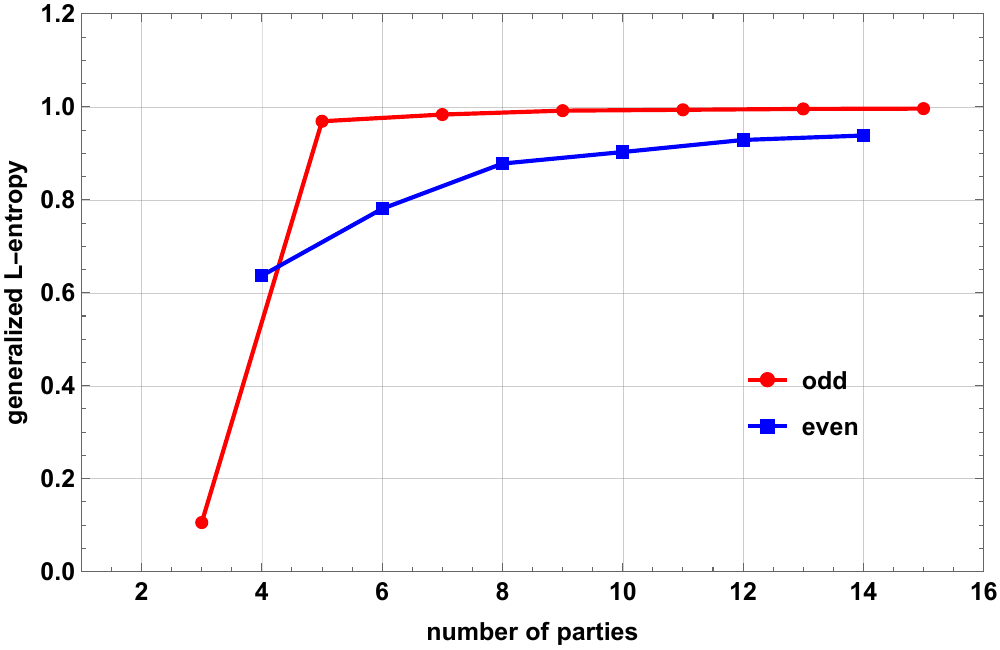}
        \caption{Using the analytical results, the figure shows the generalized $\widetilde{L}_{gen}$ values as a function of the number of parties for $d = 15$. In the large-$d$ regime, the generalized L-entropy for systems with an odd number of parties converges rapidly to $1$, while for even numbers of parties, it approaches $1$ more slowly. This behavior highlights the distinct scaling of multipartite entanglement between even- and odd-$n$ systems in high-dimensional Hilbert spaces.}
        \label{random_highd}
    \end{subfigure}

    \caption{(a) and (b) illustrate the scaling behavior of generalized latent entropy for Haar-random states. The left panel compares systems with different party numbers, while the right panel highlights the even–odd contrast at high dimensions.}
    \label{fig:randomgenl}
\end{figure}

The above results reveal an intriguing feature of multipartite entanglement in random states shown in \Cref{random_l_analytic}. Specifically, only the states with odd $n$ exhibit maximal genuine multipartite entanglement (GME) in the large-$d$ limit. In contrast, for even $n$, the generalized L-entropy converges to values below the upper bound. The analytical results also indicate that the maximal value of the generalized L-entropy for even $n$ approaches the upper bound as the number of parties increases. We reinforce this finding through a complementary numerical analysis on multi-qubit states with varying $n$. Interestingly, as shown in \Cref{random_highd}, for both even and odd $n$, the generalized L-entropy tends toward the upper bound as $n$ increases, with the odd-$n$ cases reaching it more rapidly.

\begin{figure}[H]
    \centering
    \begin{subfigure}[t]{0.45\linewidth}
        \centering
        \includegraphics[width=\linewidth]{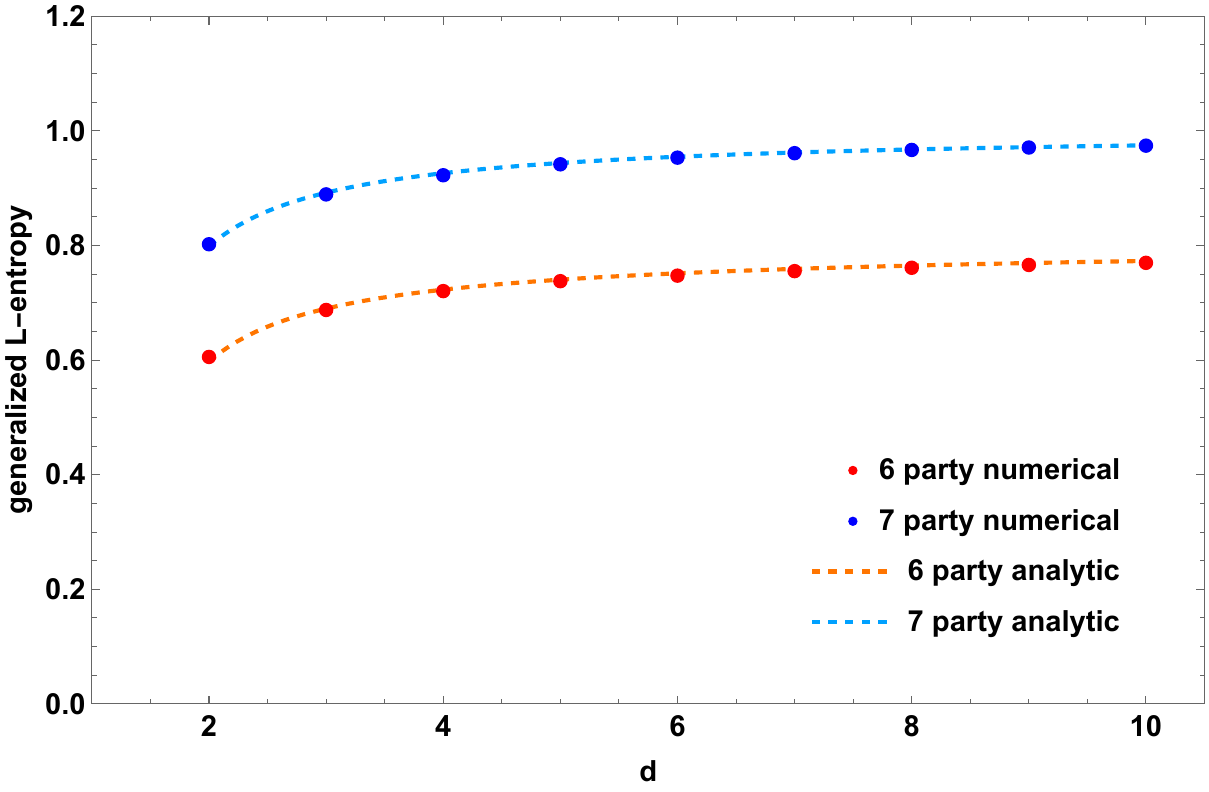}
        \caption{The results for 6-party and 7-party random qudit states are shown as functions of the subsystem dimension. The light-green dashed line represents the analytic result for the 6-party case, while the red dots correspond to numerical calculations. Similarly, the orange dashed line denotes the analytic result for 7 parties, and the blue dots show the corresponding numerical data. Although the analytic results were derived under a large-$d$ approximation, they agree well with the numerical results even for small $d$.}
        \label{num_l_67}
    \end{subfigure}
    \hfill
    \begin{subfigure}[t]{0.45\linewidth}
        \centering
        \includegraphics[width=\linewidth]{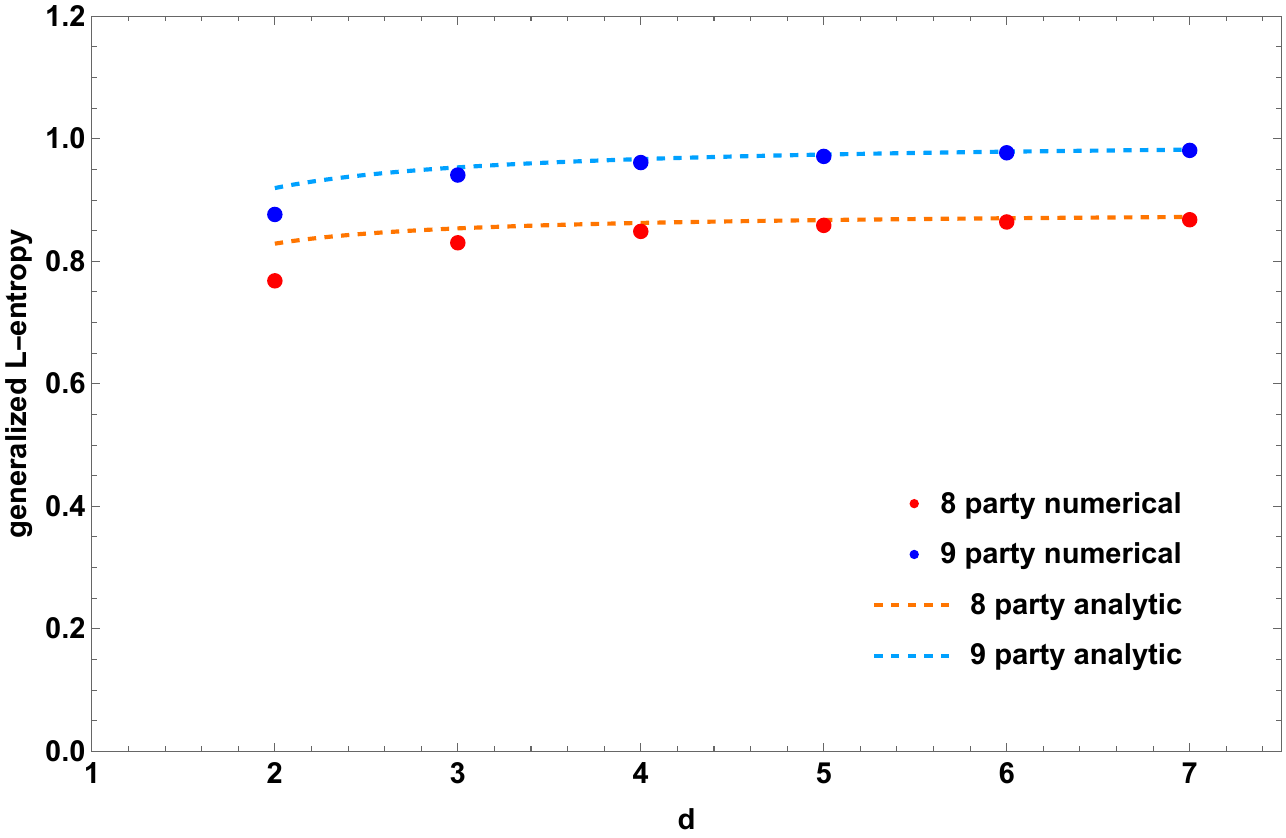}
        \caption{The corresponding results for 8-party and 9-party random qudit states are displayed. The light-green dashed line and red dots represent, respectively, the analytic and numerical results for 8 qudits, while the orange dashed line and blue dots correspond to those for 9 qudits. Note that, the results match exactly in region where $d$ is moderately high. However, we notice a mismatch in the low $d$ region which is expected as the analytic result has only been computed in large $d$ limit.}
        \label{num_l_89}
    \end{subfigure}

    \caption{(a) and (b) illustrate the generalized L-entropy for random qudit states with different numbers of parties.}
    \label{num_l}
\end{figure}

Subsequently, we numerically compute the average generalized L-entropy for random pure states sampled from the Haar random distribution and compare the results with the corresponding analytical predictions in \Cref{num_l}. Although the numerical computations become increasingly demanding with larger $n$—and hence with the total Hilbert-space dimension—we find excellent agreement with the analytical results for moderately large $d$.

\section{\textbf{Examples: SYK Model and its Variants}}\label{sec5}

In this section, we investigate the multipartite entanglement structure of the Sachdev–Ye–Kitaev (SYK) model~\cite{Sachdev:1992fk,Polchinski:2016xgd,Jevicki:2016bwu,Maldacena:2016hyu}, which describes a system of strongly interacting Majorana fermions  with couplings drawn from a Gaussian random distribution. The $SYK$ model involves a $q$-body all-to-all interaction Hamiltonian given by
\begin{align}
    H_{\mathrm{SYK}_q} = i^{q/2} \!\! \sum_{1 \le i_1 < i_2 < \cdots < i_q \le N} 
    J_{i_1 i_2 \cdots i_q}\, \chi_{i_1} \chi_{i_2} \cdots \chi_{i_q},
\end{align}
where $\chi_i$ are Majorana fermions satisfying $\{\chi_i, \chi_j\} = \delta_{ij}$, 
and the couplings $J_{i_1 i_2 \cdots i_q}$ are independent random variables drawn 
from a Gaussian distribution with zero mean and variance
\begin{align}\label{var}
    \langle J_{i_1 i_2 \cdots i_q}^2 \rangle = \frac{(q-1)! J^2}{N^{q-1}}.
\end{align}

Our primary focus is on examining the behavior of the generalized L-entropy in a qubit realization of the SYK model and its various deformations. The qubit formulation allows for a direct numerical evaluation of multipartite entanglement in finite-size systems (see the appendix of \cite{Basak:2024uwc} for details of the qubit mapping). We numerically study the time evolution and saturation of the generalized L-entropy, and its distribution in energy eigenstates across different SYK variants. We utilize the results  to understand how interaction structure and model parameters influence the structure of genuine multipartite entanglement. 

\subsection{\textbf{SYK$_2$ Model}}
\begin{figure}[H]
  \centering
  \begin{subfigure}{.3\linewidth}
    \includegraphics[height=3.5cm,width=\linewidth]{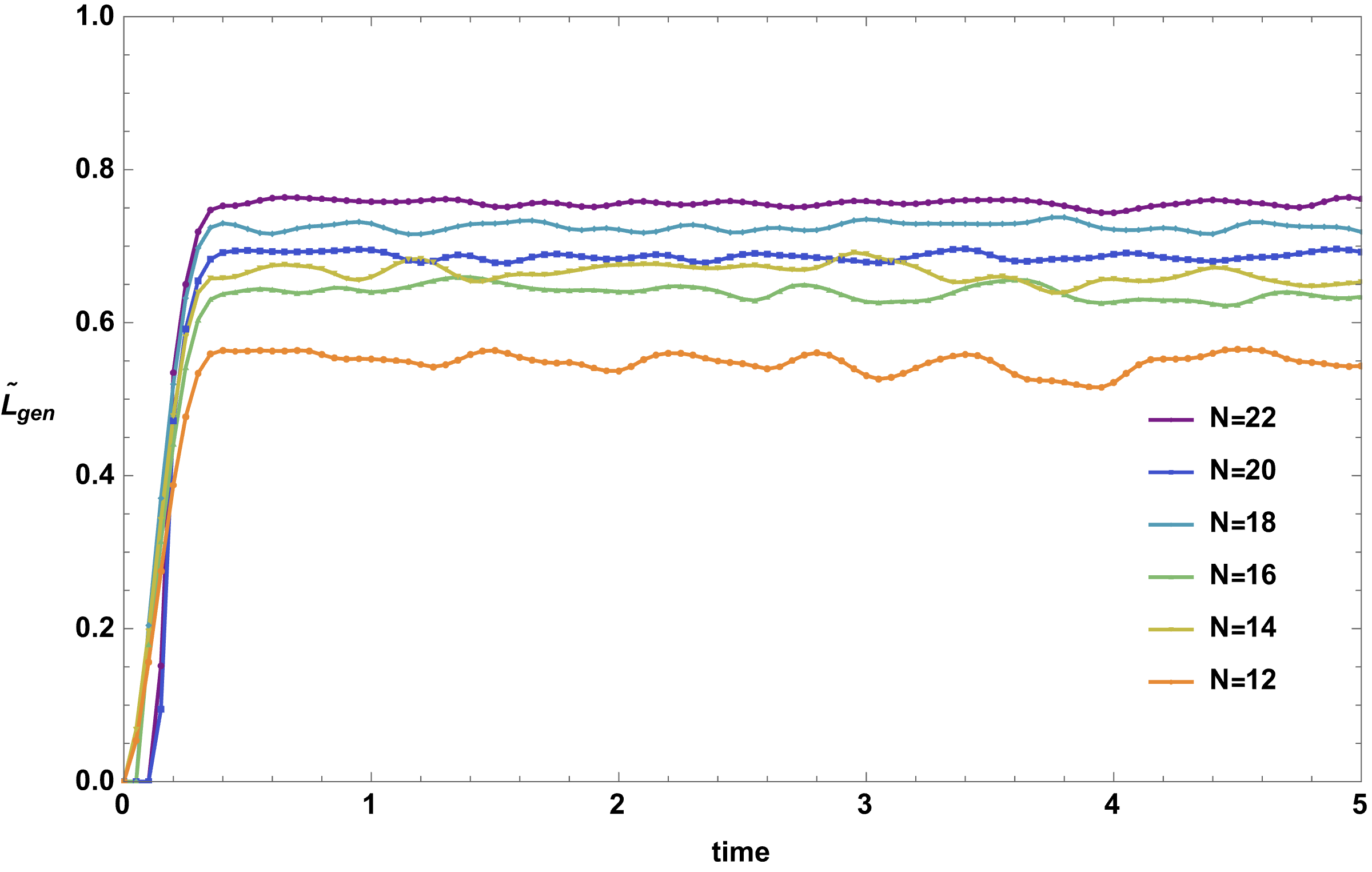}
    \subcaption{$N=12$–22 (6–11 qubits)}\label{SYK2all}
  \end{subfigure}\quad
  \begin{subfigure}{.3\linewidth}
\includegraphics[height=3.5cm,width=\linewidth]{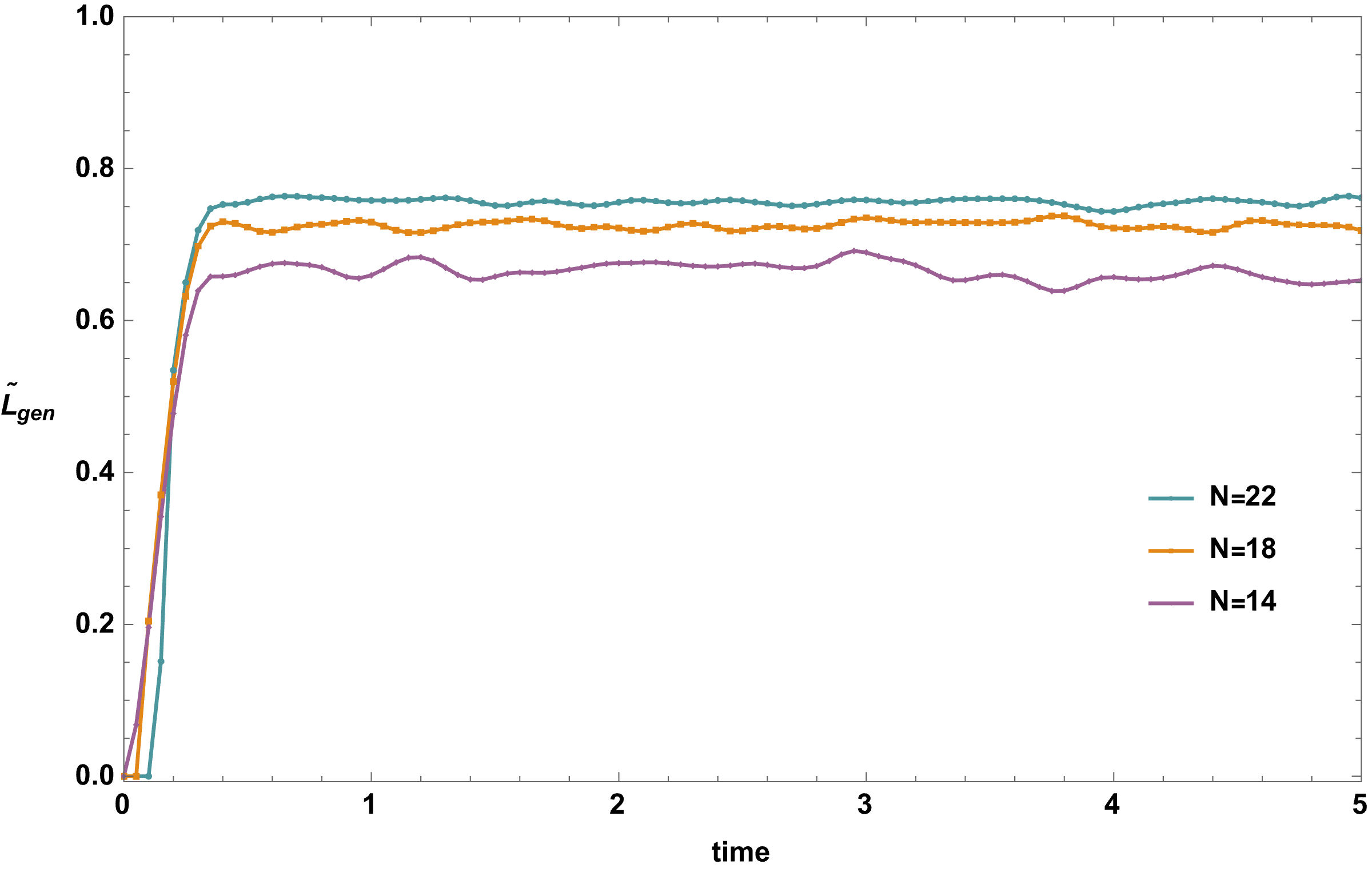}
    \caption{Odd qubits (7, 9, 11)}\label{SYK2Odd}
  \end{subfigure}\quad
  \begin{subfigure}{.3\linewidth}
    \includegraphics[height=3.5cm,width=\linewidth]{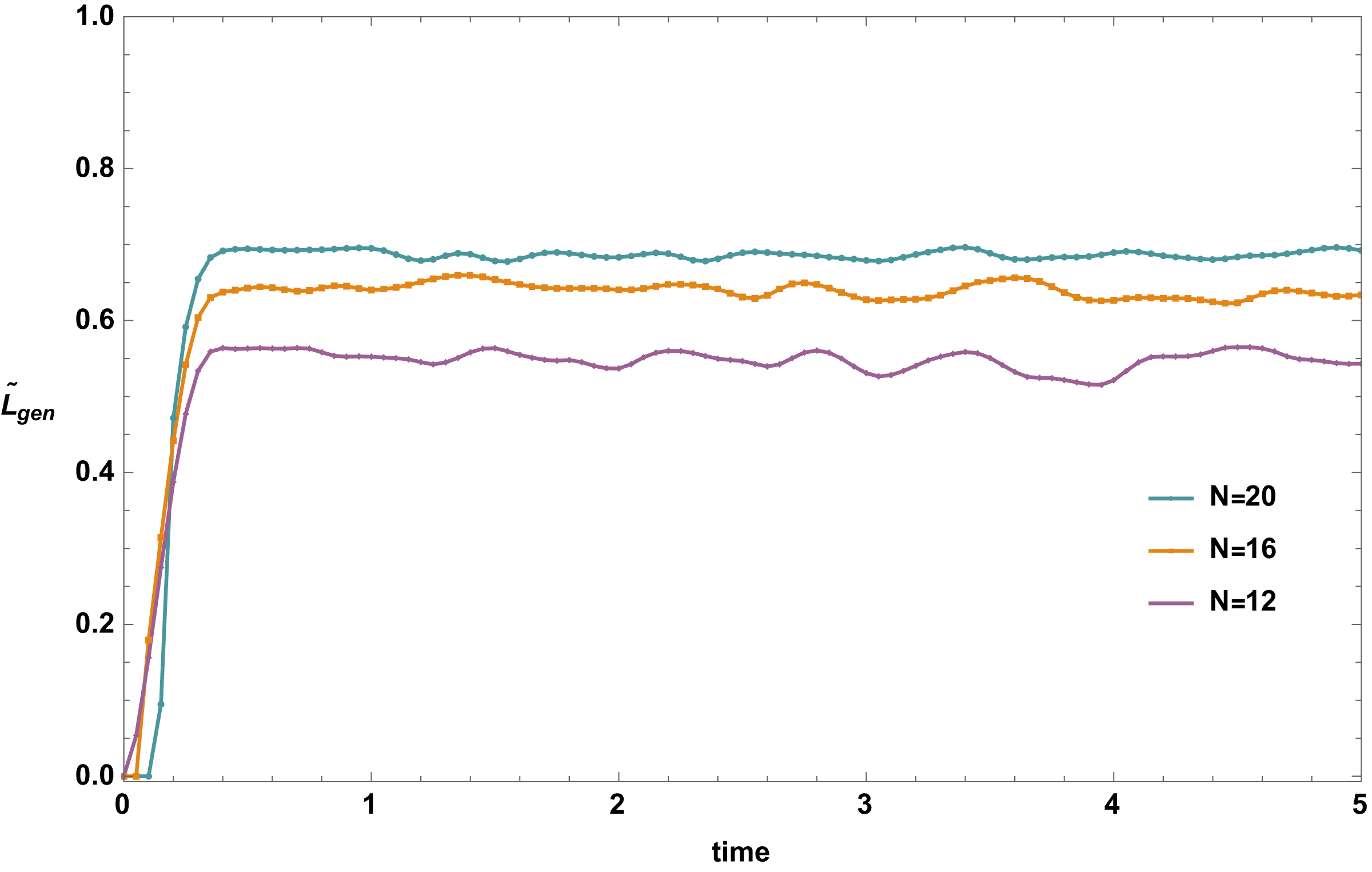}
    \caption{Even qubits (6, 8, 10)}\label{SYK2even}
  \end{subfigure}
   \caption{\footnotesize{Time evolution of the generalized L-entropy for the $SYK_2$ model starting from $\ket{000\ldots0}$, averaged over 25 samples.  The plots show the variation in late-time saturation with system size.}}
\end{figure}

\begin{figure}[H]
\centering
 \begin{subfigure}{.45\linewidth}
    \includegraphics[height=4.5cm,width=\linewidth]{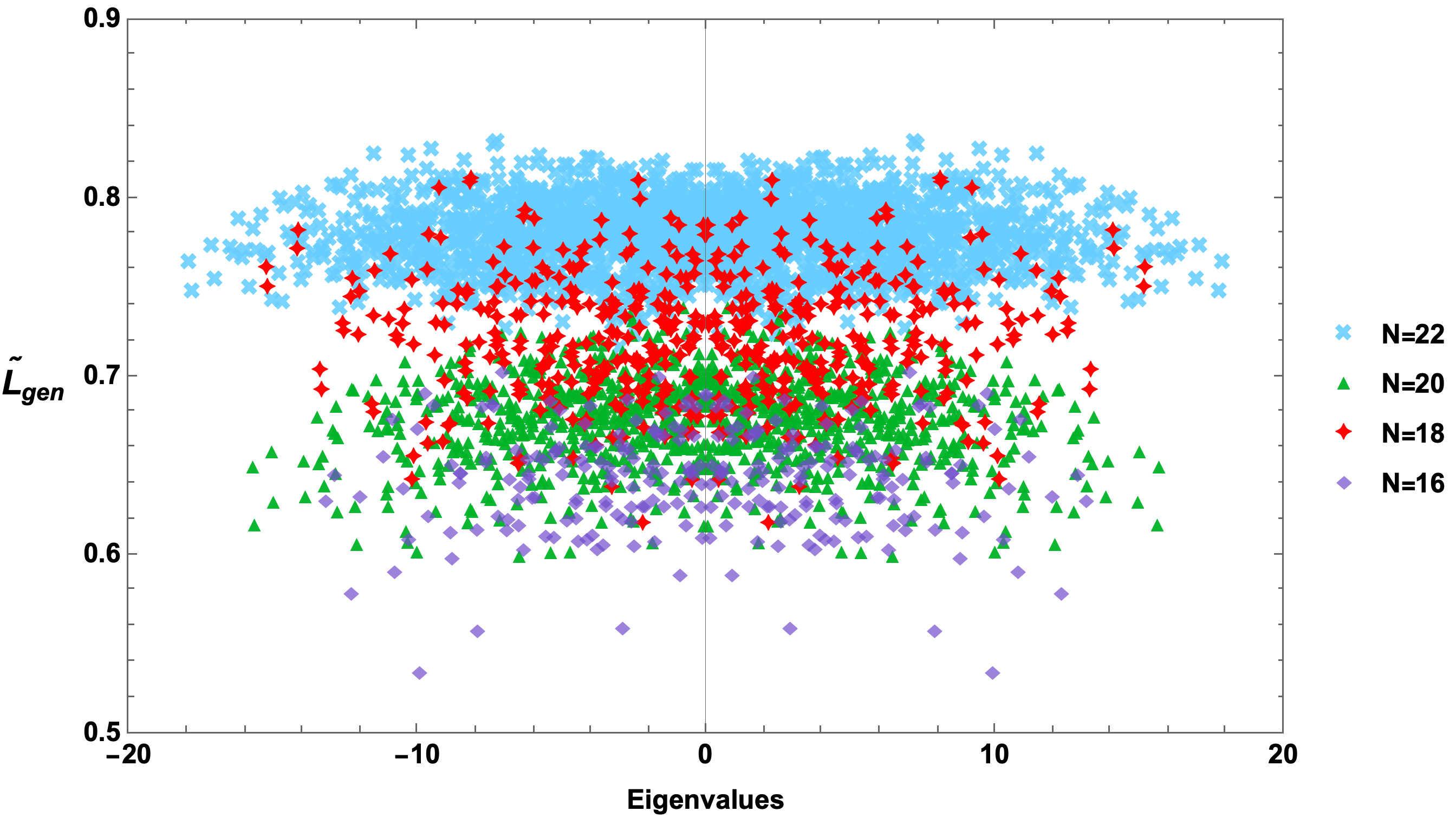}
    \caption{L-entropy of eigenstates}\label{Lenteig11SYK2}
    \end{subfigure}\quad
     \begin{subfigure}{.45\linewidth}
\includegraphics[width=\linewidth]{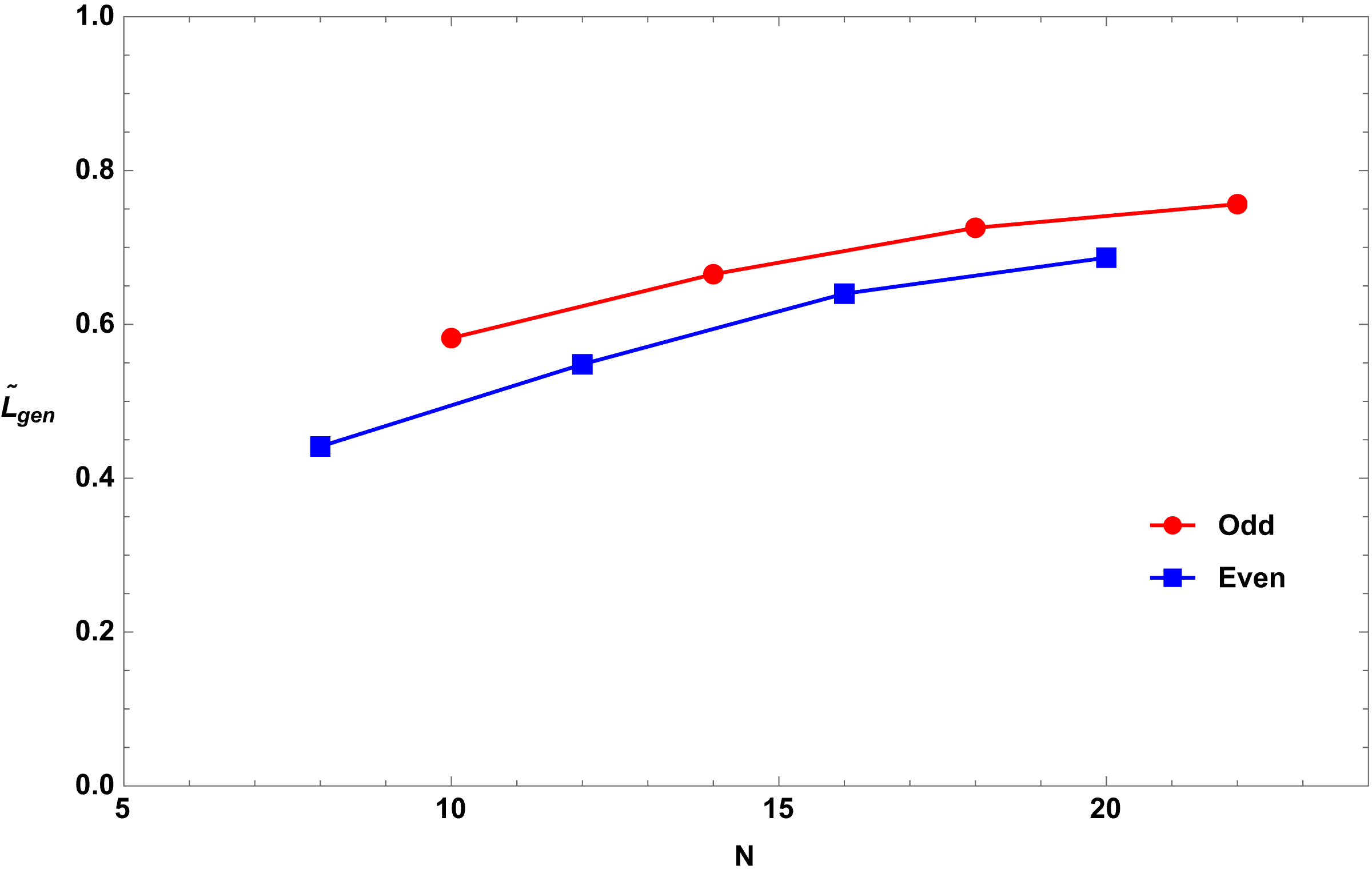}
    \caption{Saturation Values vs $N$}\label{SYK2sat}
  \end{subfigure}\quad
  \caption{\footnotesize{Generalized L-entropy distribution across energy eigenstates of the $SYK_2$ model for $N = 18, 20,$ and $22$ (left), and the saturation value of the generalized L-entropy versus $N$ (right).}}

\end{figure}

We begin our analysis of the generalized L-entropy with the $SYK_2$ model, whose Hamiltonian is given by
\begin{align}\label{SYK2ham}
H_{SYK_2} = i\sum_{i<j} J_{ij}  \chi^i \chi^j,
\end{align}
where the coupling constants $J_{ij}$ are drawn from a Gaussian distribution with zero mean and variance given in \Cref{var} for $q=2$.
\Cref{SYK2all} illustrates the time evolution of the generalized L-entropy, starting from a fully separable initial state $\ket{000\ldots0}$. The generalized L-entropy rises rapidly and saturates to a value below unity, indicating partial multipartite entanglement. Similar to the behavior observed in Haar-random states, the evolution exhibits a clear distinction between odd and even values of $n = N/2$, as shown in \Cref{SYK2Odd} and \Cref{SYK2even}. When comparing systems with different numbers of parties, the saturation value of generalized L-entropy grows with $N$, indicating enhanced multipartite entanglement for larger systems. However, a subtle dependence on the number of parties being odd or even is observed: while both odd and even $n=\frac{N}{2}$ exhibit the same qualitative growth, the odd-$n$ values consistently lie slightly above the neighboring even-$n$ ones. Thus, when plotted together, the sequence of saturation values shows an alternating but overall increasing trend which is depicted in \Cref{SYK2sat}. 

\begin{figure}[H]
  \centering
  \begin{subfigure}{.45\linewidth}
    \includegraphics[width=\linewidth]{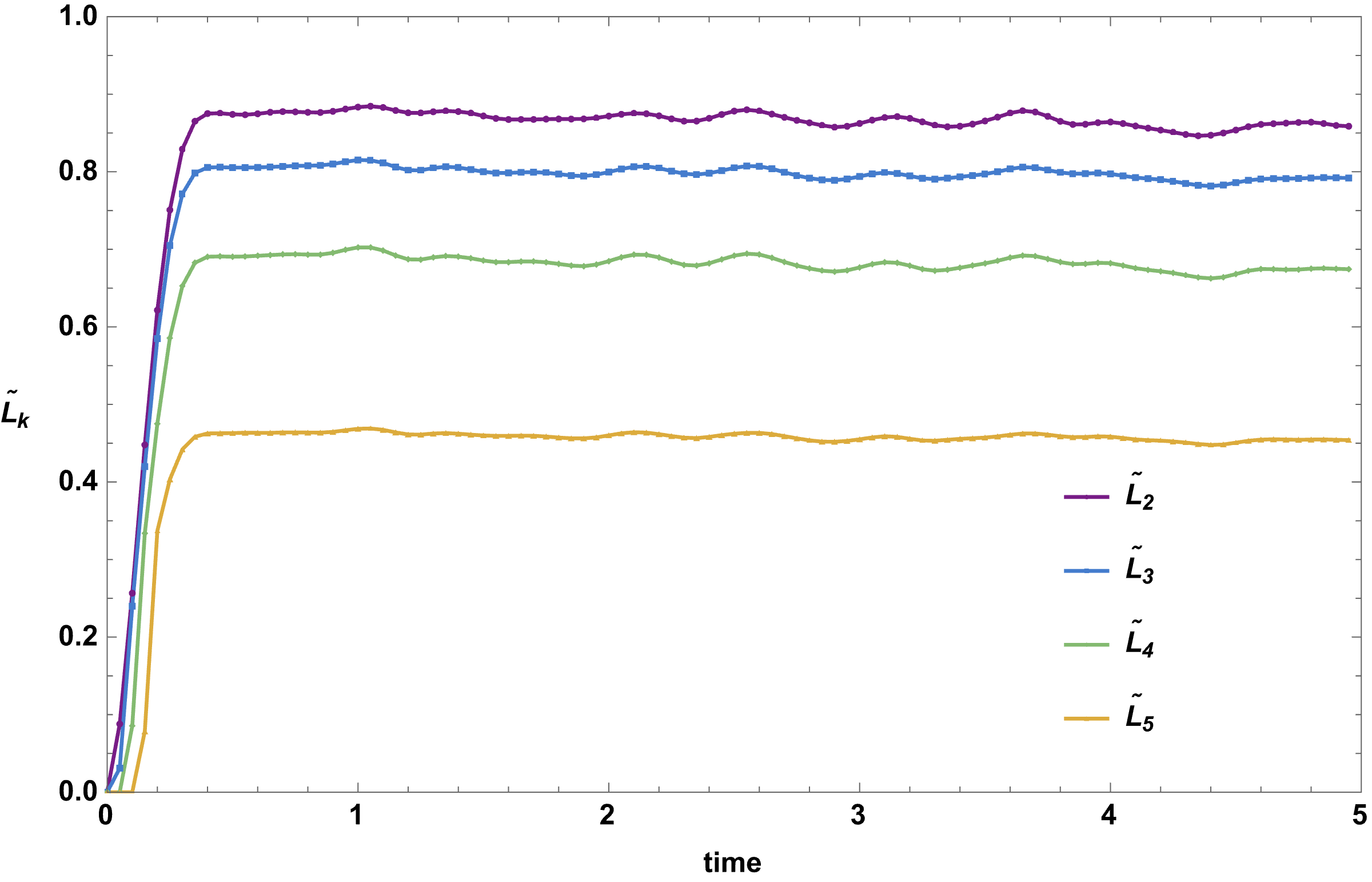}
    \subcaption{$N=20$ (10 qubits)}\label{SYK2lower10}
  \end{subfigure}\quad
  \begin{subfigure}{.45\linewidth}
\includegraphics[width=\linewidth]{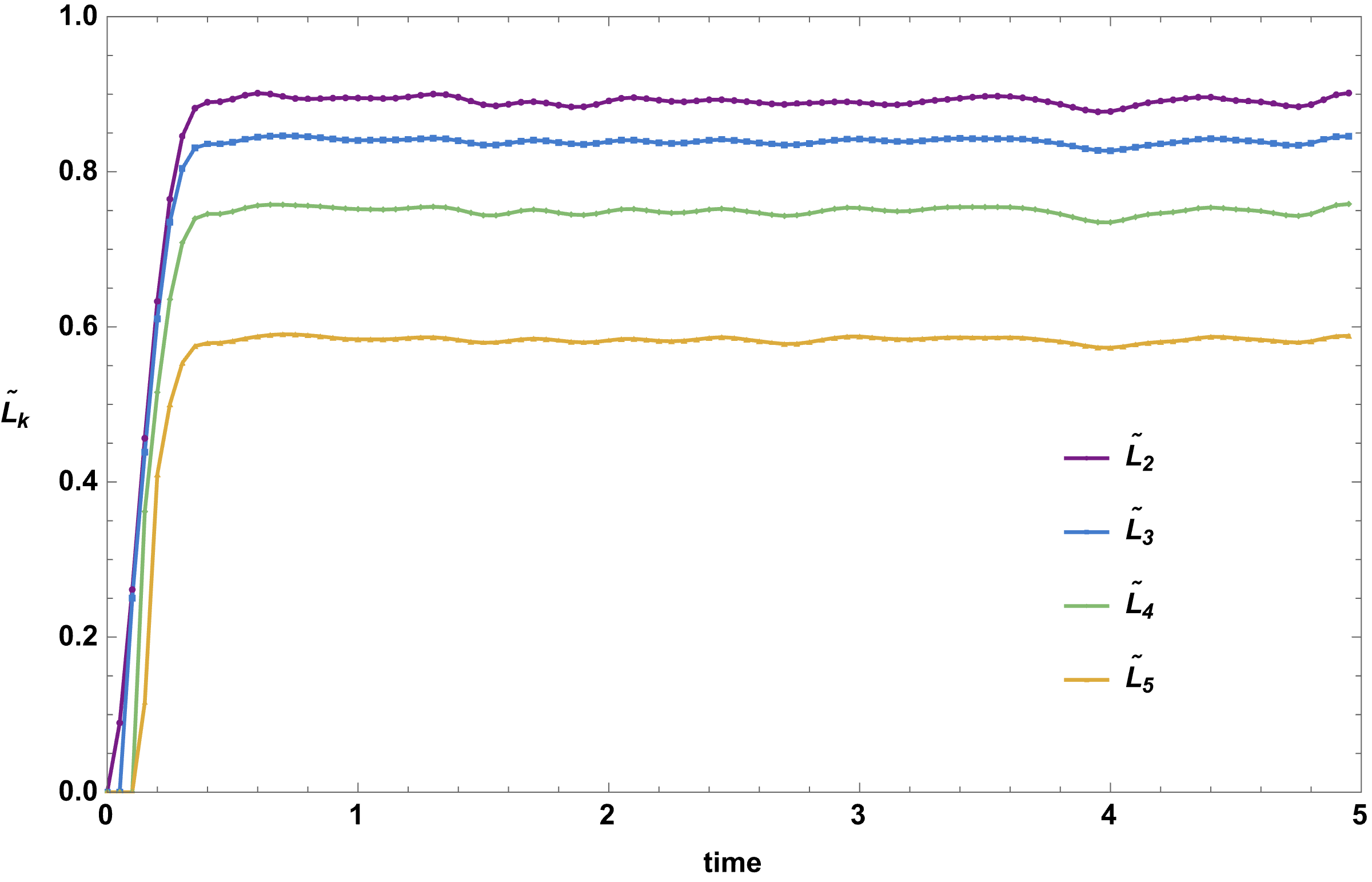}
    \caption{$N=22$ (11 qubits)}\label{SYK2lower11}
  \end{subfigure}\quad
 
   \caption{\footnotesize{ Time evolution of the lower L-entropies (characterizing $k$-uniform states) for the $SYK_2$ model with initial state $\ket{000\ldots0}$, averaged over 25 samples.  }}
\end{figure}
In \Cref{SYK2lower10} and \Cref{SYK2lower11}, we plot all the components $\tilde{L}_k$, where $k$ denotes the number of parties in the reduced density matrices being purified ($2 \leq k \leq \lfloor N/2 \rfloor$). We observe that higher-$k$ L-entropies saturate at progressively lower values. \Cref{Lenteig11SYK2} shows the distribution of the generalized L-entropy across the eigenstates of the $SYK_2$ Hamiltonian, revealing the characteristic spread of multipartite entanglement in the energy eigenbasis.

\subsection{\textbf{SYK$_4$ Model}}
We now turn to the analysis of the generalized L-entropy in the $SYK_4$ model, whose Hamiltonian is given by
\begin{align}\label{HSYK4}
H_{SYK_4} = -\sum_{i<j<k<l} J_{ijkl}\, \chi^i \chi^j \chi^k \chi^l,
\end{align}
where the coupling coefficients $J_{ijkl}$ are drawn from a Gaussian distribution with zero mean and variance given in \Cref{var} for $q=4$.

\begin{figure}[H]
  \centering
  \begin{subfigure}{.3\linewidth}
    \includegraphics[height=3.5cm,width=\linewidth]{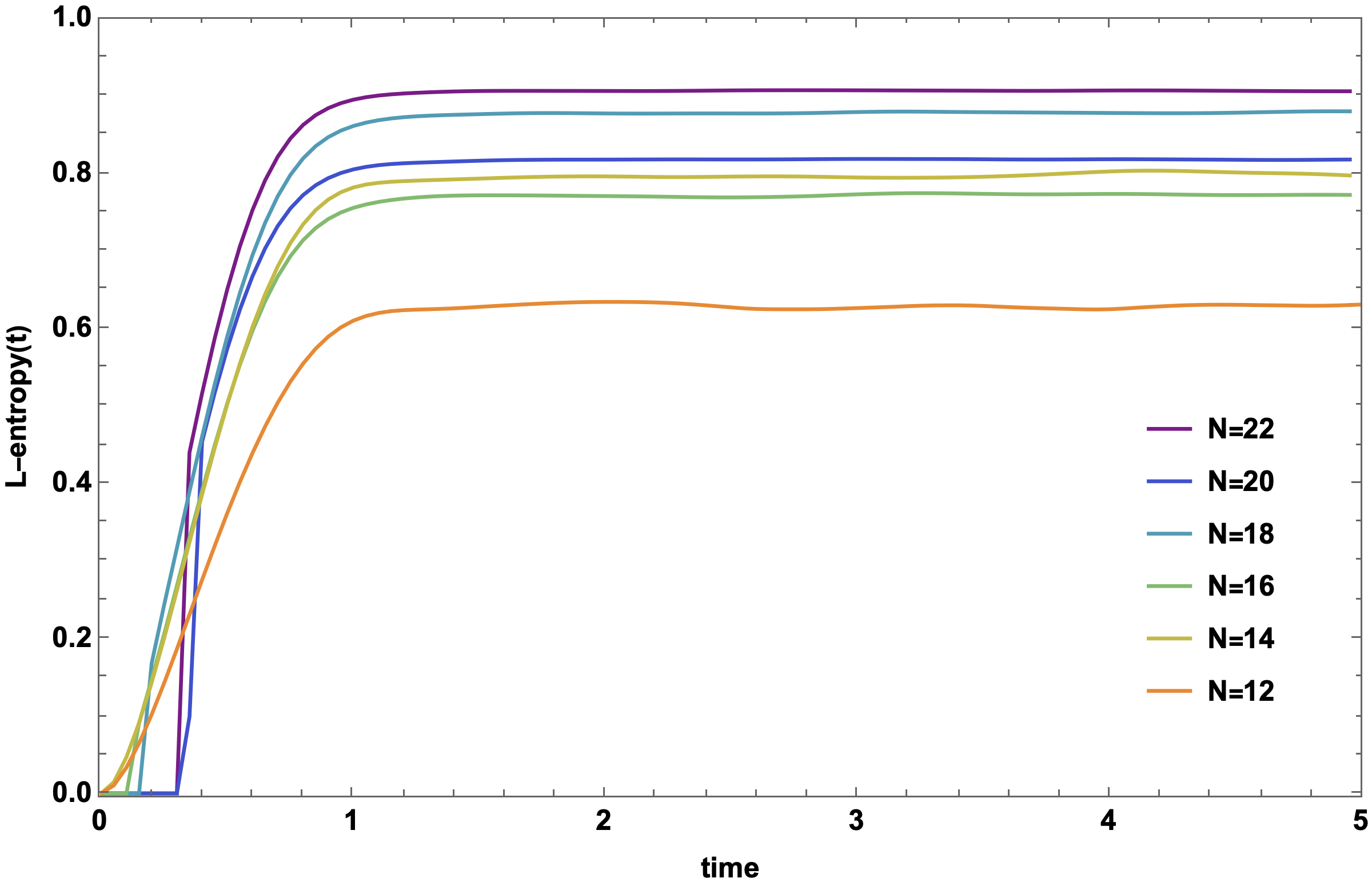}
    \subcaption{}\label{SYKall}
  \end{subfigure}\quad
  \begin{subfigure}{.3\linewidth}
\includegraphics[height=3.5cm,width=\linewidth]{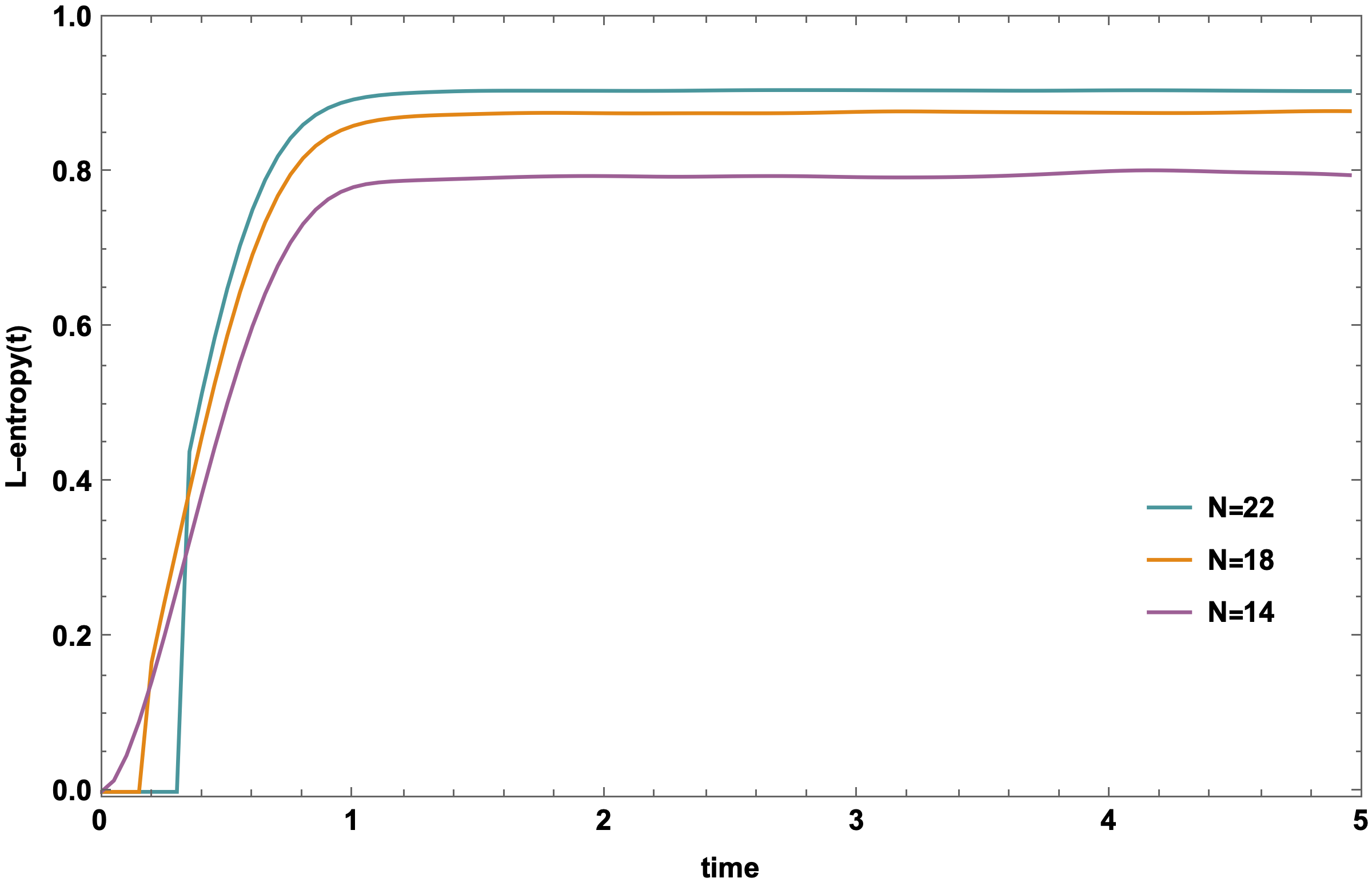}
    \caption{}\label{SYKOdd}
  \end{subfigure}\quad
  \begin{subfigure}{.35\linewidth}
    \includegraphics[height=3.5cm,width=\linewidth]{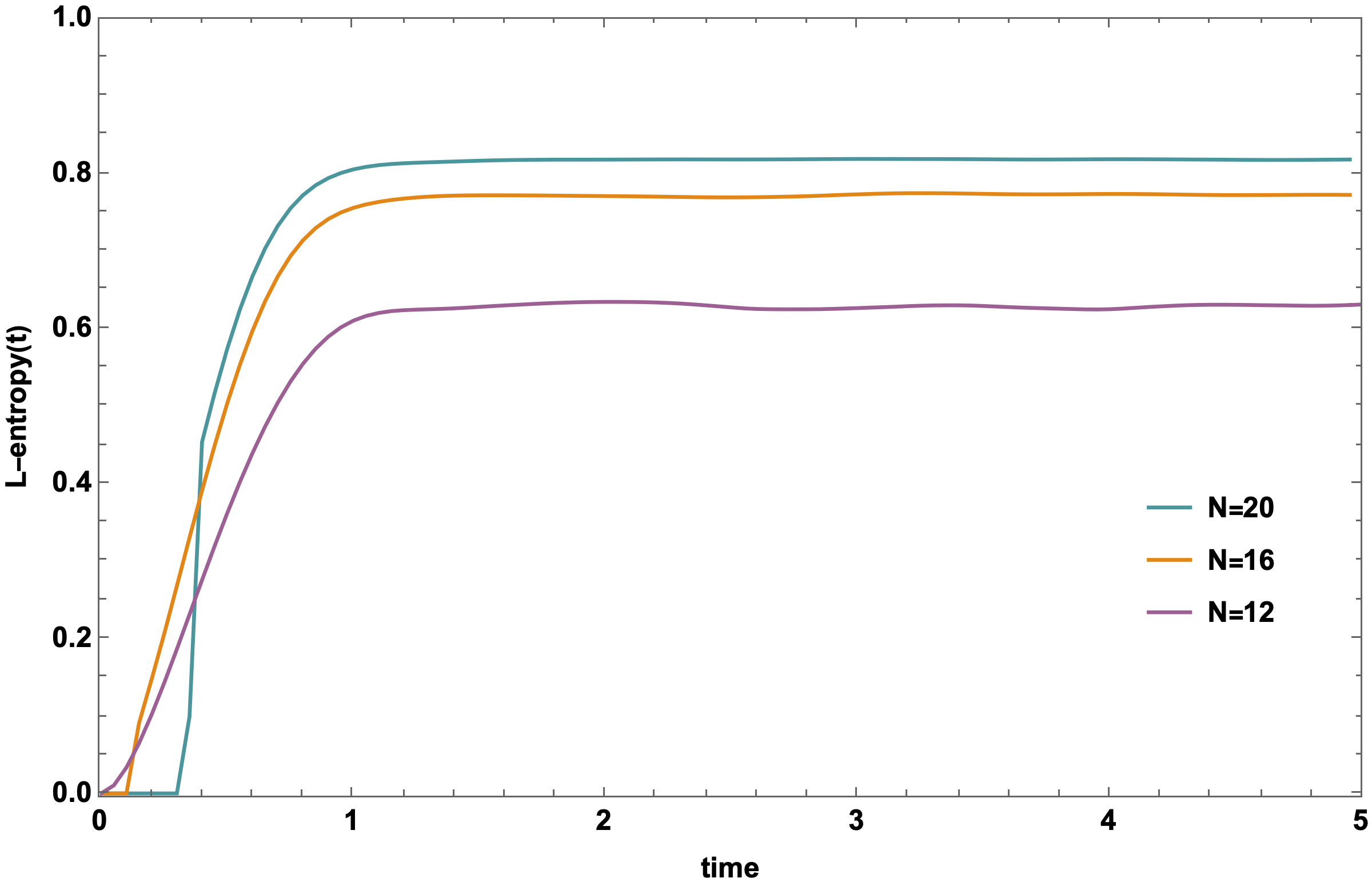}
    \caption{}\label{SYKeven}
  \end{subfigure}
  \caption{\footnotesize{
Panels (a)–(c) show the time evolution of the generalized L-entropy for the $SYK_4$ model with the initial state $\ket{000\cdots0}$, averaged over 25 samples. 
Panel (a) corresponds to $N = 12$–$22$ (6–11 qubits), (b) to $n$-even qubits , and (c) to odd qubits. 
The plots illustrate the variation of the late-time saturation value of the generalized L-entropy with system size.}}

\end{figure}

\begin{figure}[H]
\centering
 \begin{subfigure}{.45\linewidth}
 \centering \includegraphics[width=\linewidth]{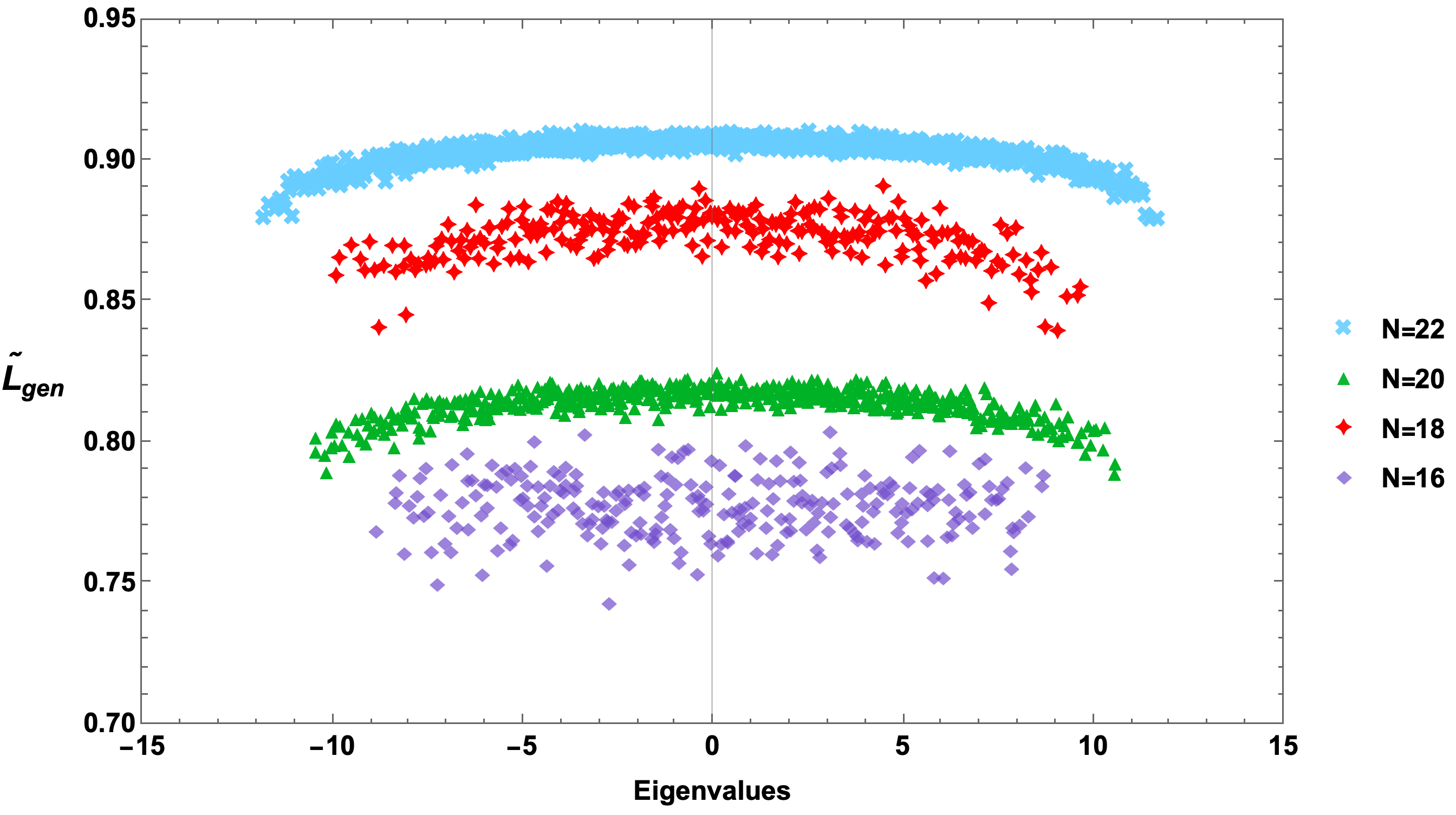}
    \caption{L-entropy of eigenstates}\label{Lenteig11SYK4}
    \end{subfigure}\quad
     \begin{subfigure}{.45\linewidth}
\includegraphics[width=\linewidth]{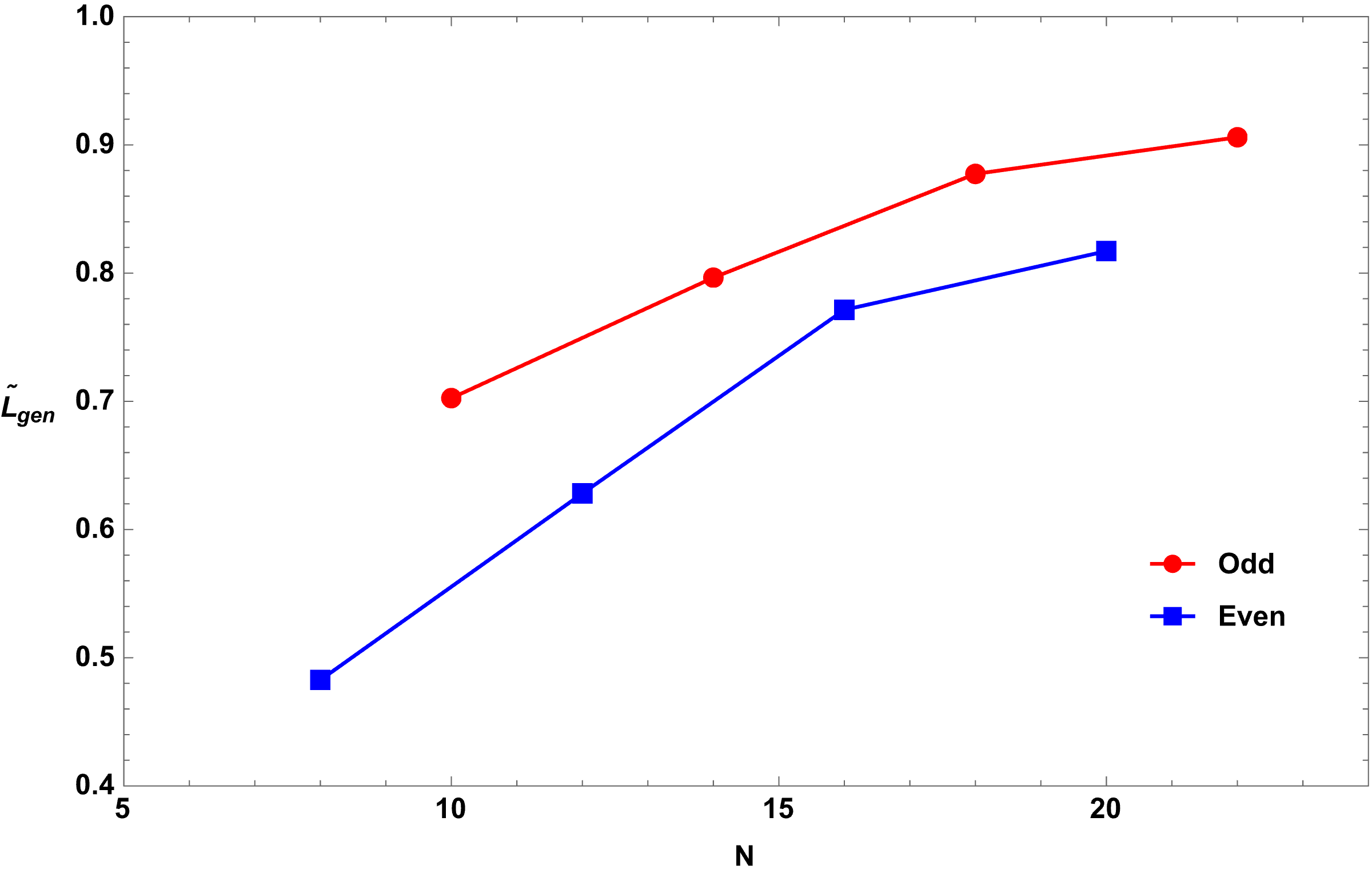}
    \caption{$\tilde{L}_{gen}^{sat}$vs $N$ }\label{SYK4sat}
  \end{subfigure}\quad
  
\caption{\footnotesize{
(a) Normalized L-entropy distribution across the energy eigenstates of the $SYK_4$ model for $N = 18, 20,$ and $22$ (9, 10, and 11 qubits) ;  (b) The saturation value of the generalized L-entropy versus $N$;  (c) The difference in the saturation values between the $SYK_4$ and $SYK_2$ models.}}
\end{figure}

Figures~\Cref{SYKall} show the time evolution of the generalized L-entropy starting from the fully separable initial state $\ket{000\ldots0}$, averaged over several random realizations. The generalized L-entropy increases rapidly and saturates close to unity, indicating the emergence of strong multipartite entanglement in the interacting $SYK_4$ dynamics. As in the $SYK_2$ case, dependence on the number of parties being odd or even persists: both odd and even $n = N/2$ exhibit similar qualitative behavior, but the odd-$n$ systems consistently reach slightly higher saturation values. The overall trend, shown in \Cref{SYK4sat}, reveals that the saturation value grows monotonically with $N$, approaching the generalized L-entropy characteristic of AME-like states in the large-$N$ limit.

\begin{figure}[H]
  \centering
  \begin{subfigure}{.45\linewidth}
    \includegraphics[height=3.5cm,width=\linewidth]{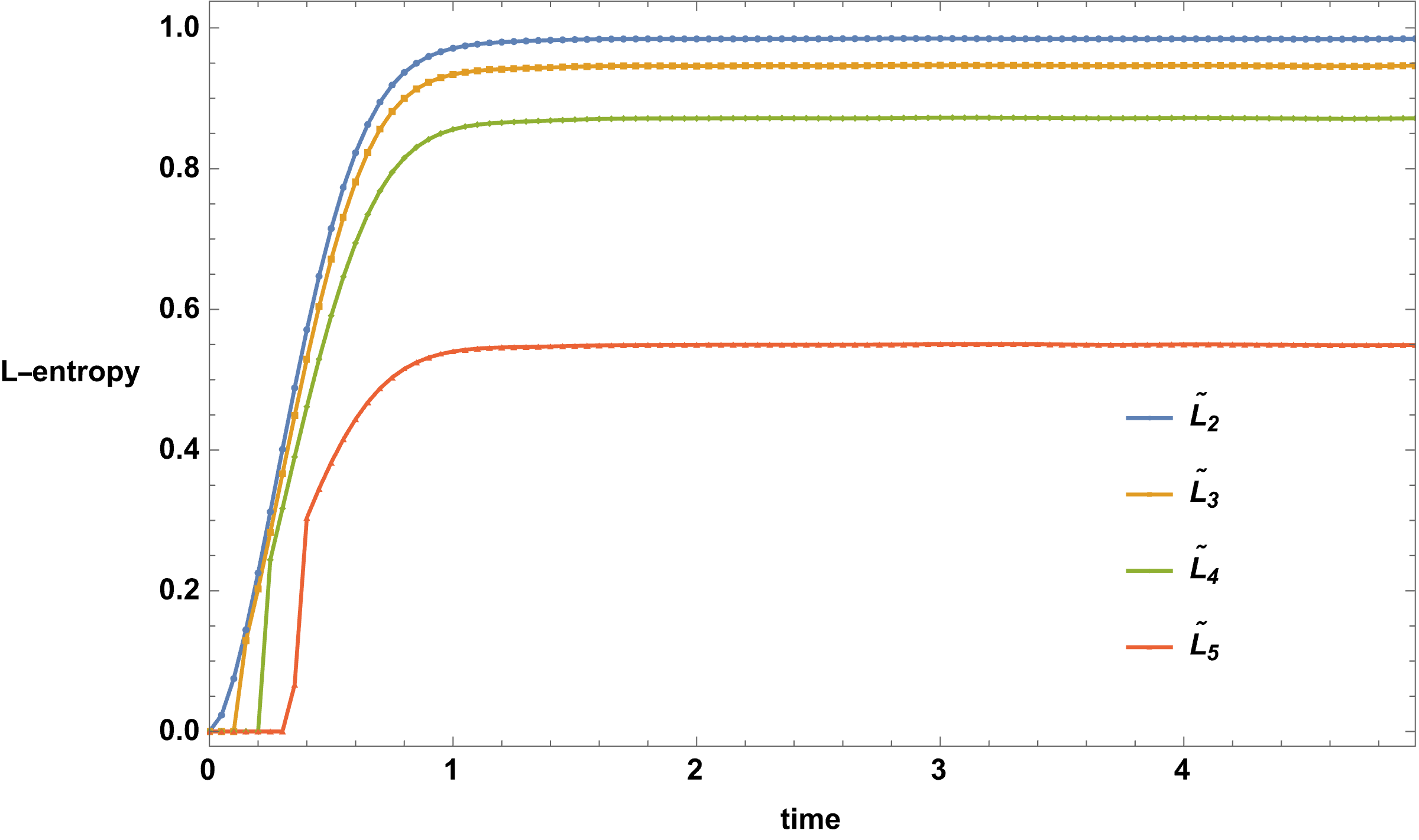}
    \subcaption{$N=20$ (10 qubits)}\label{SYKlower10}
  \end{subfigure}\quad
  \begin{subfigure}{.45\linewidth}
\includegraphics[height=3.5cm,width=\linewidth]{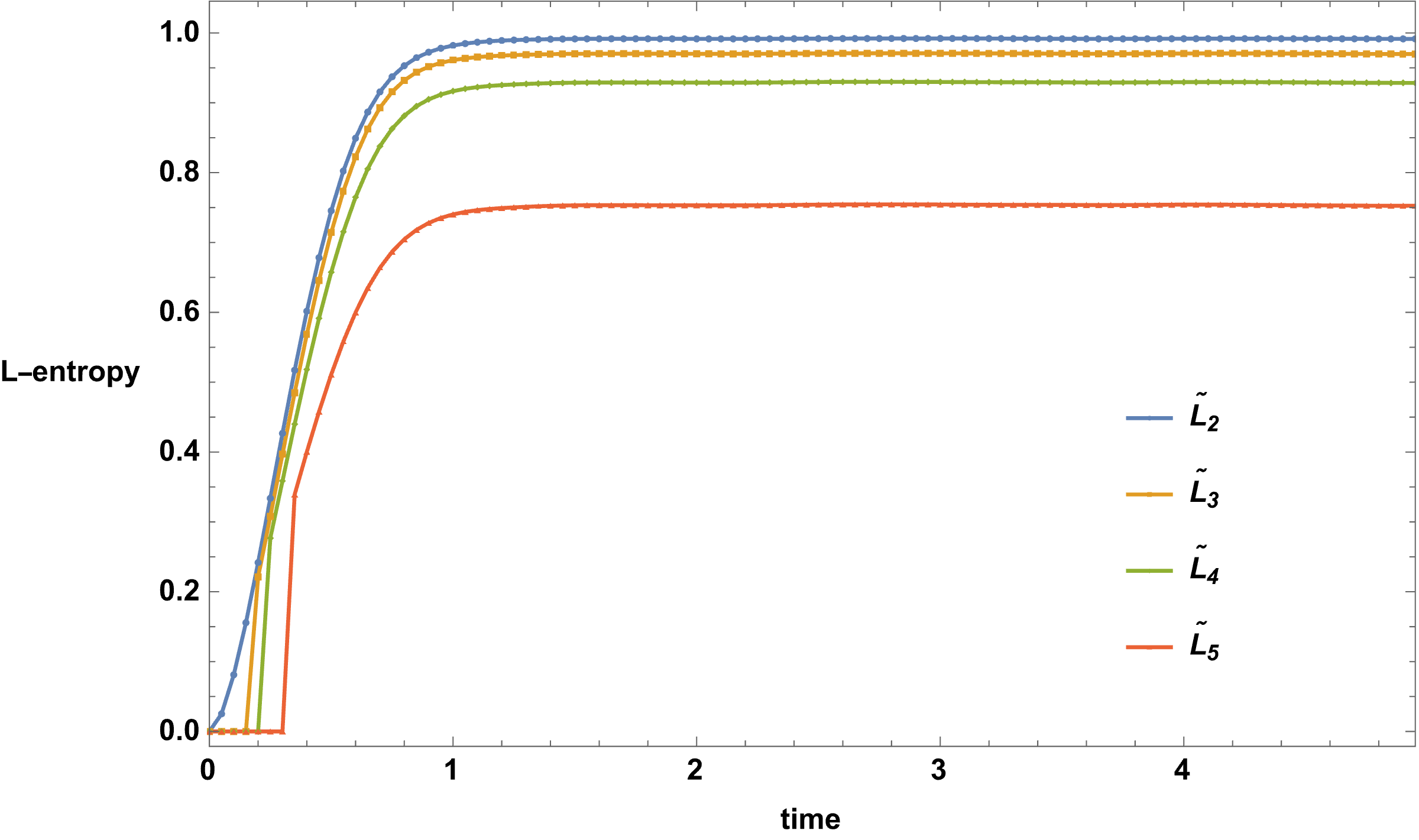}
    \caption{$N=22$ (11 qubits)}\label{SYKlower11}
  \end{subfigure}\quad
  \caption{\footnotesize{
Time evolution of all the generalized L-entropies  starting from the initial state $\ket{000\cdots0}$, averaged over 25 samples.}}

\end{figure}
In \Cref{SYKlower10} and \Cref{SYKlower11}, the behavior of the lower L-entropies for the $SYK_4$ model is shown. Similar to the $SYK_2$ case, the different $\tilde{L}_k$ components exhibit clear saturation, with higher-$k$ terms reaching lower final values. In \Cref{Lenteig11SYK4} we plot the generalized L-entropy $\tilde{L}_{\mathrm{gen}}$ for the eigenstates of the $SYK_4$ Hamiltonian for several system sizes. The data form distinct, well-separated bands for different $N$, each showing a mild variation across the spectrum. In contrast, the corresponding $SYK_2$ results shown in \Cref{Lenteig11SYK2} display a single overlapping band where data for different $N$ nearly coincide.
 Figure \ref{SYK4sat} shows the variation of the saturation value of the generalized L-entropy with system size  for the $SYK_4$ model, exhibiting an initial increase that appears to asymptote toward unity for large $N$. 
\begin{figure}[H]
\centering
\begin{subfigure}{.3\linewidth}
\includegraphics[width=\linewidth]{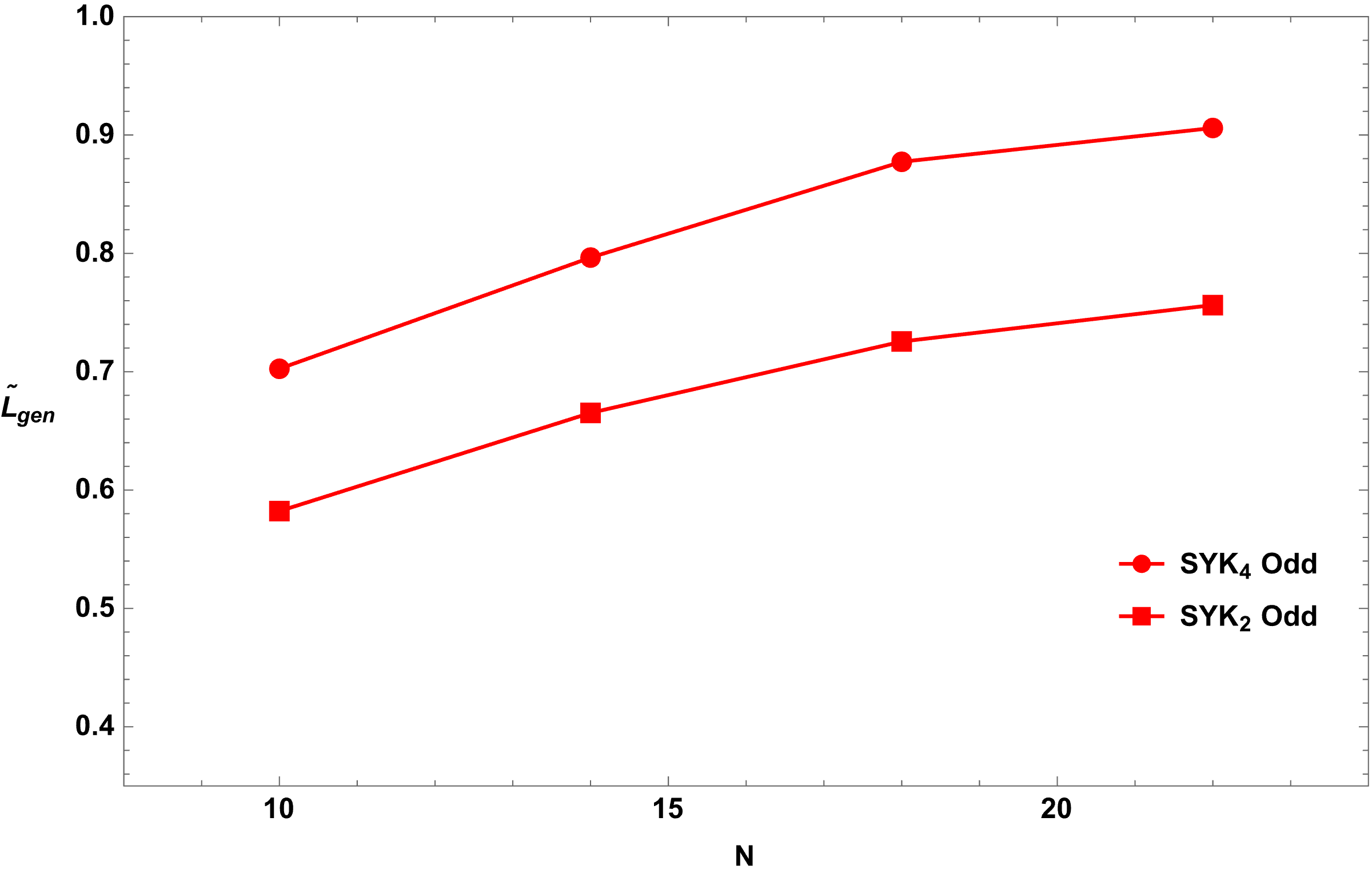}
    \caption{ $\tilde{L}_{gen}^{sat}$ vs $N$ for odd n}\label{SYK4SYK2c1}
  \end{subfigure}
  \begin{subfigure}{.3\linewidth}
\includegraphics[width=\linewidth]{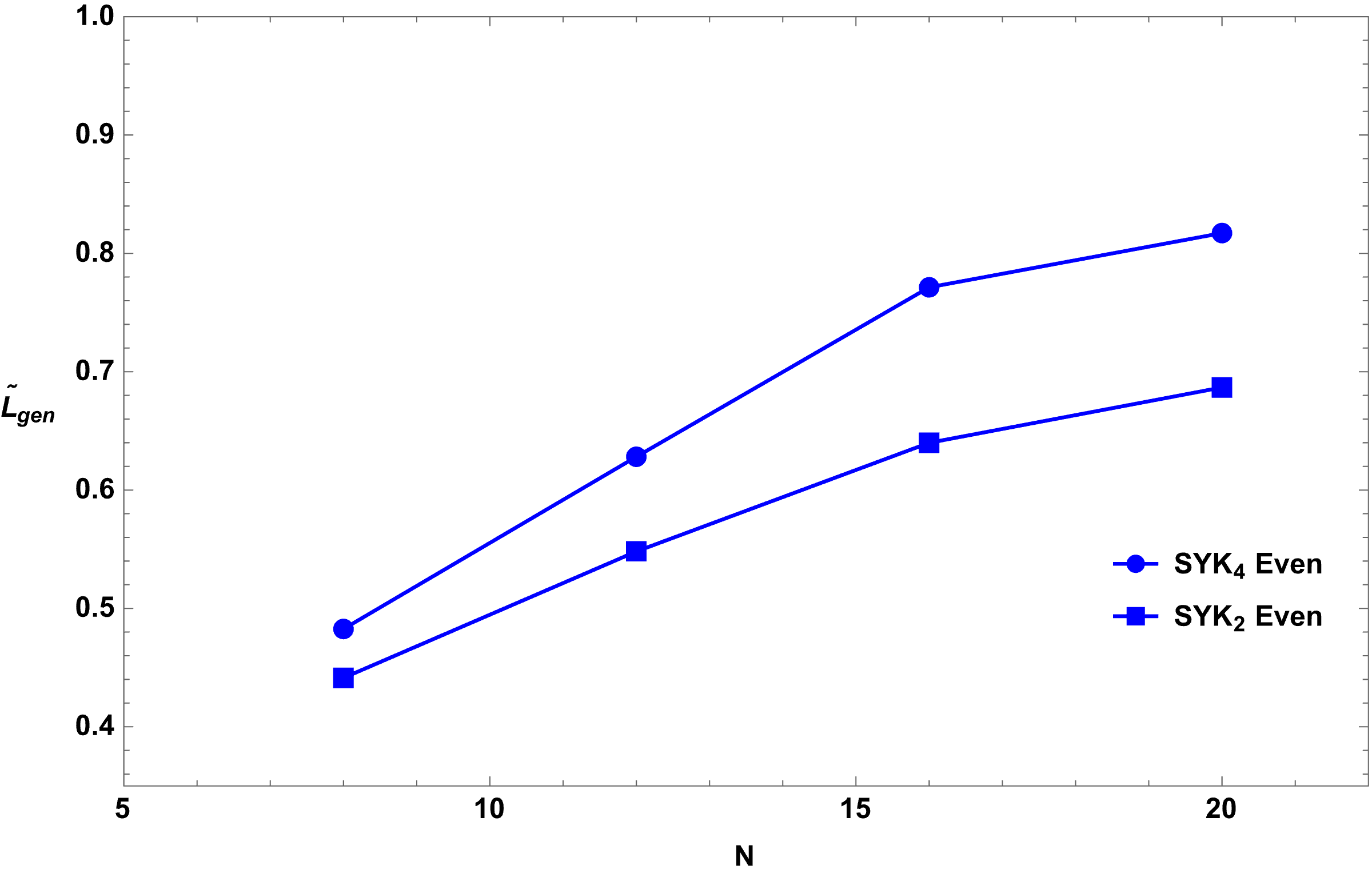}
    \caption{ $\tilde{L}_{gen}^{sat}$ vs $N$ for even n}\label{SYK4SYK2c2}
  \end{subfigure}
 \begin{subfigure}{.3\linewidth}
\includegraphics[width=\linewidth]{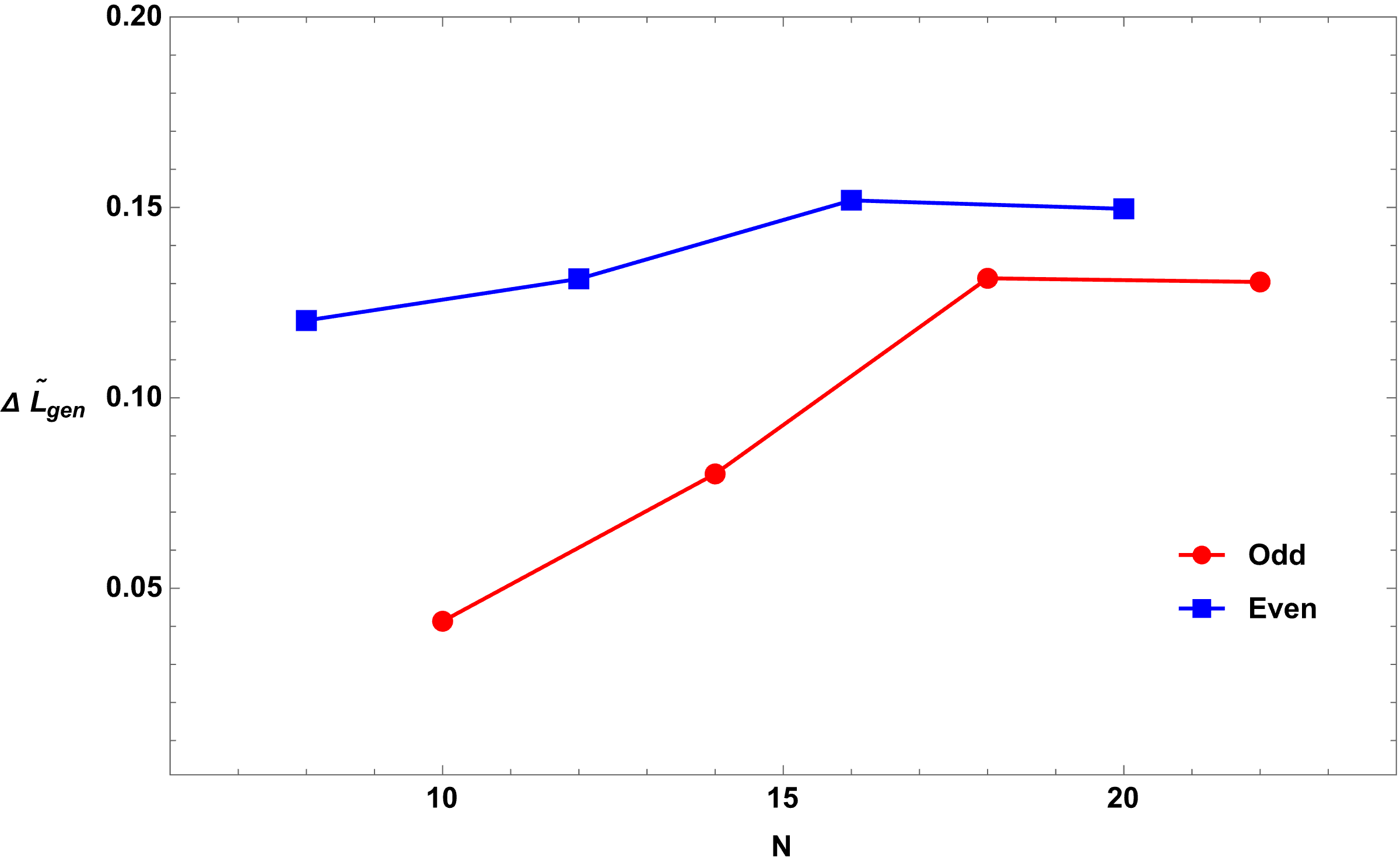}
    \caption{ $\Delta\tilde{L}_{gen}^{sat}$ vs $N$}\label{SYK4SYK2diff}
  \end{subfigure}
  \caption{\footnotesize{
(a) Normalized L-entropy distribution across the energy eigenstates of the $SYK_4$ and $SYK_2$ together;  (c) The difference in the saturation values between the $SYK_4$ and $SYK_2$ models.}}
\end{figure}
 In \Cref{SYK4SYK2c1} and \Cref{SYK4SYK2c2}, we compare the saturation values of the generalized L-entropy for the $SYK_4$ and $SYK_2$ models, and in \Cref{SYK4SYK2diff} we plot the difference between them. From the available data, it is clear that the saturation value for the $SYK_2$ model remains consistently lower than that of $SYK_4$, indicating that the asymptotic value of the generalized L-entropy in $SYK_2$ approaches a constant strictly less than unity. The gap between the saturation values of the two models, shown in \Cref{SYK4SYK2diff}, increases with $N$ for smaller system sizes and then appears to level off. However, determining its precise large-$N$ behavior would require simulations at larger system sizes beyond our current numerical reach.
\subsection{\textbf{Mass Deformed SYK}}
We next consider the mass-deformed SYK model, whose Hamiltonian includes both quadratic and quartic interactions, controlled by an interpolation parameter $g$: \cite{Garcia-Garcia:2017bkg,Nosaka:2018iat}
\begin{align}
H = (1-g)\, H_{\mathrm{SYK}4} + g\, H_{\mathrm{SYK}_2}.
\end{align}
Since the $SYK_2$ model is integrable while the $SYK_4$ model exhibits quantum chaos, this deformation provides a convenient framework to study the crossover from integrable to chaotic behavior as $g$ is varied. 

\begin{figure}[H]
  \centering
  \begin{subfigure}{.3\linewidth}
    \includegraphics[height=3.5cm,width=\linewidth]{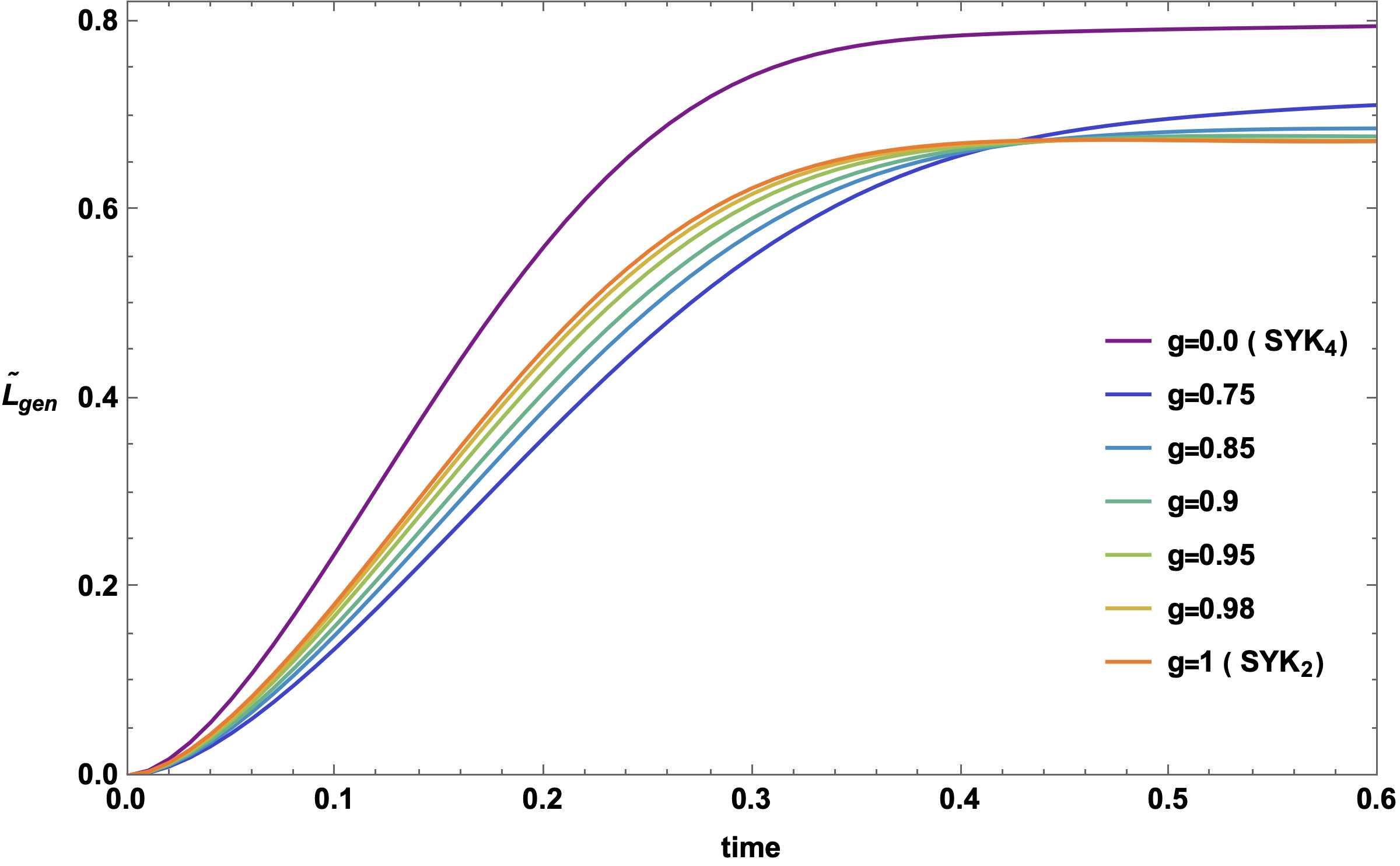}
    \subcaption{}\label{SYKMD1}
  \end{subfigure}\quad
  \begin{subfigure}{.3\linewidth}
    \includegraphics[height=3.5cm,width=\linewidth]{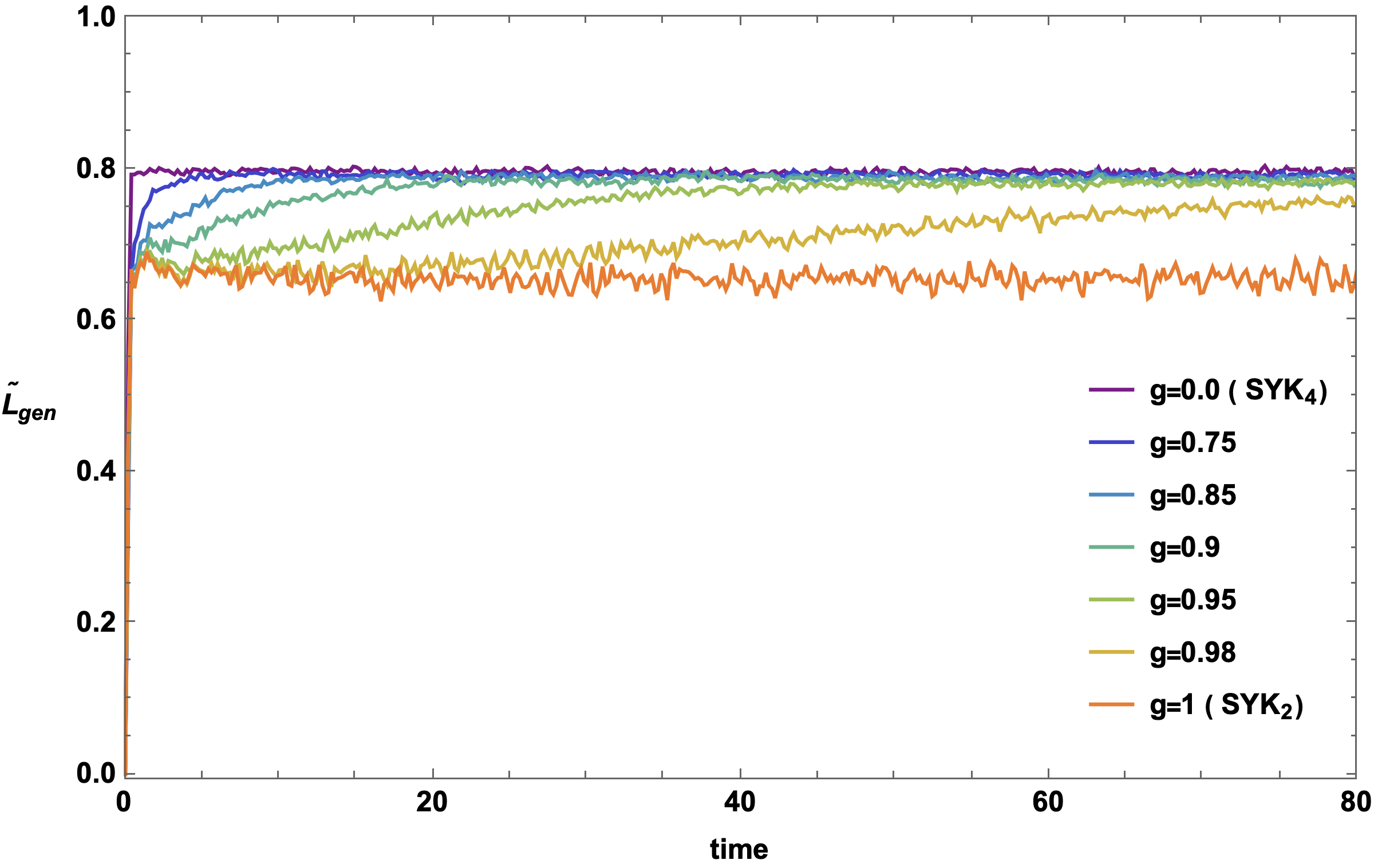}
    \caption{}\label{SYKMD2}
  \end{subfigure}
  \begin{subfigure}{.35\linewidth}
    \includegraphics[height=3.5cm,width=\linewidth]{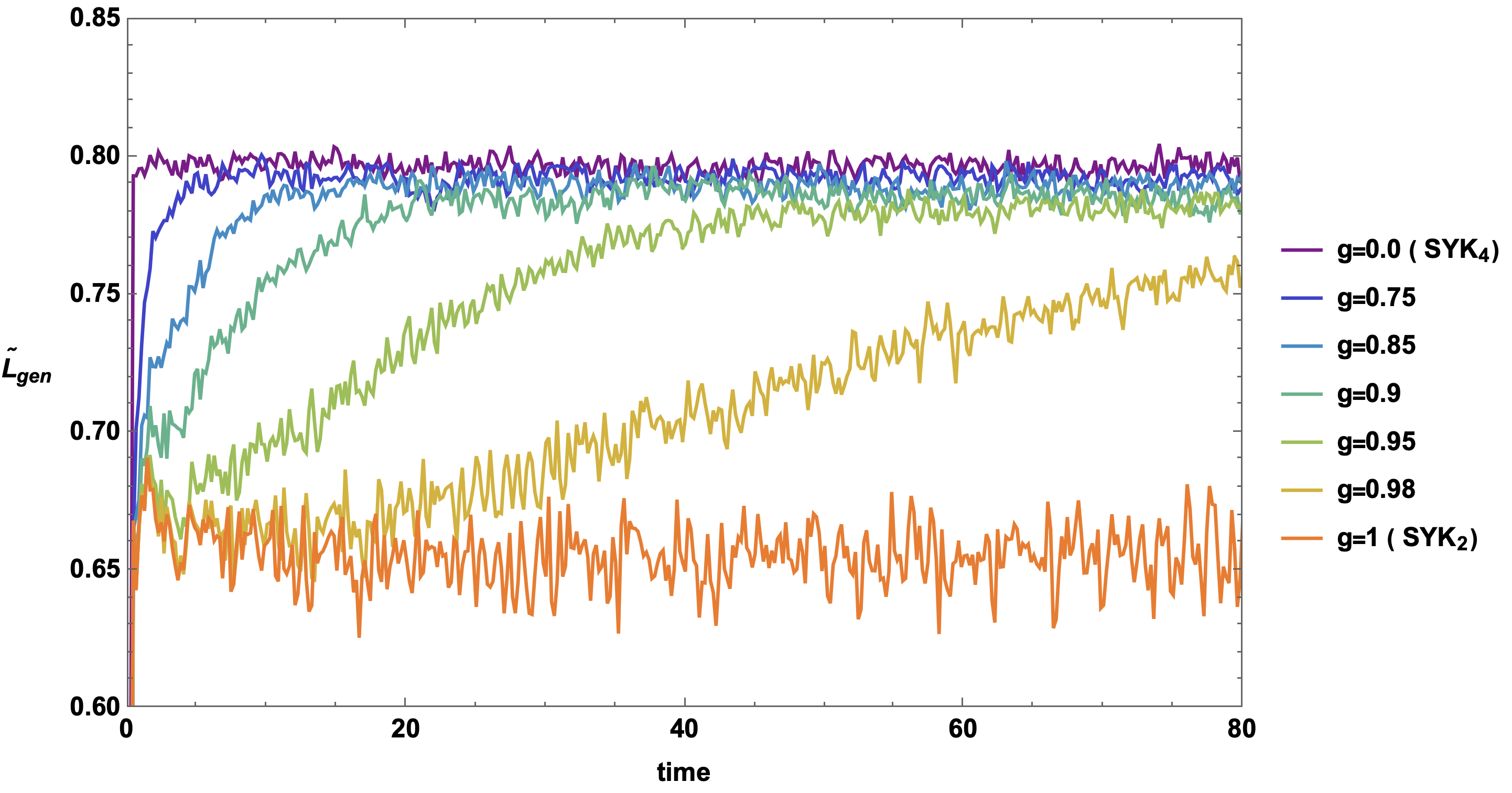}
    \caption{}\label{SYKMD3}
  \end{subfigure}
 \caption{\footnotesize{
Panels for $N = 14$ (7 qubits) show the time evolution of the generalized L-entropy starting from the initial state $\ket{000\cdots0}$, averaged over 50 samples. The plots illustrate the transition of the late-time saturation value as soon as the deformation parameter $g$ varies from the $SYK_2$ value.}}
\end{figure}

\begin{figure}[H]
  \centering
  \begin{subfigure}{.3\linewidth}
\includegraphics[height=3.5cm,width=\linewidth]{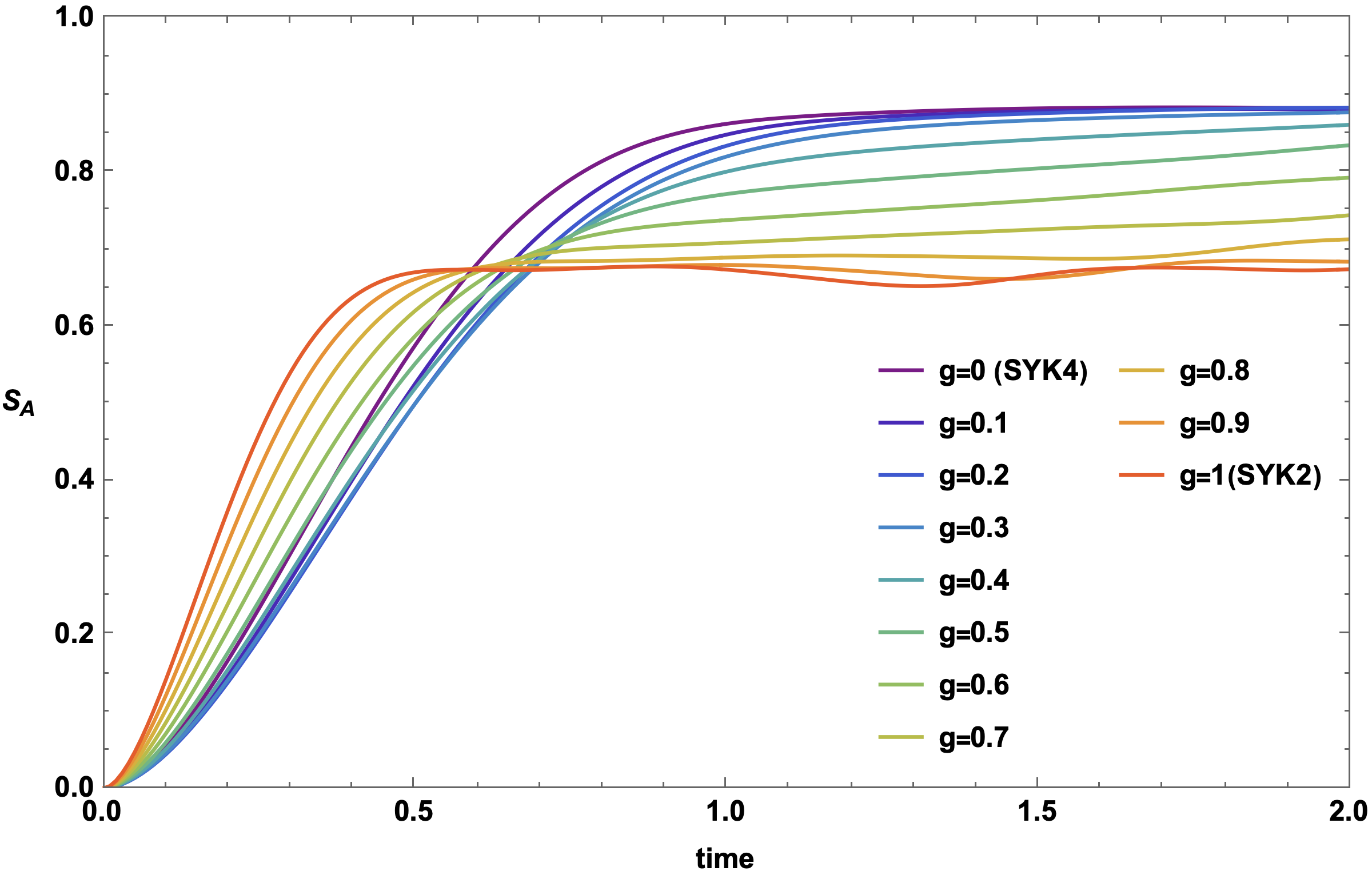}
\subcaption{}\label{EESYKMD10}
  \end{subfigure}\quad
  \begin{subfigure}{.3\linewidth}
\includegraphics[height=3.5cm,width=\linewidth]{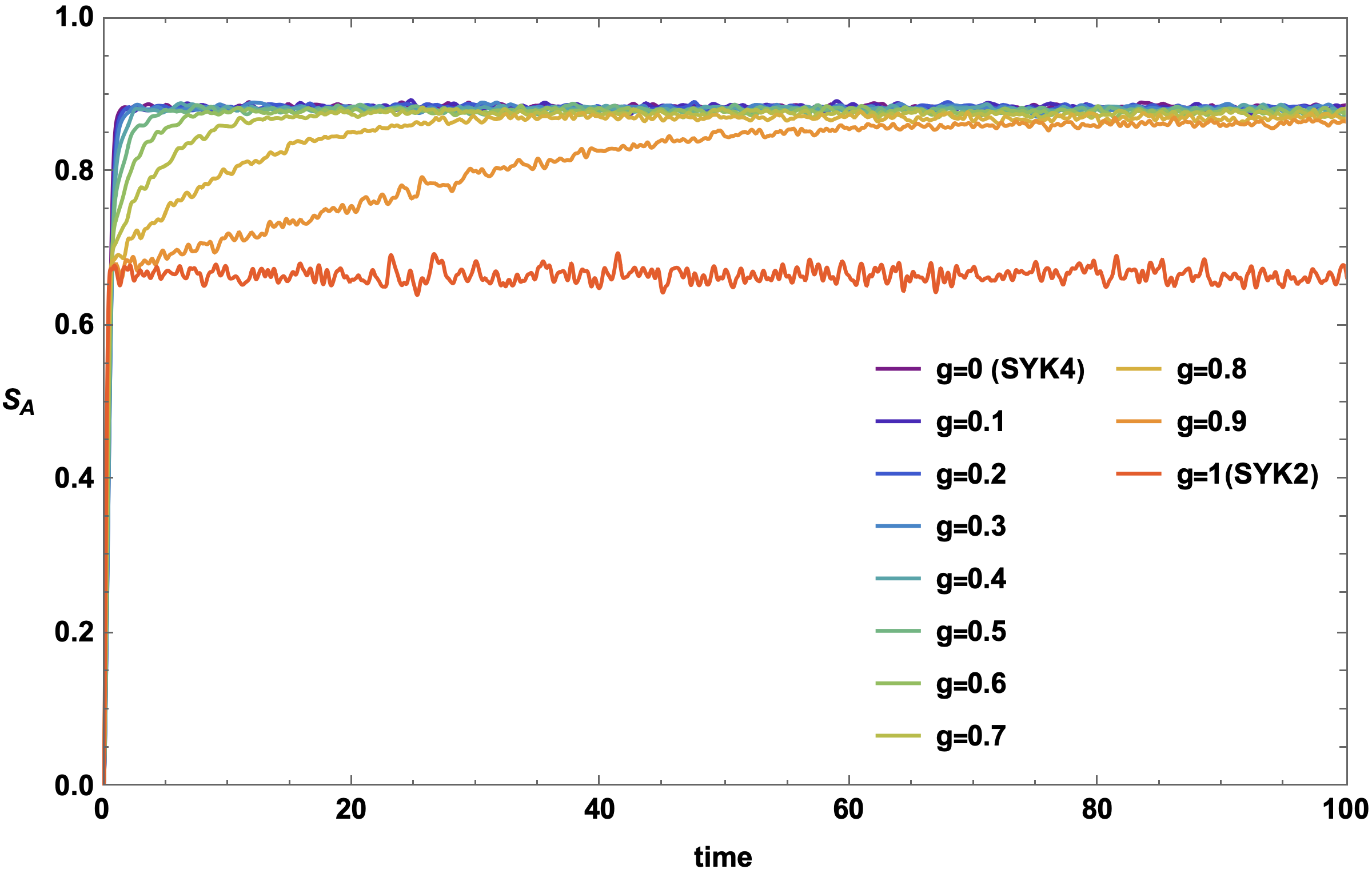}
    \caption{}\label{EESYKMD20}
  \end{subfigure}\quad
  \begin{subfigure}{.3\linewidth}
\includegraphics[height=3.5cm,width=\linewidth]{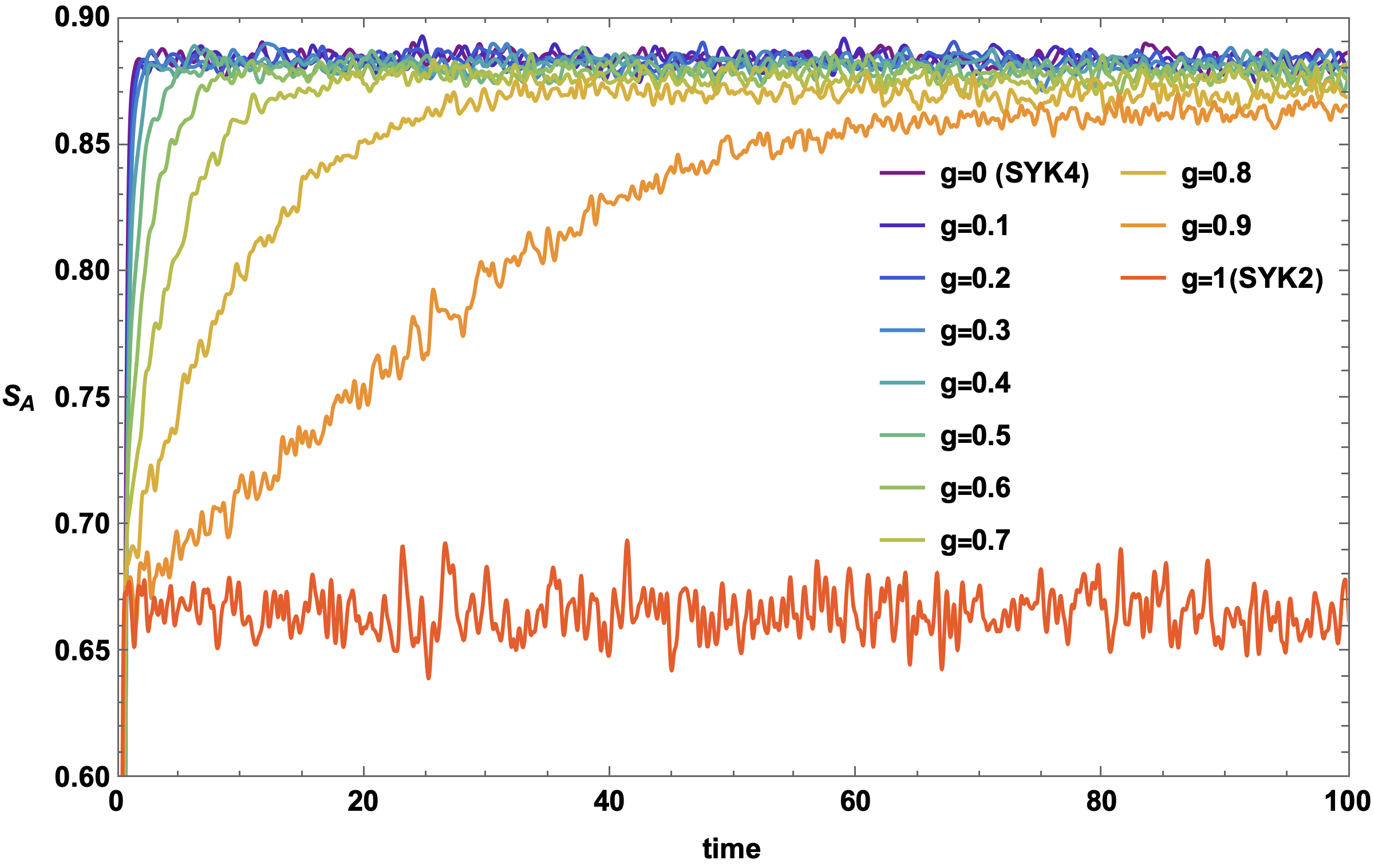}
    \caption{}\label{EESYKMD30}
  \end{subfigure}
   \caption{\footnotesize{
Panel (a) shows the time evolution of the normalized entanglement entropy (of $n_A$= 3 qubits) for $N = 14$ (7 qubits), starting from the initial state $\ket{000\cdots0}$ and averaged over 100 samples. The plots illustrate the transition of the late-time saturation value from the $SYK_2$ to the $SYK_4$ limit as the deformation parameter $g$ is varied.}}

\end{figure}

We depict that the behavior of the generalized L-entropy and the bipartite entanglement entropy in the mass-deformed SYK model, shown in \Cref{SYKMD2},  \Cref{SYKMD3}, \Cref{EESYKMD20}, and \Cref{EESYKMD30},    is qualitatively similar. The early-time evolution of the two quantities is displayed in \Cref{SYKMD1} and \Cref{EESYKMD10}, respectively. Remarkably, even a minute addition of the $SYK_4$ interaction is sufficient to drive both the entanglement entropy and the generalized L-entropy to saturate at values characteristic of the $SYK_4$ model, as illustrated in \Cref{SYKMD1}. The deformation parameter $g$ thus governs the rate of late-time growth and determines the time scale at which saturation is reached.

\begin{figure}[H]
  \centering
    \begin{subfigure}{.3\linewidth}
\includegraphics[height=3.5cm,width=\linewidth]{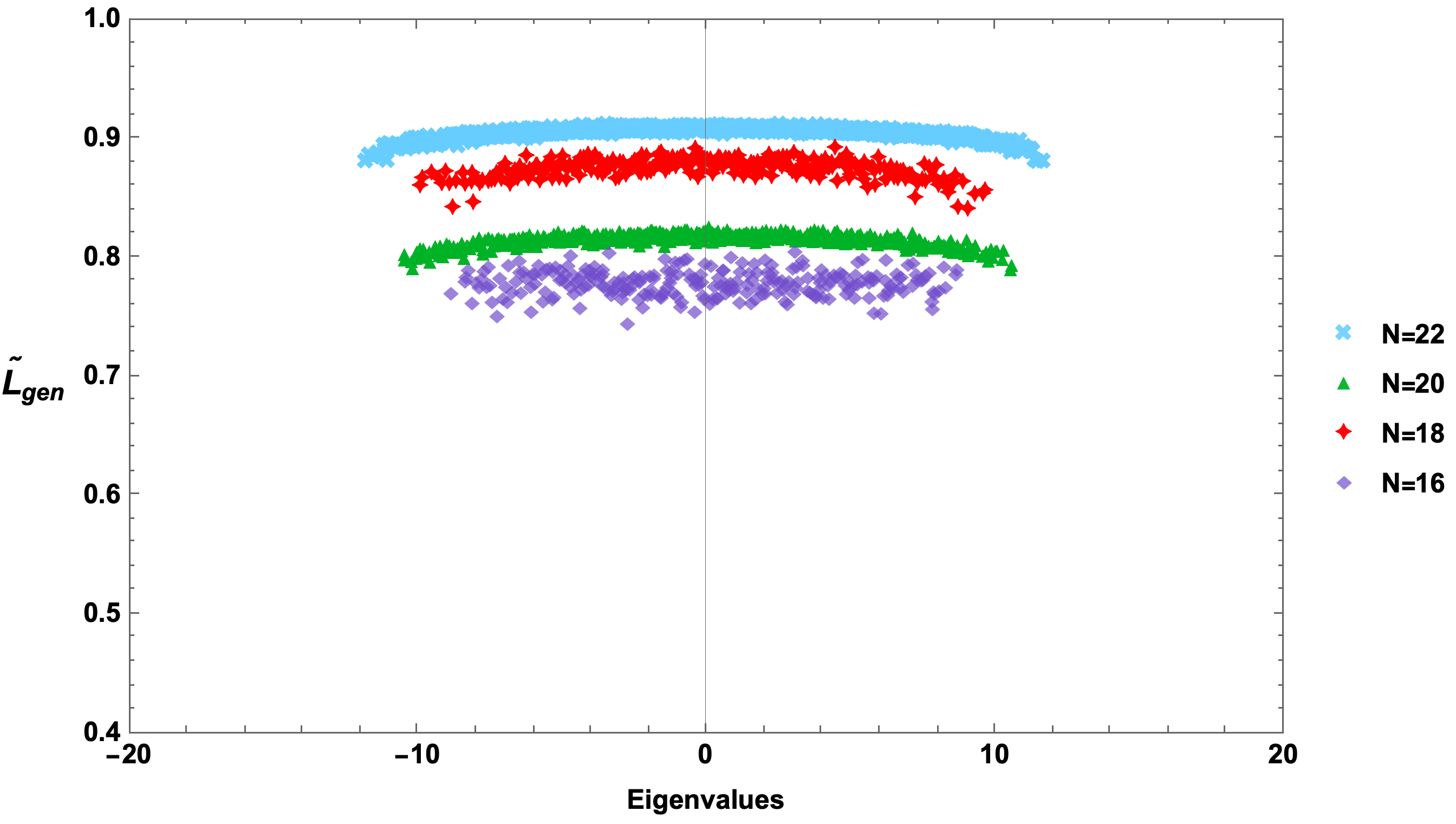}
    \caption{$g=0$ ($SYK_4$)}\label{SYKeig6}
  \end{subfigure}\quad
   \begin{subfigure}{.3\linewidth}
    \includegraphics[height=3.5cm,width=\linewidth]{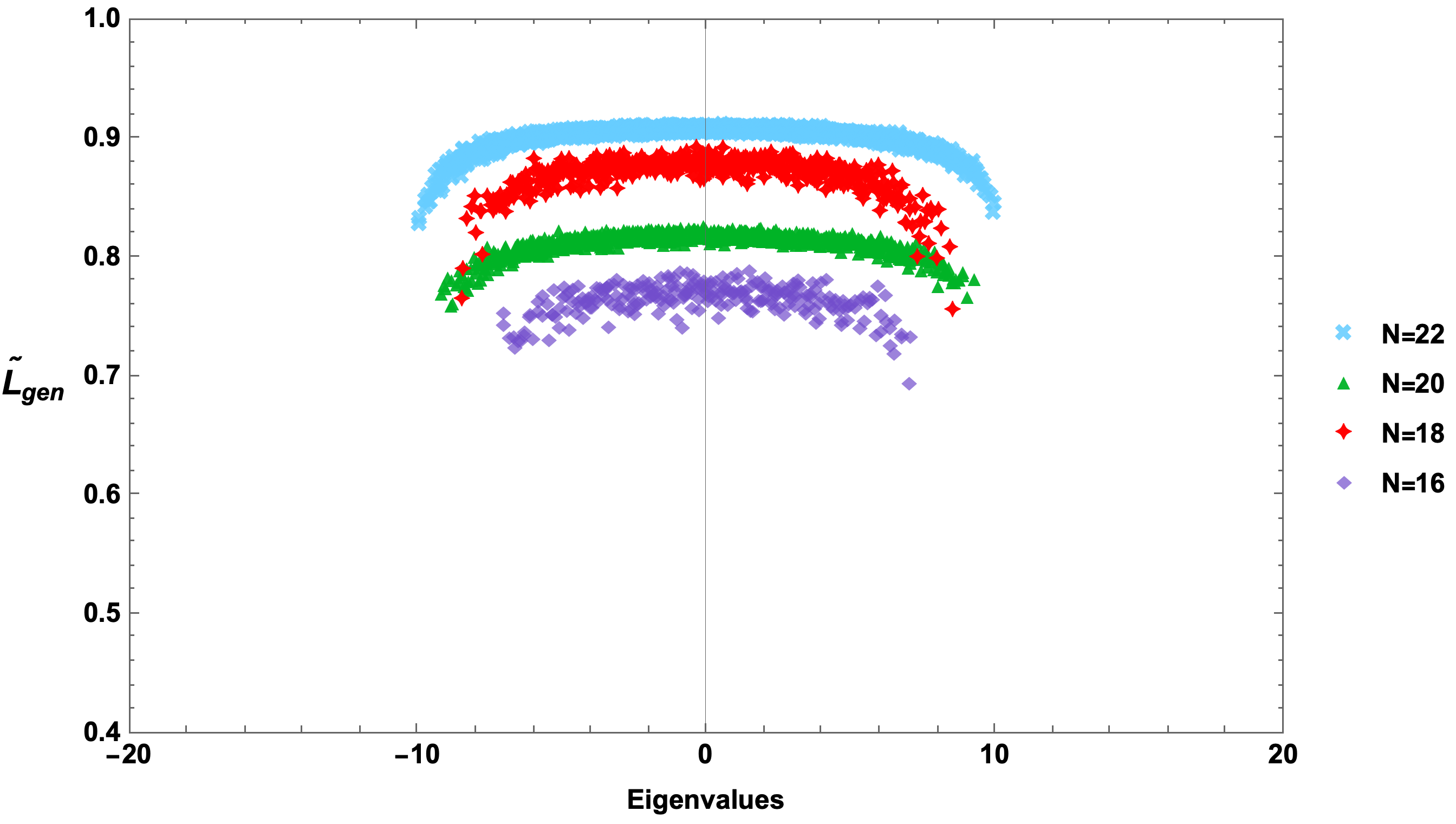}
    \caption{$g=0.25$ }\label{SYKeig5}
  \end{subfigure}\quad
    \begin{subfigure}{.3\linewidth}
    \includegraphics[height=3.5cm,width=\linewidth]{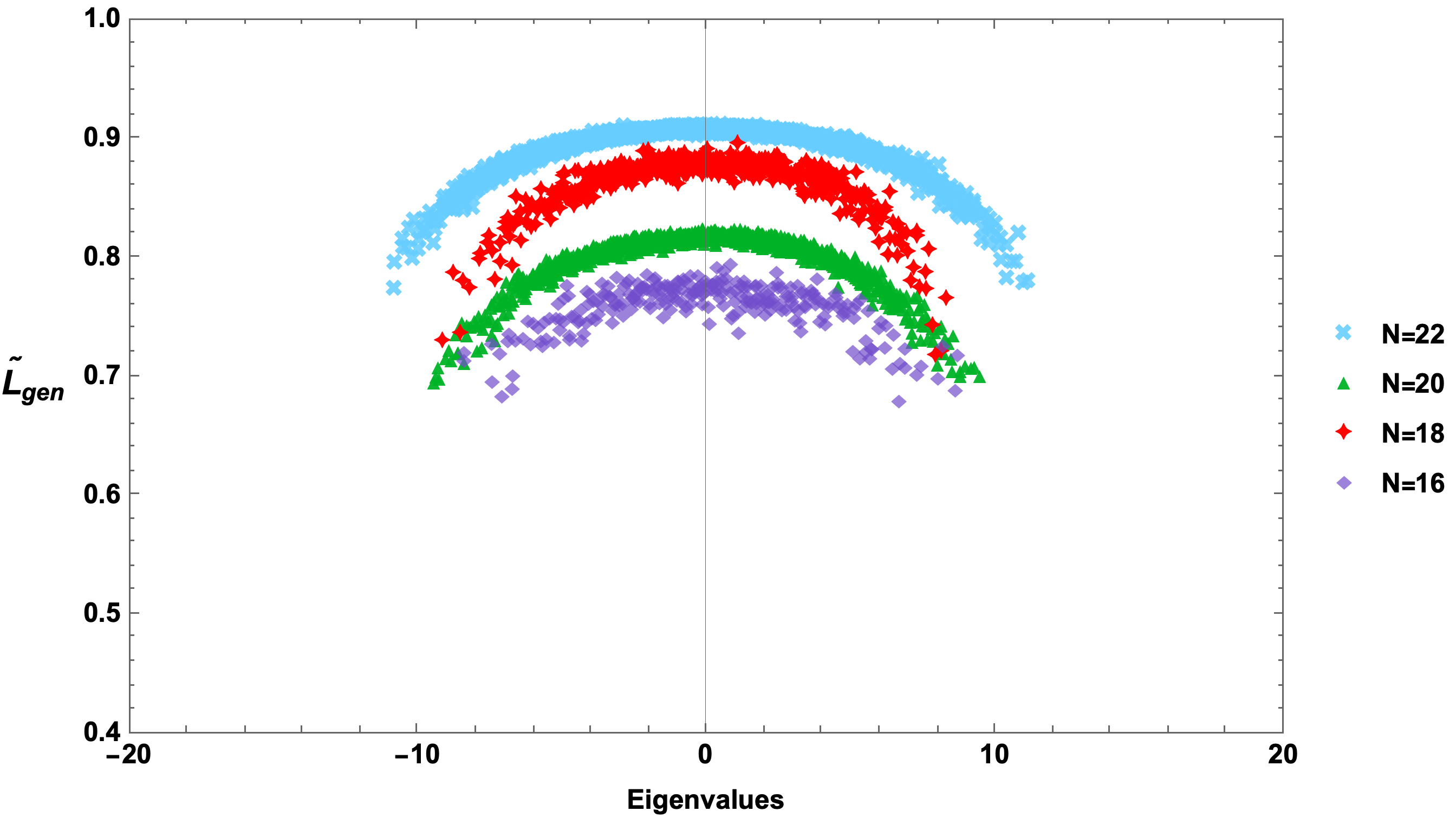}
    \caption{$g=0.5$ }\label{SYKeig4}
  \end{subfigure}\quad
   \begin{subfigure}{.3\linewidth}
    \includegraphics[height=3.5cm,width=\linewidth]{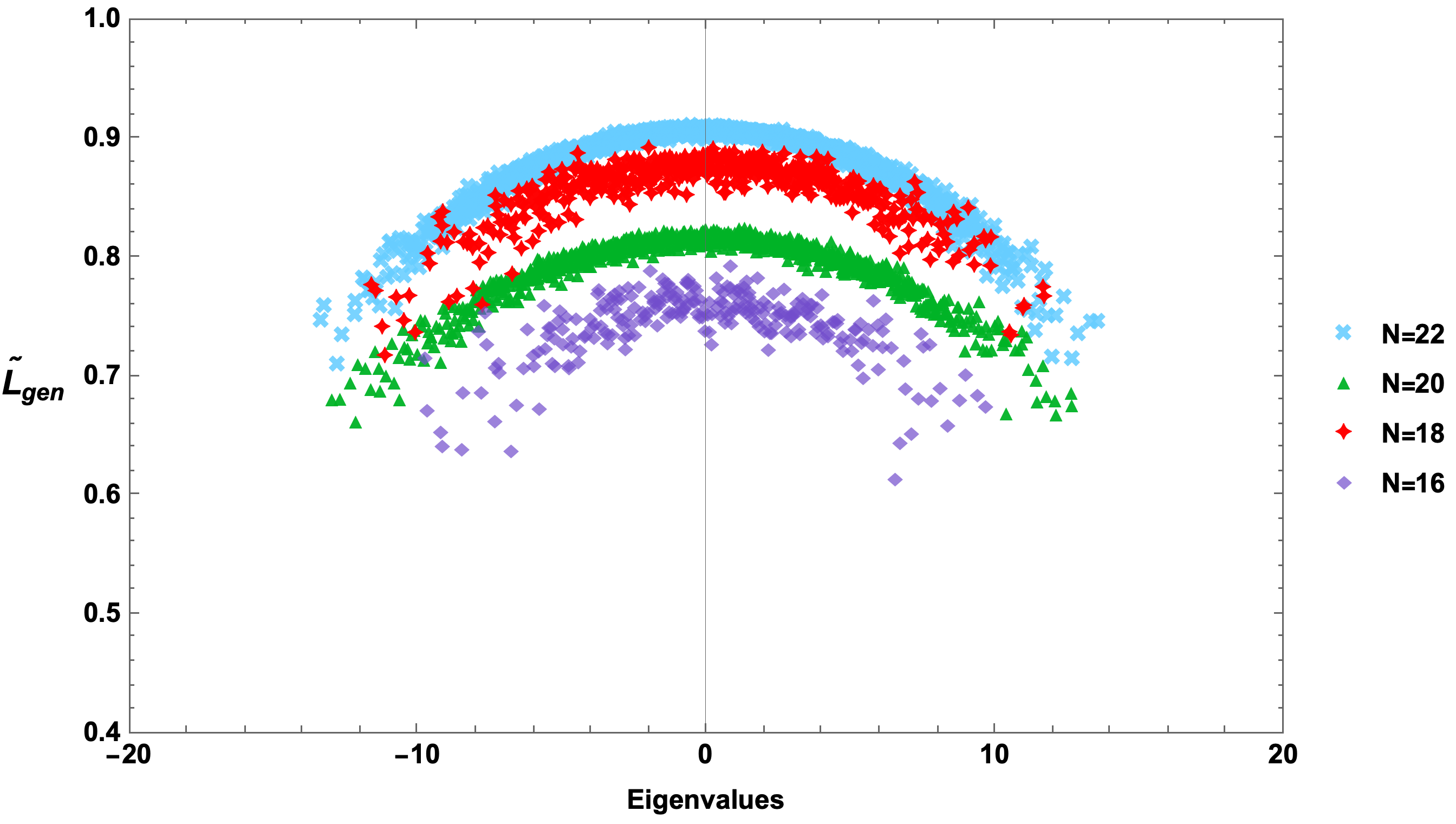}
    \caption{$g=0.75$ }\label{SYKeig3}
  \end{subfigure}\quad
  \begin{subfigure}{.3\linewidth}
    \includegraphics[height=3.5cm,width=\linewidth]{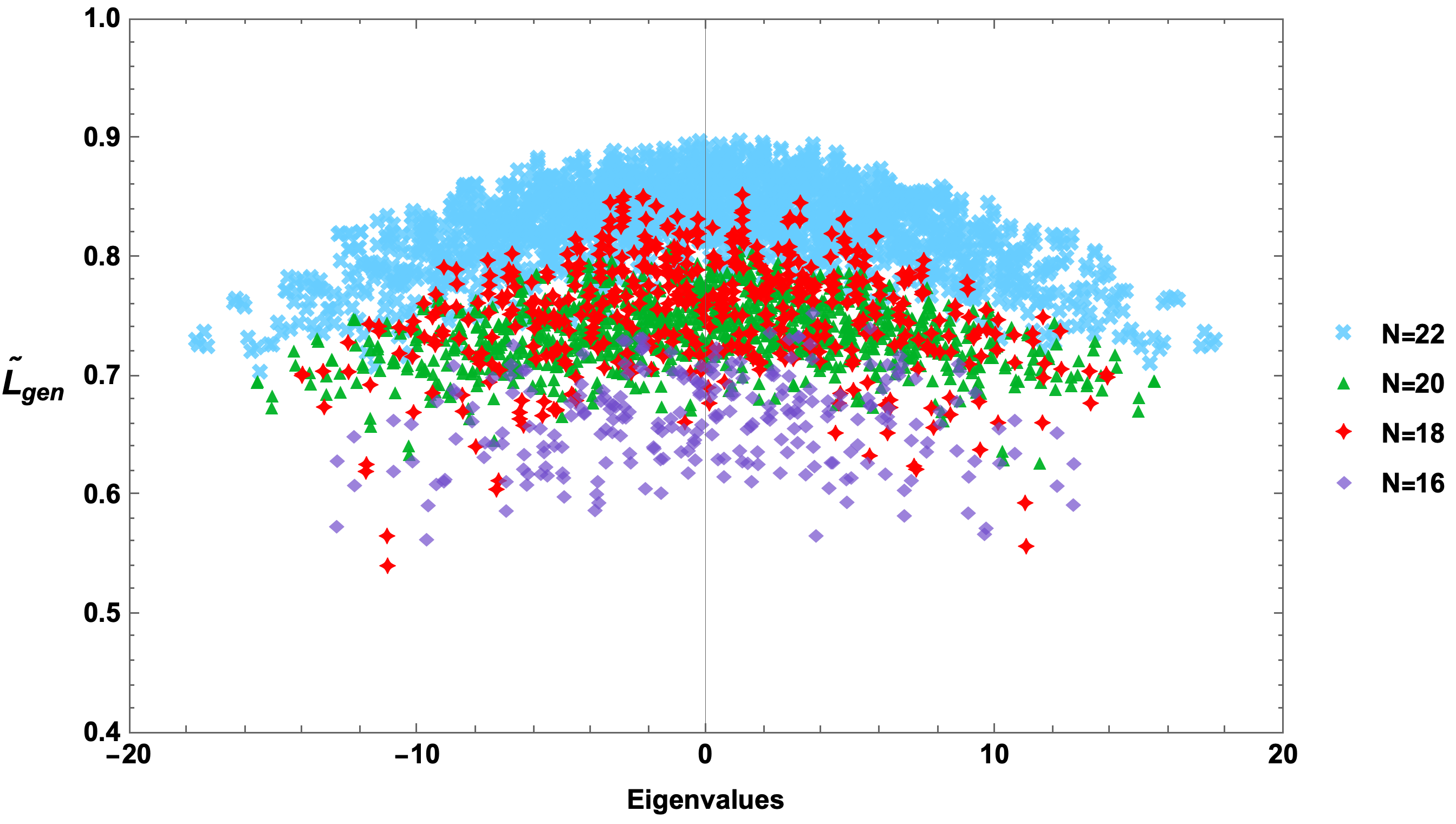}
    \caption{$g=0.95$ }\label{SYKeig2}
  \end{subfigure}\quad
  \begin{subfigure}{.3\linewidth}
    \includegraphics[height=3.5cm,width=\linewidth]{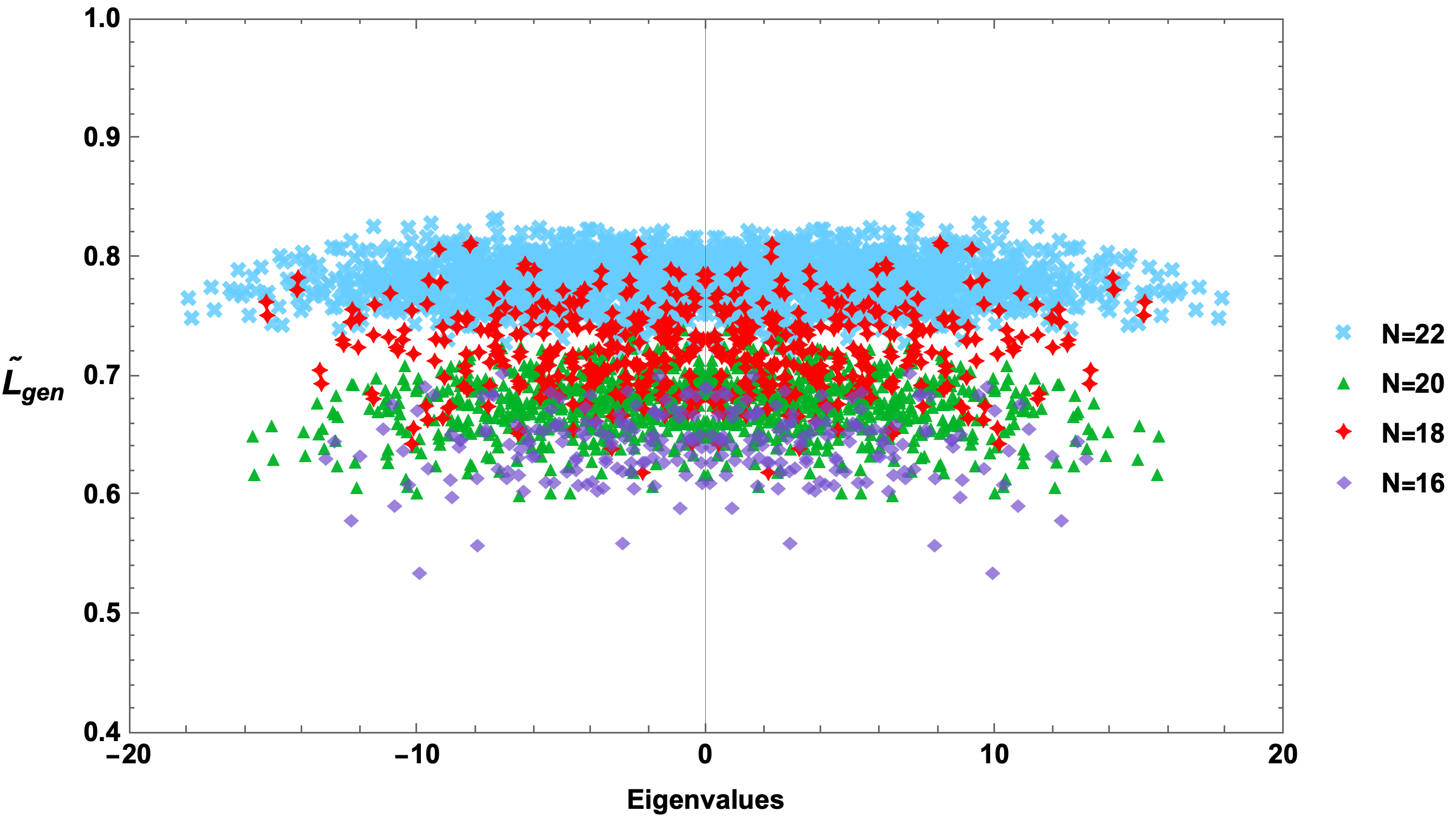}
    \subcaption{$g=1$ ($SYK_2$)}\label{SYKeig1}
  \end{subfigure}\quad

  \caption{\footnotesize{Eigenstate distribution of the generalized L-entropy across the mass-deformed SYK model. As the deformation parameter increases from the chaotic $SYK_4$ limit ($g=0$) to the integrable $SYK_2$ limit ($g=1$), the distinct narrow bands characteristic of $SYK_4$ evolve into a single overlapping band.}}\label{LenteigSYKMD}
\end{figure}

In \Cref{LenteigSYKMD}, we show the eigenstate distribution of the generalized L-entropy across the mass-deformed SYK Hamiltonian. As the deformation parameter is varied from the chaotic $SYK_4$ limit toward the integrable $SYK_2$ limit, the distinct narrow bands observed for $SYK_4$ gradually merge into a single overlapping band, indicating a  transition in the multipartite entanglement structure of the eigenstates.

\subsection{\textbf{Sparse SYK Model}}

\begin{figure}[H]
  \centering
  \begin{subfigure}{.3\linewidth}
    \includegraphics[height=3.5cm,width=\linewidth]{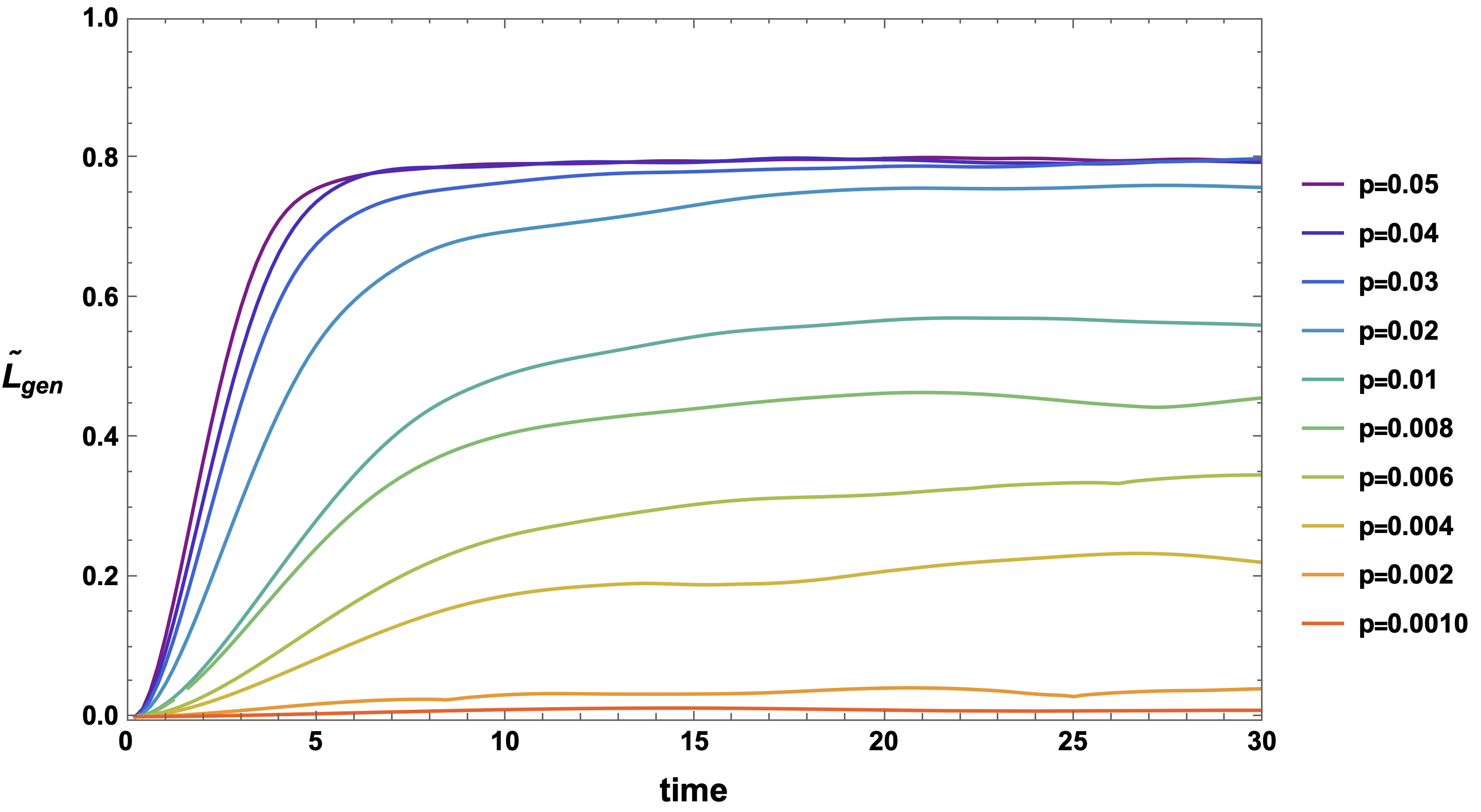}
    \caption{$N=14$ (7 qubits)}\label{SPSYK6q}
  \end{subfigure}\quad
  \begin{subfigure}{.3\linewidth}
\includegraphics[height=3.5cm,width=\linewidth]{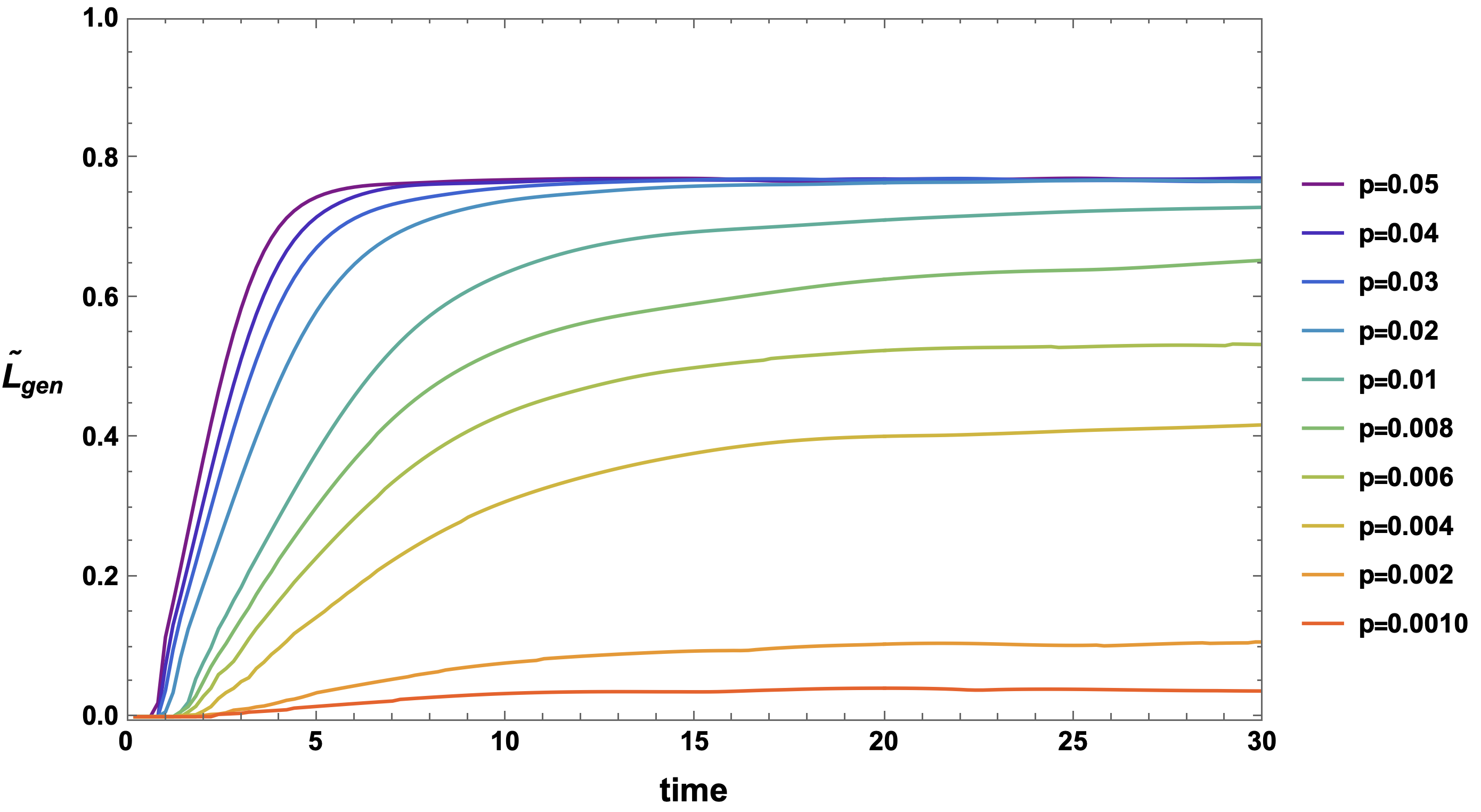}
    \caption{$N=16$ (8 qubits) }\label{SPSYK7q}
  \end{subfigure}\quad
  \begin{subfigure}{.3\linewidth}
\includegraphics[height=3.5cm,width=\linewidth]{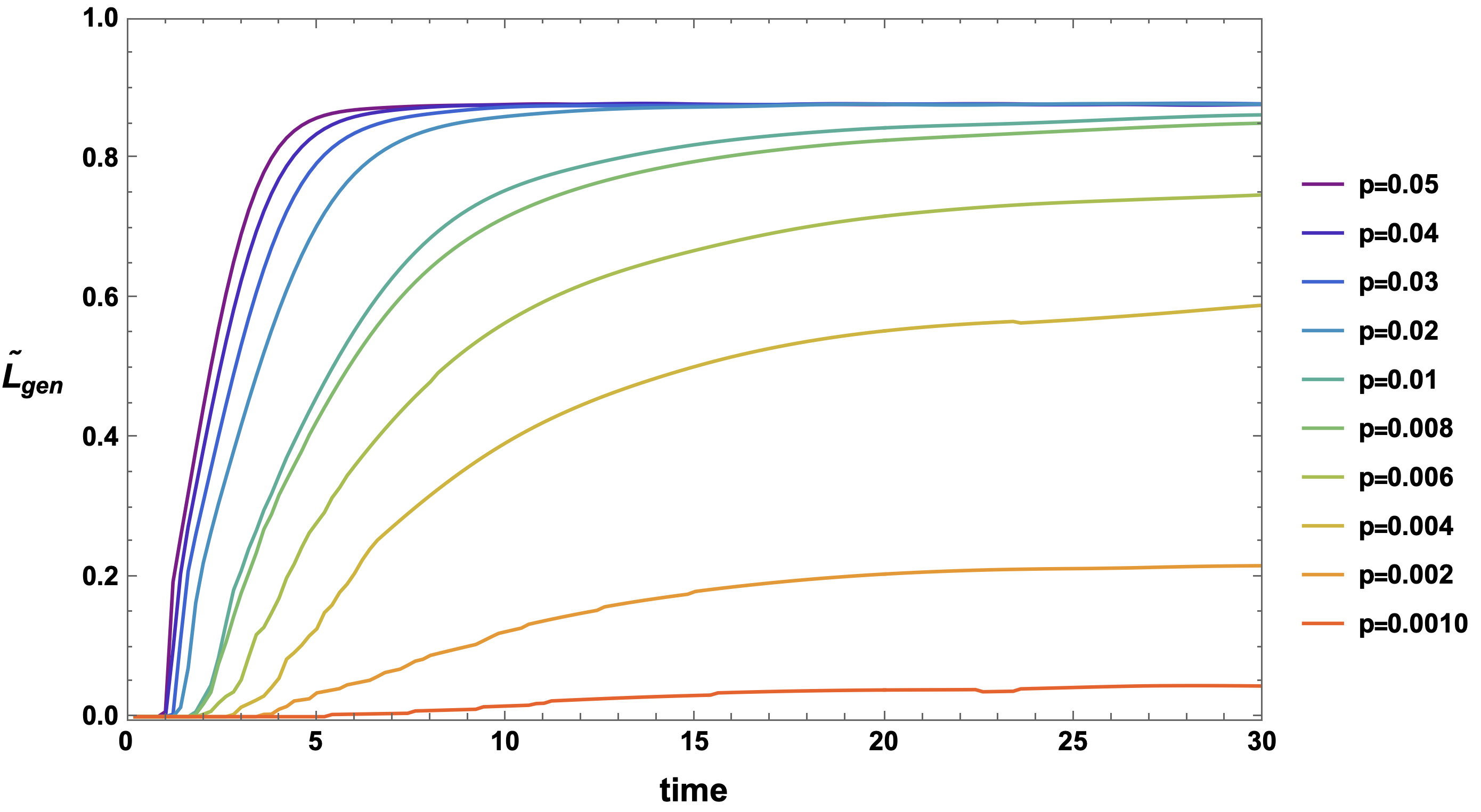}
    \caption{$N=18$ (9 qubits) }\label{SPSYK8q}
  \end{subfigure}
  \caption{\footnotesize{
The above panels show the time evolution of the L-entropy starting from the initial state $\ket{000\cdots0}$, averaged over 50 samples. The plots illustrate the change in the late-time saturation value as the sparsity parameter $p$, controlling the probability of nonzero couplings, is varied toward the less sparse regime.}}
\label{SPSYK}
\end{figure}

We next investigate a sparse variant of the $SYK_4$ model, as introduced in \cite{Xu:2020shn,Orman:2024mpw,Tezuka:2022mrr,Pathak:2025udi}. 
In this version, the all-to-all random four-fermion couplings of the standard $SYK_4$ Hamiltonian are diluted by introducing a sparsity parameter \( p \), which controls the fraction of non-vanishing interaction terms. 
Specifically, each coupling \( J_{ijkl} \), originally drawn from a Gaussian distribution 
, is retained with probability \( p \) and set to zero otherwise. 
The resulting Hamiltonian takes the form 
\begin{align}
    H_{\mathrm{sparse}} = \sum_{i<j<k<l} J_{ijkl}^{\mathrm{sparse}}\, \chi^i \chi^j \chi^k \chi^l,
\end{align}
with 
\begin{align}
J_{ijkl}^{\mathrm{sparse}} =
\begin{cases}
J_{ijkl}, & \text{with probability } p, \\[4pt]
0, & \text{with probability } 1 - p.
\end{cases}
\end{align}
The parameter \( p \) therefore interpolates between the fully connected $SYK_4$ model (\(p=1\)) and a highly diluted, nearly integrable limit (\(p \ll 1\)), allowing us to study how the structure of interactions influences the entanglement properties of the system. In \Cref{SPSYK}, we show the time evolution of the generalized latent entropy in the sparse $SYK$ model. A smooth transition is observed in the saturation value of the generalized $L$-entropy, which increases from values close to zero to significantly higher ones as the sparseness parameter $p$ is increased.

\begin{figure}[H]
  \centering
  \begin{subfigure}{.3\linewidth}
    \includegraphics[height=3.5cm,width=\linewidth]{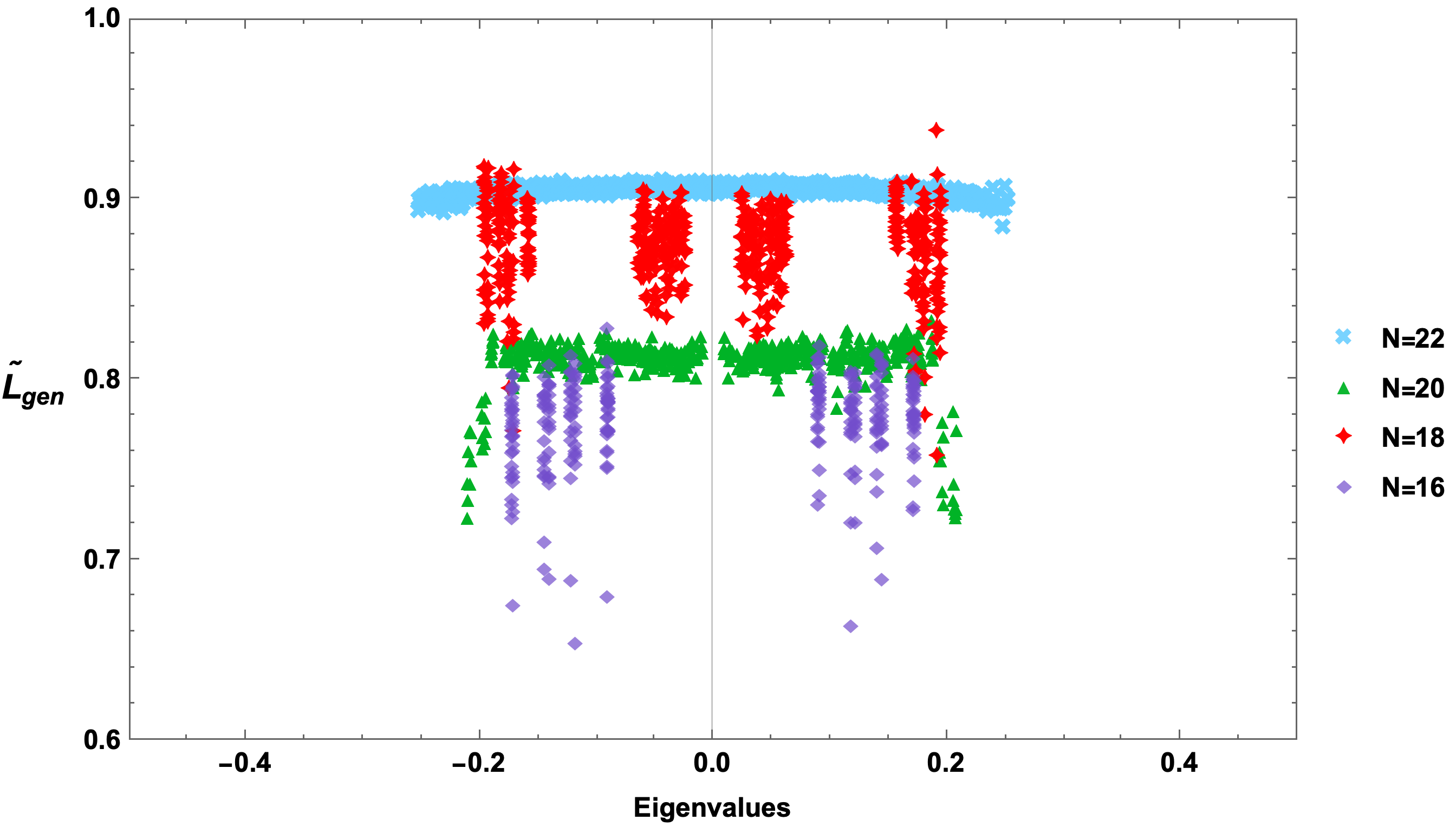}
    \caption{$p=0.005$}\label{SPSYKeig1}
  \end{subfigure}\quad
  \begin{subfigure}{.3\linewidth}
\includegraphics[height=3.5cm,width=\linewidth]{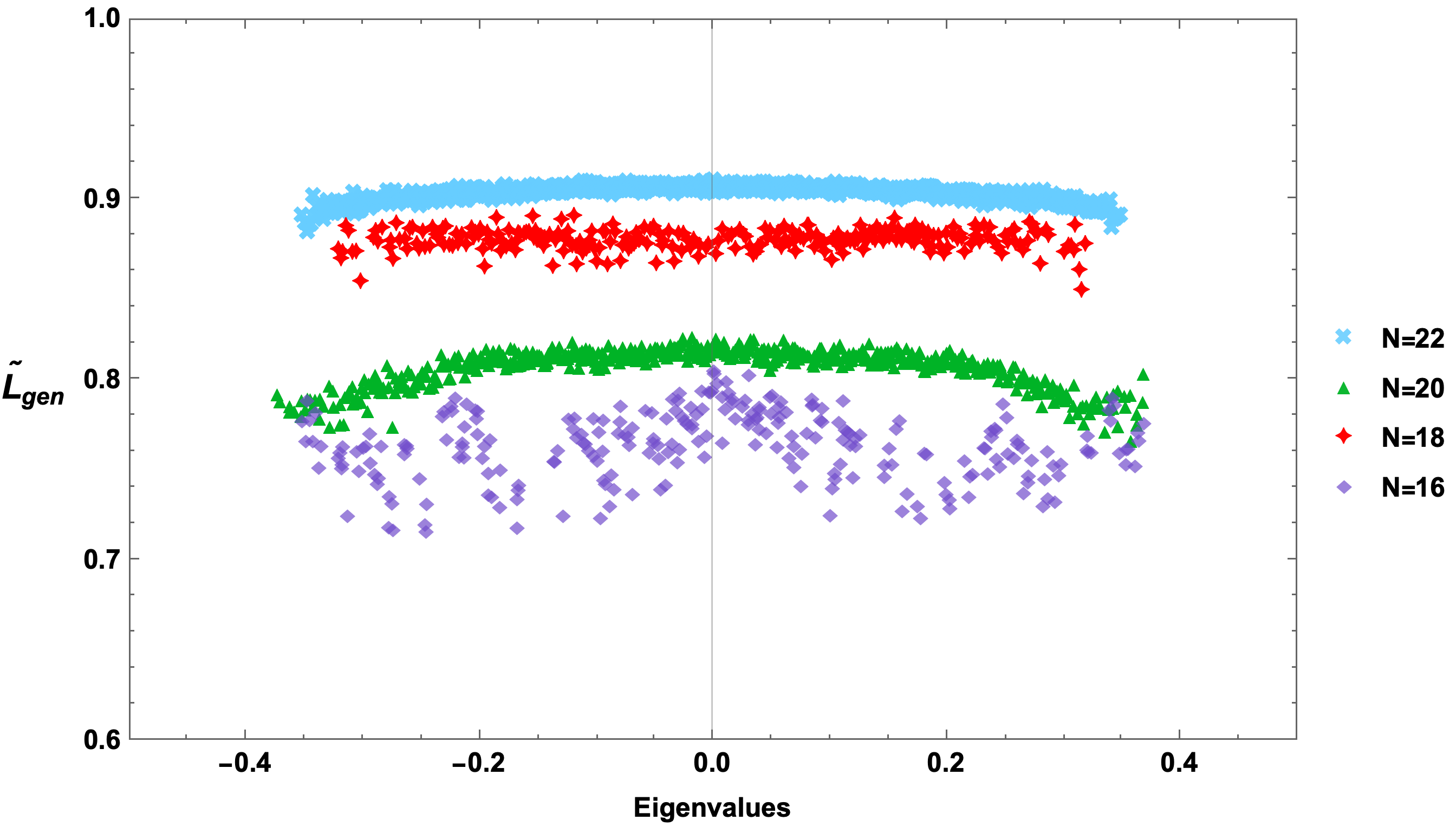}
    \caption{$p=0.01$ }\label{SPSYKeig2}
  \end{subfigure}\quad
  \begin{subfigure}{.3\linewidth}
\includegraphics[height=3.5cm,width=\linewidth]{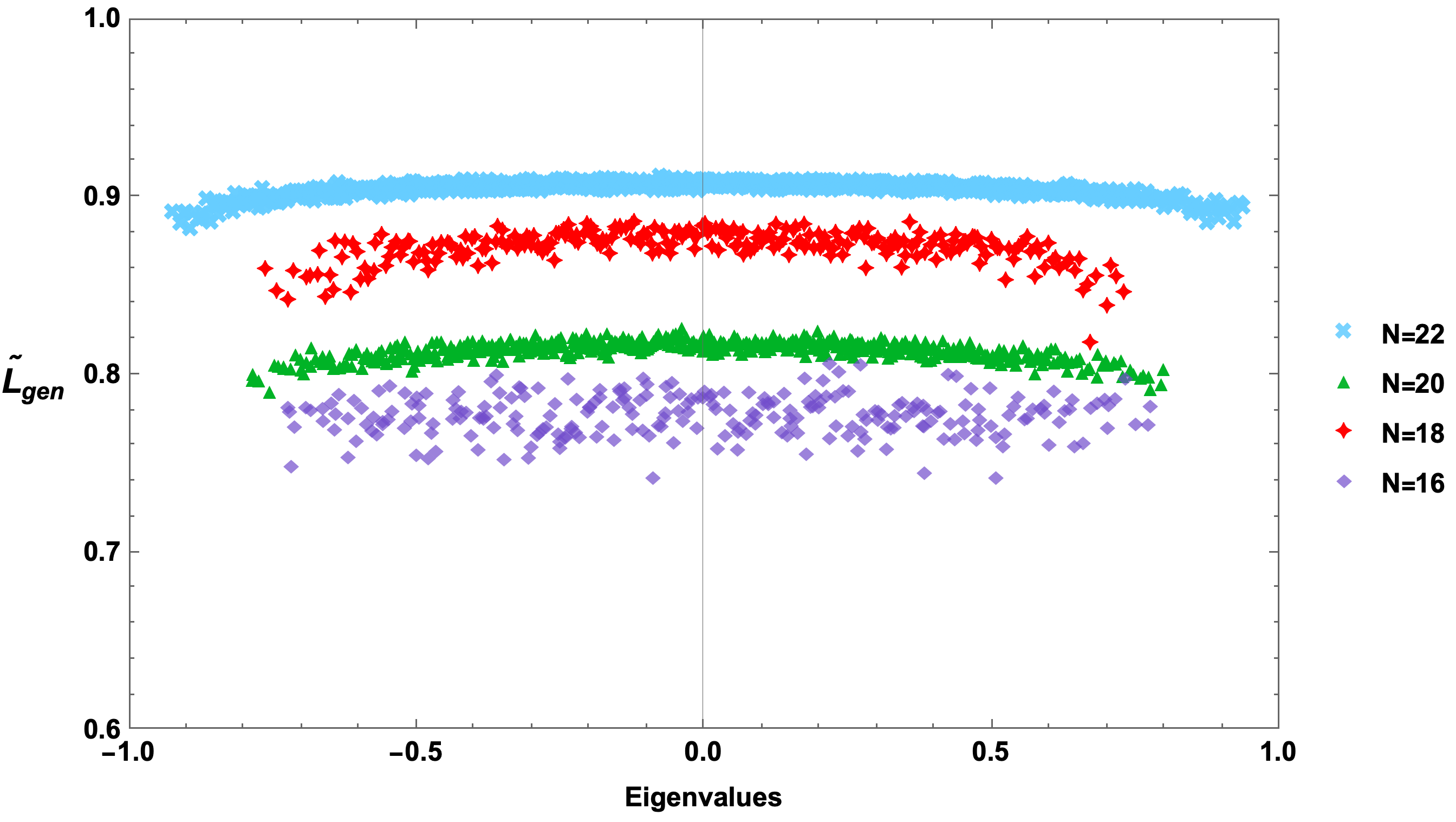}
    \caption{$p=0.05$ }\label{SPSYKeig3}
  \end{subfigure}
\caption{\footnotesize{
Distribution of the generalized L-entropy across the energy eigenstates of the sparse $SYK_4$ model for $N = 16, 18,$ and $20$ (8, 9, and 10 qubits). The plots show the transition in the distribution shape—from a dip to a central peak—as the sparsity parameter $p$, controlling the probability of nonzero couplings, is varied.}}
  \label{SPSYKeig}
\end{figure}

Finally, \Cref{SPSYKeig} illustrates the evolution of the eigenstate distribution of the generalized L-entropy with varying deformation parameter $p$. For very small $p$ and small system size $n$, the generalized L-entropy exhibits a larger vertical spread, as shown in \Cref{SPSYKeig1}, even though the eigenvalues and eigenvectors appear more clustered. We attribute this behavior to the combined effect of the small deformation parameter and the limited Hilbert-space dimension, in contrast to larger $n$ where such features are absent. As $p$ increases, the bunching gradually disappears, and the vertical spread transitions into a horizontal one, resulting in a broader distribution band, as evident from \Cref{SPSYKeig2} and \Cref{SPSYKeig3}. 

\subsection{\textbf{$\mathcal{N}=2$ Supersymmetric SYK Model}}

In this section, we evaluate the generalized L-entropy to investigate the multipartite entanglement of fortuitous states~\cite{Chang:2024lxt}. The $\mathcal{N}=2$ supersymmetric (SUSY) SYK model provides a particularly rich theoretical laboratory for this purpose, as it has been shown that all of its BPS states are fortuitous~\cite{Chang:2024lxt}. The model is defined by a pair of conjugate supercharges, $Q$ and $Q^\dag$, constructed from $N$ complex fermions, $\psi^i$ and $\bar{\psi}_i$ ($i=1, \dots, N$):
\begin{align}
    Q = i \sum_{1 \le i<j<k \le N} C_{ijk} \psi^i \psi^j \psi^k, \quad Q^\dag = i \sum_{1 \le i<j<k \le N} \overline{C}^{ijk} \bar{\psi}_i \bar{\psi}_j \bar{\psi}_k.
\end{align}
Here, the couplings $C_{ijk}$ are independent complex random variables drawn from a Gaussian distribution with zero mean and variance $\langle C_{ijk}\overline{C}^{ijk} \rangle = \frac{2J}{N^2}$. In our convention, we identify $\bar{\psi}_i$ as the fermion creation operator and $\psi^i$ as the annihilation operator. The Fock space is constructed by acting with creation operators on the vacuum state $|0\rangle$, which is defined by the condition $\psi^i|0\rangle=0$ for all $i$. A basis for this space is given by the states:
\begin{align}
|\nu_1 \cdots \nu_N\rangle = (\bar{\psi}_1)^{\nu_1} (\bar{\psi}_2)^{\nu_2} \cdots (\bar{\psi}_N)^{\nu_N} |0\rangle\;\;, \qquad (\nu_i \in \{0, 1\}).
\end{align}
The model possesses a $U(1)_R$ symmetry, whose charge $q_R$ is naturally identified with the fermion number. Consequently, the state $|\nu_1 \cdots \nu_N\rangle$ has a charge $q_R = \sum_{i=1}^N \nu_i$, and the total Hilbert space decomposes into subsectors $\mathcal{H}_{q_R}$ of fixed $R$-charge.

Our investigation centers on comparing the multipartite entanglement of these fortuitous BPS states against two other distinct ensembles of states within the same charge sectors.

\textbf{BPS (Fortuitous) States:} These are the supersymmetric ground states of the model, defined as the zero-energy eigenstates of the Hamiltonian $H = \{Q, Q^\dag\}$. Mathematically, they correspond to the non-trivial elements of the supercharge cohomology, forming the vector space $\text{Ker}(Q)/\text{Im}(Q)$. It is known that these BPS states exist only within a specific range of charge sectors, namely $\frac{N-3}{2} \le q_R \le \frac{N+3}{2}$~\cite{Fu:2016vas,Chang:2024lxt}. For even $N$, the allowed sectors are $q_R = \frac{N}{2}, \frac{N \pm 2}{2}$, while for odd $N$, they are $q_R = \frac{N \pm 1}{2}, \frac{N \pm 3}{2}$. We construct an ensemble of \textit{typical BPS states} by taking random superpositions of the basis states spanning the Q-cohomology within a given sector $\mathcal{H}_{q_R}$:
\begin{align}
|\Psi_{\text{\tiny typical BPS}}\rangle = \frac{1}{M} \sum_{i} a_i |\phi_i \rangle, \quad \text{where} \quad |\phi_i\rangle \in \text{Ker}(Q)/\text{Im}(Q).
\end{align}
Here, $\{|\phi_i\rangle\}$ is a basis for the Q-cohomology, $a_i$ are coefficients drawn from a Gaussian random distribution, and $M$ is a normalization constant.

\textbf{Q-exact States:} These are states that are cohomologically trivial, as they lie in the image of the supercharge operator, $\text{Im}(Q)$. They provide a crucial point of comparison to the BPS states. We generate an ensemble of \textit{typical $Q$-exact states} within a sector of charge $q_R$ by acting with $Q$ on random states drawn from the precursor sector with charge $q_R+3$:
\begin{align}
|\Psi_{\text{\tiny typical Q-exact}}\rangle = \frac{1}{M} \sum_{\{\nu_i\}} b_{\{\nu_i\}} Q |\nu_1 \cdots \nu_N \rangle, \quad \text{where} \quad \sum_{i=1}^N \nu_i = q_R+3.
\end{align}
The coefficients $b_{\{\nu_i\}}$ are again random numbers.

\textbf{Typical States in Charge Sector:} To establish a benchmark for generic entanglement within a given charge sector, we also consider \textit{typical states}, which are Haar-random states drawn from the full Hilbert space of that sector, $\mathcal{H}_{q_R}$.

By comparing the generalized L-entropy across these three ensembles, we can investigate how the supersymmetry influence multipartite entanglement and whether this measure is sensitive to the underlying algebraic structure of the fortuitous states.

\begin{figure}[H]
  \centering
   \begin{subfigure}{.45\linewidth}
    \includegraphics[width=\linewidth]{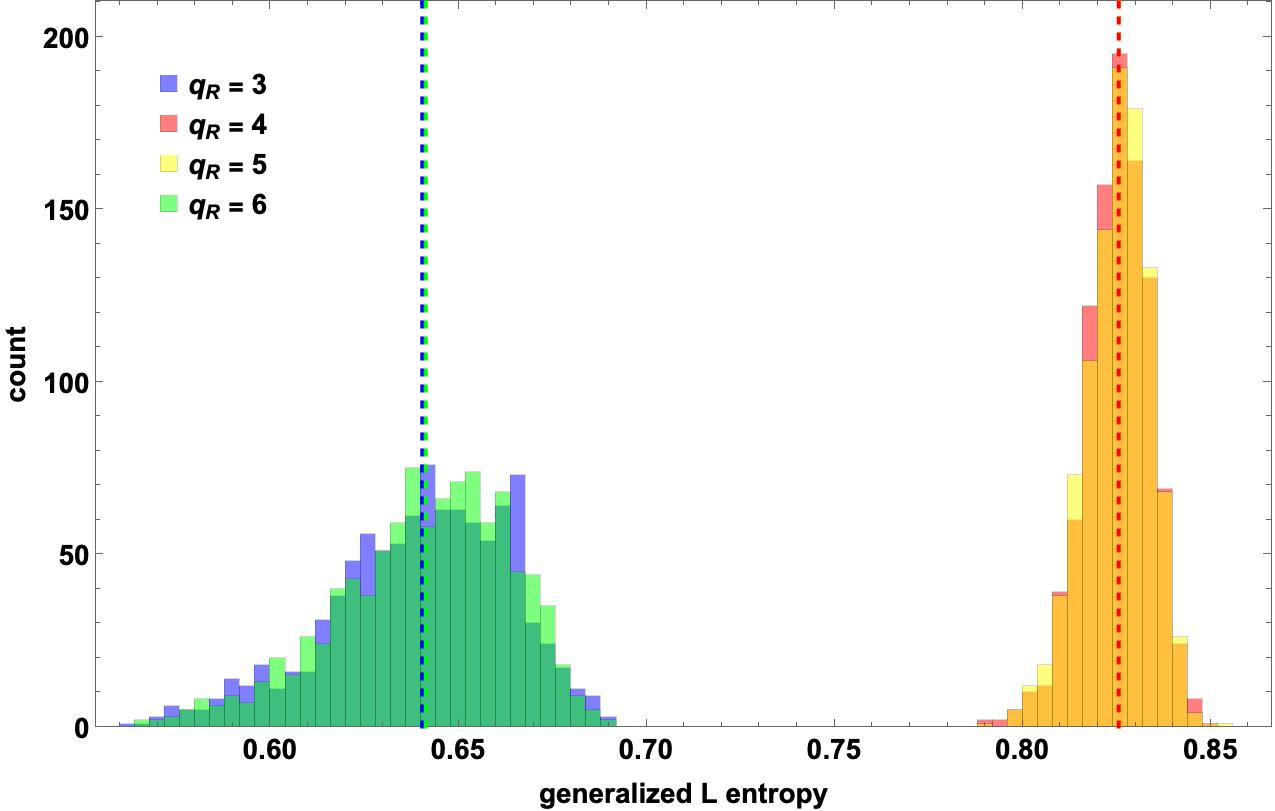}
    \caption{$N=9$}
    \end{subfigure}\quad
    \centering
    \begin{subfigure}{.45\linewidth}
    \includegraphics[width=\linewidth]{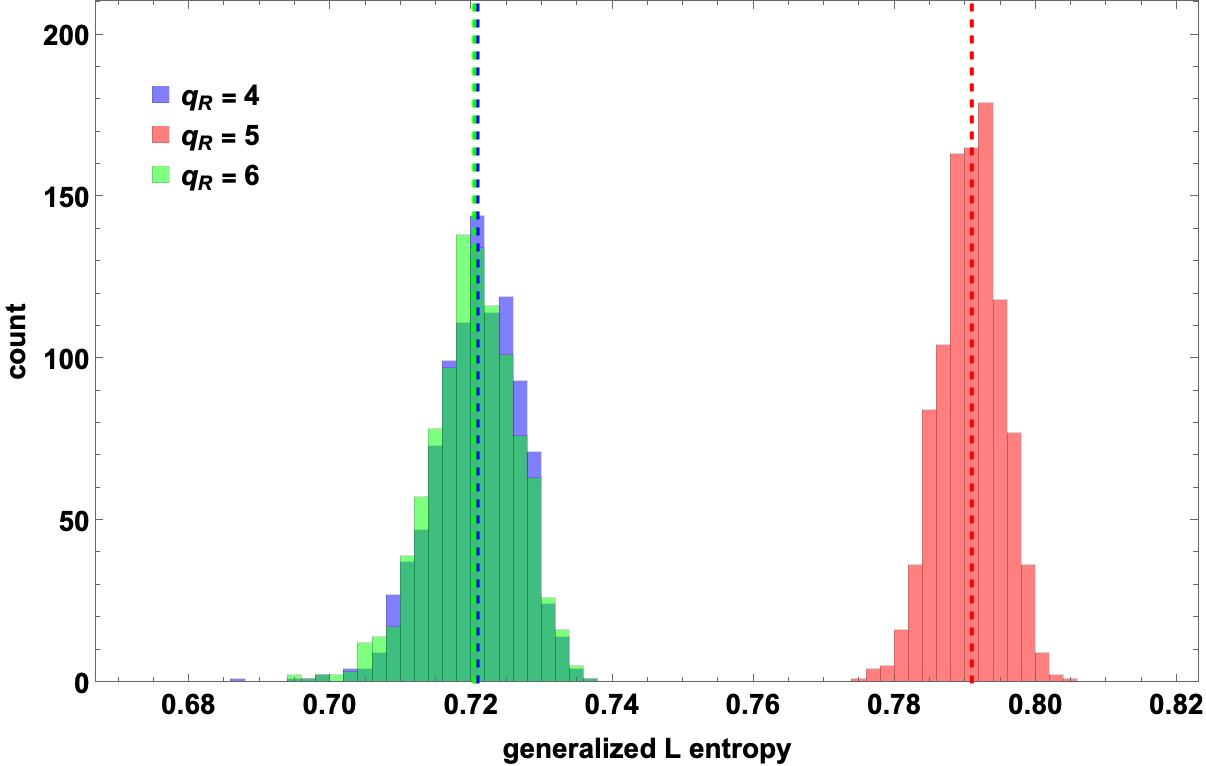}
    \caption{$N=10$}
    \end{subfigure}\quad
\caption{\footnotesize{
Distribution of the generalized L-entropy $\tilde{L}_{\text{gen}}$ for BPS states with $N$=9 and 10, shown across different $R$-charge sectors.}}\label{SUSYBPS}
\end{figure}

We begin by examining the multipartite entanglement properties of the BPS states alone. As shown in \Cref{SUSYBPS}, the distribution of the generalized L-entropy, $\tilde{L}_{\text{gen}}$, for typical BPS states exhibits a strong dependence on the $U(1)_R$ charge sector. For both $N=9$ and $N=10$, the generalized L-entropy is maximized in the central charge sectors (\textit{e.g.}, $q_R=4, 5$ for $N=9$ and $q_R=5$ for $N=10$). In contrast, BPS states in sectors towards the edges of the allowed BPS range, such as $q_R=3, 6$ for $N=9$ and $q_R=4, 6$ for $N=10$, display significantly lower values of $\tilde{L}_{\text{gen}}$. This indicates that the multipartite entanglement structure of the BPS states is highly non-uniform across the spectrum of allowed charges.

\begin{figure}[h]
  \centering
  \begin{subfigure}{.45\linewidth}
\includegraphics[height=4.5cm,width=\linewidth]{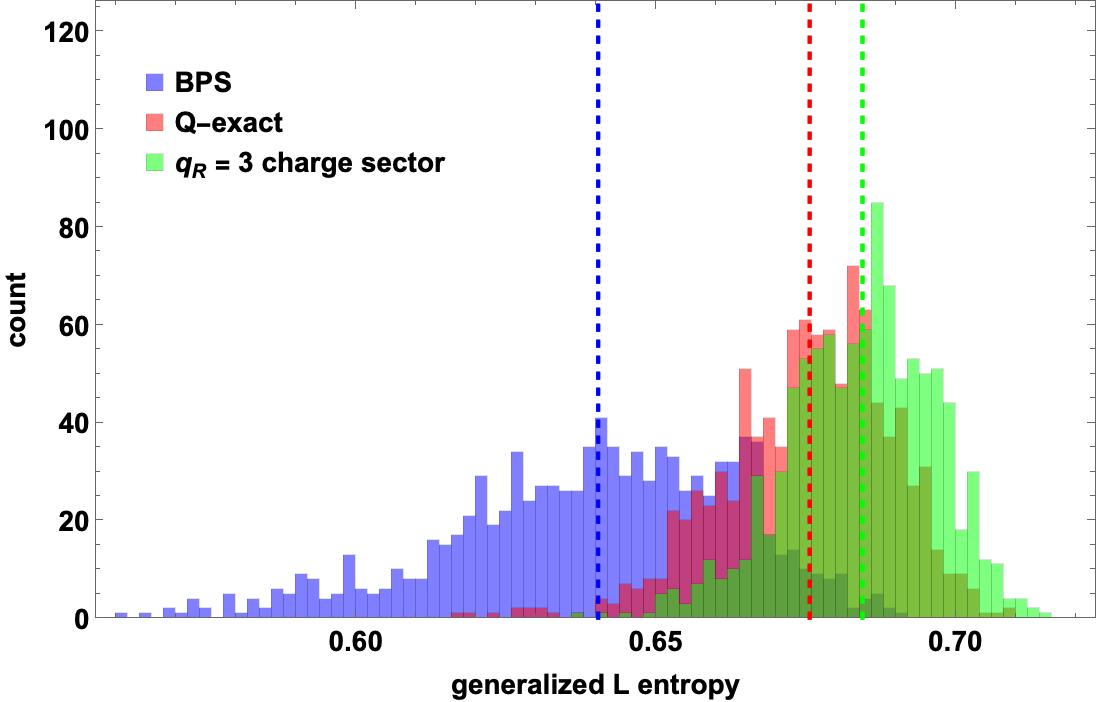}
    \caption{$N=9$, $q_R=3$}\label{SUSY93}
  \end{subfigure}\quad
  \begin{subfigure}{.45\linewidth}
    \includegraphics[height=4.5cm,width=\linewidth]{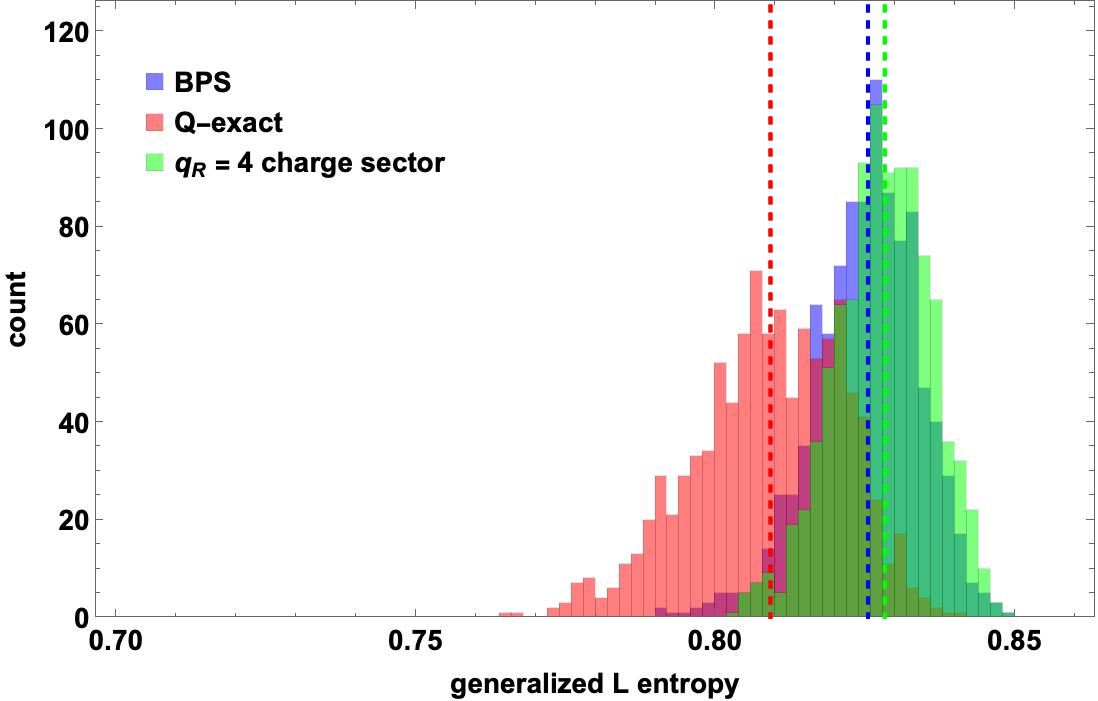}
    \subcaption{$N=9$, $q_R=4$}\label{SUSY94}
  \end{subfigure}\quad\\
  \begin{subfigure}{.45\linewidth}
\includegraphics[height=4.5cm,width=\linewidth]{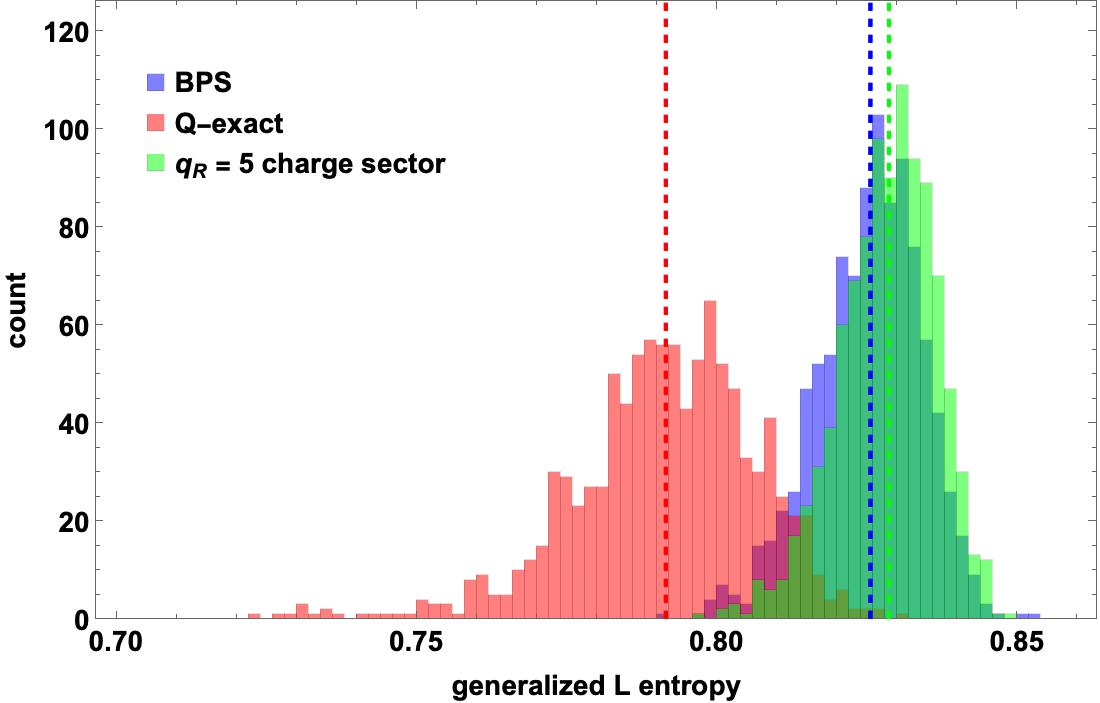}
    \caption{$N=9$, $q_R=5$}\label{SUSY95}
  \end{subfigure}\quad
   \begin{subfigure}{.45\linewidth}
    \includegraphics[height=4.5cm,width=\linewidth]{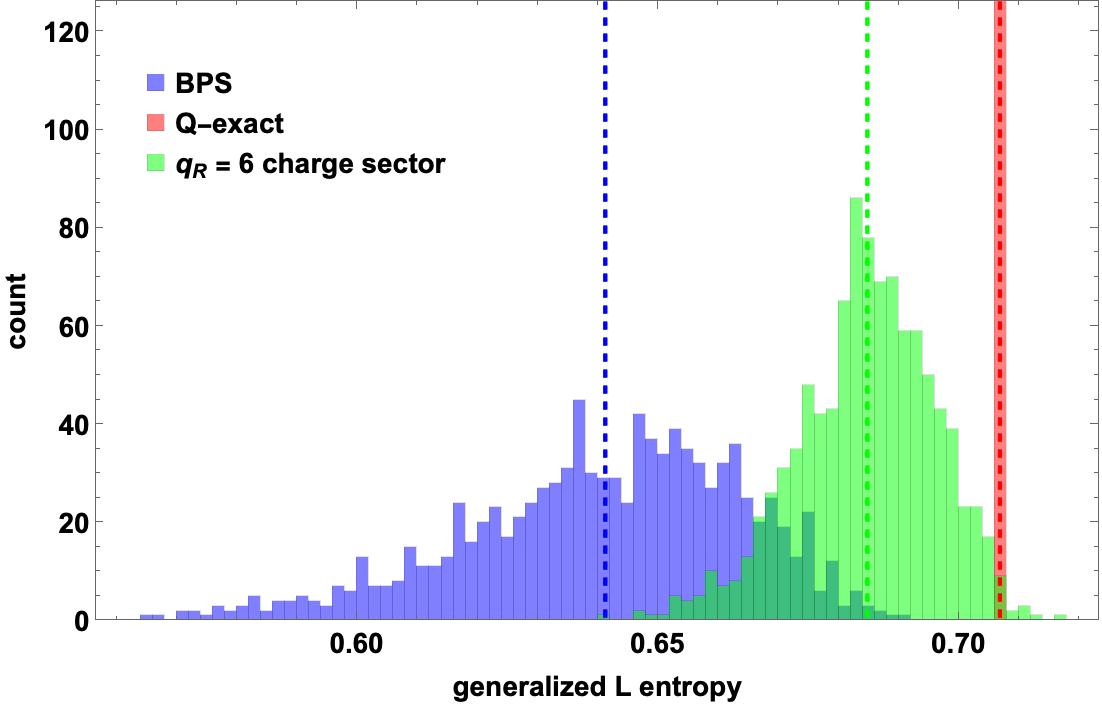}
    \subcaption{$N=9$, $q_R=6$}\label{SUSY96}
  \end{subfigure}\quad
   \caption{\footnotesize{ Distribution of the generalized L-entropy $\tilde{L}_{\text{gen}}$ of typical BPS states, typical $Q$-exact states, and typical states in the $R$-charge sector for $N=9$.}}\label{SUSY9}
\end{figure}

\begin{figure}[h]
  \centering
  \begin{subfigure}{.3\linewidth}
    \includegraphics[height=3.5cm,width=\linewidth]{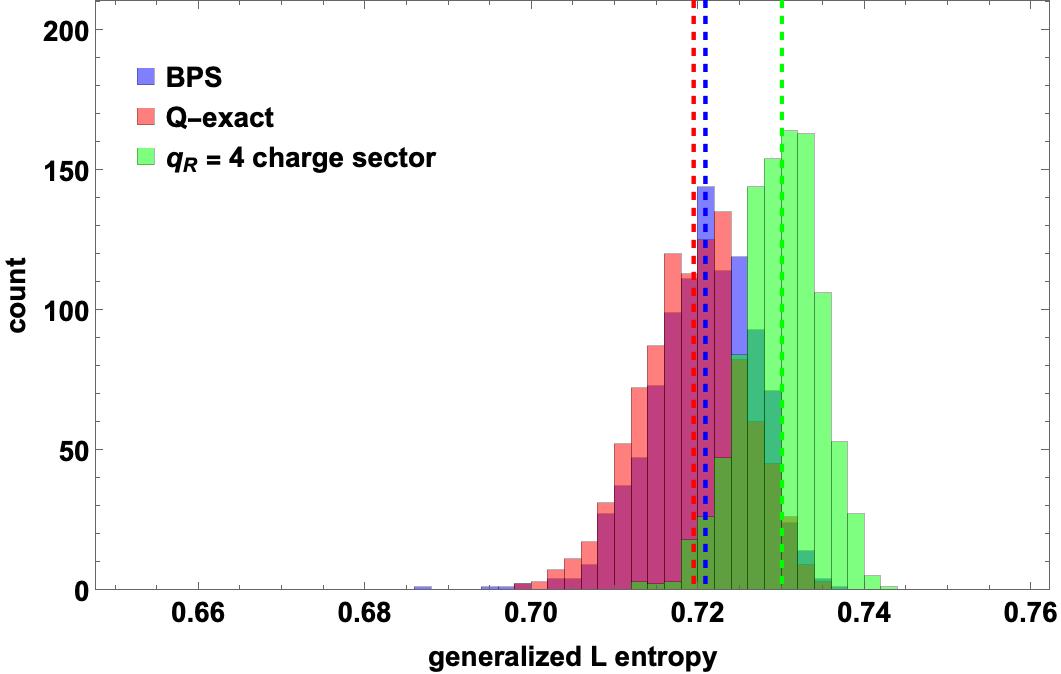}
    \caption{$N=10$, $q_R=4$}\label{SUSY104}
  \end{subfigure}\quad
  \begin{subfigure}{.3\linewidth}
 \includegraphics[height=3.5cm,width=\linewidth]{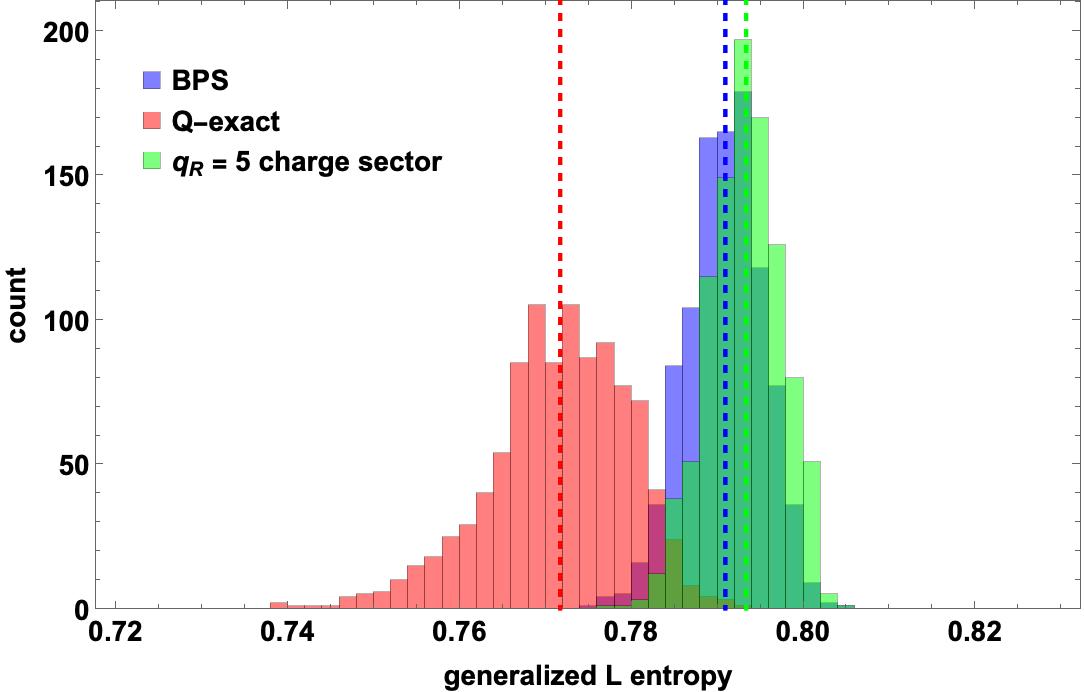}
   \caption{$N=10$, $q_R=5$}\label{SUSY105}
  \end{subfigure}\quad
  \begin{subfigure}{.3\linewidth}
 \includegraphics[height=3.5cm,width=\linewidth]{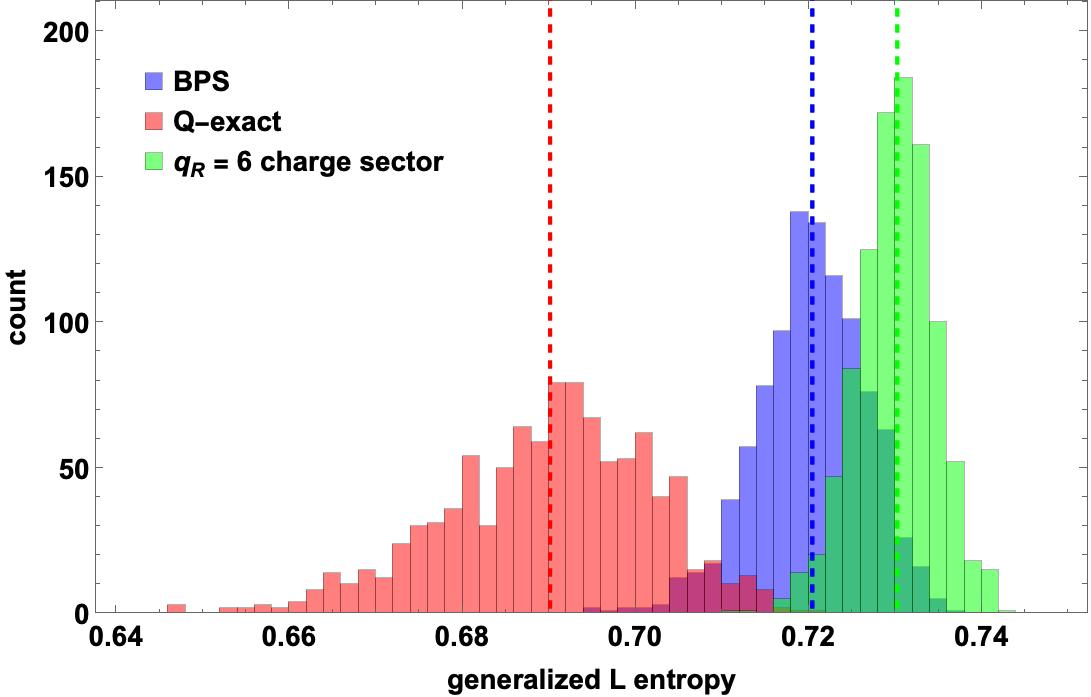}
    \caption{$N=10$, $q_R=6$}\label{SUSY106}
  \end{subfigure}
\caption{\footnotesize{
Distribution of the generalized L-entropy $\tilde{L}_{\text{gen}}$ of typical BPS states, typical $Q$-exact states, and typical states in the $R$-charge sector for $N=10$.}}\label{SUSY10}
\end{figure}

A more detailed picture emerges from a comparative analysis of all three state ensembles, presented in \Cref{SUSY9} for $N=9$ and \Cref{SUSY10} for $N=10$. A clear and consistent hierarchy in the average value of the generalized L-entropy is observed across nearly all sectors:
\begin{align}
\tilde{L}_{\text{gen}}(\text{typical}) \gtrsim \tilde{L}_{\text{gen}}(\text{BPS}) > \tilde{L}_{\text{gen}}(\text{Q-exact}).
\end{align}
The distributions for each class of states are largely distinct, suggesting that they represent structurally different ensembles of multipartite entanglement. The generalized L-entropy thus serves as an effective physical observable capable of distinguishing states based on their cohomological properties. The fact that typical states, which include all excited states within a sector, generally possess the largest value of the generalized L-entropy is expected, as ground states are often less entangled than highly excited states (See \Cref{Lenteig11SYK4}). More significantly, the generalized L-entropy robustly separates the cohomologically nontrivial BPS states from the trivial Q-exact states, with the former being consistently more entangled.

A notable exception to this hierarchy appears in the $q_R=6$ sector for $N=9$ (\Cref{SUSY96}), where the Q-exact state exhibits a larger value of the generalized L-entropy than the BPS states. This would be a finite-size artifact. The precursor states for this Q-exact state reside in the sector $\mathcal{H}_{q_R=9}$. For $N=9$, this sector is one-dimensional, containing only the fully filled state. Consequently, there is only a single $Q$-exact state in the $q_R=6$ sector, and its properties are not statistically representative of an ensemble.

The most crucial physical insight arises not just from the existence of this hierarchy, but from the magnitude of the separation between the distributions. A careful examination of \Cref{SUSY9} and \Cref{SUSY10} reveals that the gap between the average value of the generalized L-entropy of BPS states and that of typical states is smallest in the central charge sectors. For $N=10$, this gap is visibly smaller in the $q_R=5$ sector compared to the $q_R=4$ and $q_R=6$ sectors. Similarly, for $N=9$, the gap is smaller in the $q_R=4$ and $q_R=5$ sectors than in the $q_R=3$ sector.

Our analysis culminates in a key observation: the multipartite entanglement of fortuitous BPS states is not uniform but is instead strongly dependent on the $U(1)_R$ charge sector. The most striking result is found in the central charge sectors, specifically for $\frac{N-1}{2} \le q_R \le \frac{N+1}{2}$, where the generalized L-entropy of these fortuitous states approaches that of typical, Haar-random states. This indicates that, far from being simple, low-entanglement ground states, the fortuitous states in these sectors possess a strong multipartite entanglement structure that is representative of the sector as a whole. This finding has significant implications for the holographic interpretation of these states as microstates of an extremal black hole. It suggests that the black hole's entropy is predominantly accounted for by microstates that are not only supersymmetric but also exhibit near-maximal levels of genuine multipartite entanglement, mirroring the properties of a generic, highly chaotic quantum system.



\section{\textbf{Discussion}}\label{sec6}

In this work, we have explored the structure of genuine multipartite entanglement (GME) for pure states of generic quantum systems composed of an arbitrary number of subsystems. Extending the notion of latent entropy, previously proposed as a measure of GME in \cite{Basak:2024uwc}, we have formulated a consistent generalization applicable to any multipartite configuration. The generalized latent entropy of a $n$-party system is constructed from the reflected entropy obtained via  purification of $k$-party reduced density matrices ($2\leq k\leq \text{Max}\left[ 2, \lfloor\frac{n}{2}\rfloor\right]$) and their corresponding upper bounds. By systematically extending the bipartite construction of reflected entropy to appropriate multipartite partitions and their respective purifications, this framework captures the intrinsic organization of multipartite entanglement. For systems with three, four, and five parties, the generalized latent entropy reproduces the earlier latent entropy of \cite{Basak:2024uwc} exactly, thereby confirming the consistency and natural continuity of the construction. The generalized latent entropy satisfies all key axioms required of a measure of genuine multipartite entanglement: it vanishes for separable states, remains invariant under local unitary transformations, and is non-increasing under local operations and classical communication (LOCC). These properties firmly establish it as a valid and physically meaningful quantifier of multipartite entanglement.

We have examined the behavior of the generalized latent entropy for a range of highly entangled quantum states, with particular focus on absolutely maximally entangled (AME) and $k$-uniform states. AME states occupy a distinguished position in quantum information theory, as every $\lfloor n/2 \rfloor$-party reduced density matrix is maximally mixed, implying a vanishing reflected entropy across any bipartition and maximal entanglement entropy for all subsystems. Consequently, the generalized latent entropy reaches its theoretical maximum, identifying AME states as those with maximal multipartite entanglement. Extending this analysis to $k$-uniform states, we find that the generalized latent entropy increases monotonically with the degree of uniformity, faithfully capturing the hierarchy of multipartite entanglement. This hierarchical ordering has been explicitly verified for systems involving six to ten qubits, where the measure successfully distinguishes between different uniformity classes and even differentiates inequivalent states within the same class. Such fine-grained sensitivity underscores its potential as a refined tool for classifying multipartite entanglement.

Subsequently, we have undertaken both analytical and numerical evaluations of the generalized latent entropy for Haar random pure states comprising an arbitrary number of parties. Whereas earlier analyses of latent entropy were limited to systems of up to five subsystems, the present generalization enables computation for any number of parties. The analytical treatment, based on a large local Hilbert space dimension $d$, allows the measure to be systematically expanded in powers of $1/d$. Within this framework, we identify two distinct regimes: for Haar random states with an odd number of parties, the generalized latent entropy saturates its theoretical upper bound, while for even party configurations it remains below this limit but approaches it asymptotically as the number of parties increases. These analytical predictions are fully supported by extensive numerical simulations, wherein ensemble-averaged values computed from Haar random samples exhibit excellent agreement and rapid convergence in the large $d$ regime.

Finally, we have employed the generalized latent entropy to investigate multipartite correlations across several variants of the Sachdev–Ye–Kitaev (SYK) model. In the interacting $SYK_4$ model, the saturation value of the measure increases with system size and approaches that of absolutely maximally entangled (AME) states, whereas in the quadratic $SYK_2$ model it remains consistently lower and appears to asymptote to a smaller constant. The gap between their ensemble-averaged saturation values widens with $N$ and tends to stabilize at a finite value for large systems. In the mass-deformed SYK model, the generalized L-entropy rapidly attains the saturation value characteristic of the $SYK_4$ model as soon as the deformation parameter departs from the purely $SYK_2$ limit. The corresponding saturation time and late-time behavior show a pronounced dependence on the deformation strength.  In contrast, the sparse SYK model displays a smooth crossover in the saturation values of the generalized latent entropy as sparseness is varied, accompanied by a qualitative reorganization in its eigenstate distribution. Finally, in the $\mathcal{N}=2$ SUSY SYK model, typical fortuitous BPS states in the central $R$-charge sector exhibits higher multipartite entanglement compared to typical $Q$-exact states while the BPS state in other charge sector has relatively small L-entropy. This potentially indicates the hierarchy of multipartite entanglement structure of the holographic dual BPS black holes.

The construction of generalized latent entropy suggests several compelling directions for future research spanning quantum information, many-body physics, and holography. Since the generalized latent entropy naturally detects absolutely maximally entangled and $k$-uniform states, developing optimization techniques based on it could provide an analytic route to discovering new classes of highly entangled states with potential utility in quantum technologies. Extending the construction to conformal field theories may uncover universal structures in multipartite correlations, while a holographic realization could reveal deep connections with the entanglement cone or identify a bulk dual quantity capturing genuine multipartite entanglement. Moreover, it would be interesting to investigate potential connections between the generalized latent entropy and other multipartite entanglement or correlation measures in holographic settings \cite{Gadde:2022cqi,Mori:2024gwe,Mori:2025gqe}. Together, these possibilities highlight the broader role of the generalized latent entropy as a bridge between quantum information and holographic principles.

\section*{\textbf{Acknowledgment}}

B.A., V.M., G.K. and J.Y. was supported by the National Research Foundation of Korea (NRF) grant funded by the Korean government (MSIT) (RS-2022-NR069038) and by the Brain Pool program funded by the Ministry of Science and ICT through the National Research Foundation of Korea (RS-2023-00261799).
B. A. was supported by Basic Science Research Program through the National Research Foundation of Korea funded by the Ministry of Education (NRF-2022R1I1A1A01064342). 
The research work of JKB is supported by the Brain Pool program funded by the Ministry of Science and ICT through the National Research Foundation of Korea (RS-2024-00445164).
This work was supported by the Basic Science Research Program through the National Research Foundation of Korea (NRF) funded by the Ministry of Science, ICT \& Future Planning (NRF-2021R1A2C1006791), the framework of international cooperation program managed by the NRF of Korea (RS-2025-02307394), the Creation of the Quantum Information Science R\&D Ecosystem (Grant No. RS-2023-NR068116) through the National Research Foundation of Korea (NRF) funded by the Korean government (Ministry of Science and ICT), the Gwangju Institute of Science and Technology (GIST) research fund (Future leading Specialized Resarch Project, 2025) and the Al-based GIST Research Scientist Project grant funded by the GIST in 2025. This research was also supported by the Regional Innovation System \& Education(RISE) program through the (Gwangju RISE Center), funded by the Ministry of Education(MOE) and the (Gwangju Metropolitan City), Republic of Korea.(2025-RISE-05-001)

\appendix
\section{\textbf{k-uniform states}} \label{app_A}
The states used in \Cref{tab:entropy-uniform} are expressed as follows~\cite{Goyeneche:2014xcq}.

\subsection*{\textbf{5 qubit states}}

\begin{align}
\left|\psi^{(5)}_1\right\rangle
&= \frac{1}{\sqrt{2}}\left(|00000\rangle+|11111\rangle\right) \nonumber\\
\left|\psi^{(5)}_2\right\rangle
&= \frac{1}{4}\left(
|00000\rangle+|10010\rangle+|01001\rangle+|10100\rangle+|01010\rangle-|11011\rangle-|00110\rangle-|11000\rangle \right. \nonumber\\
&\qquad\left.
-|11101\rangle-|00011\rangle-|11110\rangle-|01111\rangle-|10001\rangle-|01100\rangle-|10111\rangle+|00101\rangle
\right) \nonumber
\end{align}

\begin{align}
\left|\psi^{(5)}_3\right\rangle
&= \frac{1}{\sqrt{8}}\left(
|00000\rangle+|00011\rangle+|01101\rangle+|01110\rangle+|10101\rangle-|10110\rangle+|11000\rangle-|11011\rangle
\right) \nonumber
\end{align}

\subsection*{\textbf{6 qubit states}}

\begin{align}
\left|\psi^{(6)}_1\right\rangle
&= \frac{1}{\sqrt{2}}\left(|000000\rangle+|111111\rangle\right) \nonumber\\
\left|\psi^{(6)}_2\right\rangle
&= \frac{1}{\sqrt{8}}\left(
|000000\rangle+|001111\rangle+|110011\rangle+|111100\rangle+|101010\rangle+|100101\rangle+|011001\rangle+|010110\rangle
\right) \nonumber
\end{align}

\begin{align}
\left|\psi^{(6)}_3\right\rangle
&= \frac{1}{\sqrt{8}}\left(
|111111\rangle+|110000\rangle+|001100\rangle+|000011\rangle+|101010\rangle+|100101\rangle+|011001\rangle+|010110\rangle
\right) \nonumber
\end{align}

\begin{align}
\left|\psi^{(6)}_4\right\rangle
&= \frac{1}{4}\left(
|000110\rangle+|011100\rangle+|100000\rangle+|111010\rangle-|001001\rangle-|010011\rangle-|101111\rangle-|110101\rangle
\right. \nonumber\\
&\qquad\left.
+i\left(
|000101\rangle+|010000\rangle+|101100\rangle+|111001\rangle-|001010\rangle-|011111\rangle-|100011\rangle-|110110\rangle
\right)
\right) \nonumber
\end{align}
\begin{align}
\left|\psi^{(6)}_5\right\rangle
&= \frac{1}{4}\left(
|000000\rangle-|110000\rangle-|001100\rangle-|000011\rangle-|001111\rangle-|110011\rangle-|111100\rangle+|111111\rangle
\right. \nonumber\\
&\qquad\left.
+|010101\rangle+|010110\rangle+|011010\rangle+|100110\rangle-|011001\rangle-|100101\rangle-|101001\rangle-|101010\rangle
\right) \nonumber
\end{align}

\subsection*{\textbf{7 qubit states}}

\begin{align}
\left|\psi^{(7)}_1\right\rangle
&= \frac{1}{\sqrt{2}}\left(|0000000\rangle+|1111111\rangle\right) \nonumber\\
\left|\psi^{(7)}_2\right\rangle
&= \frac{1}{\sqrt{8}}\left(
|1111111\rangle+|0101010\rangle+|1001100\rangle+|0011001\rangle
\right. \nonumber\\
&\qquad\left.
+|1110000\rangle+|0100101\rangle+|1000011\rangle+|0010110\rangle
\right) \nonumber
\end{align}

\subsection*{\textbf{8 qubit states}}

\begin{align}
\left|\psi^{(8)}_1\right\rangle
&= \frac{1}{\sqrt{2}}\left(|00000000\rangle+|11111111\rangle\right) \nonumber\\
\left|\psi^{(8)}_2\right\rangle
&= \frac{1}{\sqrt{12}}\left(
|00000000\rangle+|00001111\rangle+|01010101\rangle+|10100110\rangle
\right. \nonumber\\
&\qquad
+|10011100\rangle+|10110001\rangle+|11000011\rangle+|01101000\rangle \nonumber\\
&\qquad\left.
+|00111011\rangle+|01110110\rangle+|11011010\rangle+|11101101\rangle
\right) \nonumber\\
\left|\psi^{(8)}_3\right\rangle
&= \frac{1}{4}\left(
|00000000\rangle+|00110011\rangle+|11001100\rangle+|00111100\rangle
\right. \nonumber\\
&\qquad
+|11000011\rangle+|00001111\rangle+|11110000\rangle+|11111111\rangle \nonumber\\
&\qquad
+|01010101\rangle+|10101010\rangle+|01011010\rangle+|10100101\rangle \nonumber\\
&\qquad\left.
+|01101001\rangle+|10010110\rangle+|01100110\rangle+|10011001\rangle
\right) \nonumber
\end{align}

\begin{align}
\left|\psi^{(8)}_4\right\rangle
&= \frac{1}{8}\left(
|00000000\rangle+|00011100\rangle-|00100100\rangle+|00111000\rangle
\right. \nonumber\\
&\qquad
+|00000011\rangle+|00011111\rangle-|00100111\rangle+|00111011\rangle \nonumber\\
&\qquad
-|11000001\rangle-|11011101\rangle+|11100101\rangle-|11111001\rangle \nonumber\\
&\qquad
+|11000010\rangle+|11011110\rangle-|11100110\rangle+|11111010\rangle \nonumber\\
&\qquad
-|00000101\rangle-|00011001\rangle-|00100001\rangle+|00111101\rangle \nonumber\\
&\qquad
+|00000110\rangle+|00011010\rangle+|00100010\rangle-|00111110\rangle \nonumber\\
&\qquad
+|11000100\rangle+|11011000\rangle+|11100000\rangle-|11111100\rangle \nonumber\\
&\qquad
+|11000111\rangle+|11011011\rangle+|11100011\rangle-|11111111\rangle \nonumber\\
&\qquad
-|01001100\rangle+|01010000\rangle+|01101000\rangle+|01110100\rangle \nonumber\\
&\qquad
+|01001111\rangle-|01010011\rangle-|01101011\rangle-|01110111\rangle \nonumber\\
&\qquad
-|10001101\rangle+|10010001\rangle+|10101001\rangle+|10110101\rangle \nonumber\\
&\qquad
-|10001110\rangle+|10010010\rangle+|10101010\rangle+|10110110\rangle \nonumber\\
&\qquad
-|01001001\rangle+|01010101\rangle-|01101101\rangle-|01110001\rangle \nonumber\\
&\qquad
-|01001010\rangle+|01010110\rangle-|01101110\rangle-|01110010\rangle \nonumber\\
&\qquad
-|10001000\rangle+|10010100\rangle-|10101100\rangle-|10110000\rangle \nonumber\\
&\qquad\left.
+|10001011\rangle-|10010111\rangle+|10101111\rangle+|10110011\rangle
\right) \nonumber
\end{align}

\subsection*{\textbf{9 qubit states}}

\begin{align}
\left|\psi^{(9)}_1\right\rangle
&= \frac{1}{\sqrt{2}}\left(|000000000\rangle+|111111111\rangle\right) \nonumber\\
\left|\psi^{(9)}_2\right\rangle
&= \frac{1}{\sqrt{12}}\left(
|000000000\rangle+|100011101\rangle+|010001110\rangle+|101000111\rangle
\right. \nonumber\\
&\qquad
+|110100011\rangle+|011010001\rangle+|101101000\rangle+|111011010\rangle \nonumber\\
&\qquad\left.
+|011101101\rangle+|001110110\rangle+|000111011\rangle+|110110100\rangle
\right) \nonumber
\end{align}

\begin{align}
\left|\psi^{(9)}_3\right\rangle
&= \frac{1}{4}\left(
|100000000\rangle+|111111111\rangle+|111000011\rangle+|100111100\rangle
\right. \nonumber\\
&\qquad
+|101100110\rangle+|110011001\rangle+|101011010\rangle+|110100101\rangle \nonumber\\
&\qquad
+|011001100\rangle+|000110011\rangle+|011110000\rangle+|000001111\rangle \nonumber\\
&\qquad\left.
+|010101010\rangle+|001010101\rangle+|010010110\rangle+|001101001\rangle
\right) \nonumber
\end{align}

\subsection*{\textbf{10 qubit states}}

\begin{align}
\left|\psi^{(10)}_1\right\rangle
&= \frac{1}{\sqrt{2}}\left(|0000000000\rangle+|1111111111\rangle\right) \nonumber\\
\left|\psi^{(10)}_2\right\rangle
&= \frac{1}{\sqrt{12}}\left(
|0000000000\rangle+|0100011101\rangle+|1010001110\rangle+|1101000111\rangle
\right. \nonumber\\
&\qquad
+|0110100011\rangle+|1011010001\rangle+|1101101000\rangle+|1110110100\rangle \nonumber\\
&\qquad\left.
+|0111011010\rangle+|0011101101\rangle+|0001110110\rangle+|1000111011\rangle
\right) \nonumber\\
\left|\psi^{(10)}_3\right\rangle
&= \frac{1}{4}\left(
|1111111111\rangle+|1010101010\rangle+|0011001100\rangle+|0110011001\rangle
\right. \nonumber\\
&\qquad
+|0011110000\rangle+|0110100101\rangle+|1111000011\rangle+|1010010110\rangle \nonumber\\
&\qquad
+|1100000000\rangle+|1001010101\rangle+|0000110011\rangle+|0101100110\rangle \nonumber\\
&\qquad\left.
+|0000001111\rangle+|0101011010\rangle+|1100111100\rangle+|1001101001\rangle
\right) \nonumber
\end{align}

\end{document}